\documentclass[altaffilletter,10pt,tightenlines,english,nofootinbib,eqsecnum,superscriptaddress,prd,showpacs,showkeys,floats,aps,amsmath,amssymb]{revtex4}
\usepackage[utf8]{inputenc}
\usepackage{babel}
\usepackage{color}
\usepackage{float}
\usepackage{amsmath}
\usepackage{amssymb}
\usepackage{amsfonts}
\usepackage{graphicx}
\usepackage{enumerate}
\usepackage{esint}
\usepackage{epstopdf}
\usepackage{tabularx}
\usepackage{multirow}
\usepackage[unicode=true,pdfusetitle,
 bookmarks=true,bookmarksnumbered=false,bookmarksopen=false,
 breaklinks=false,pdfborder={0 0 1},backref=false,colorlinks=false]
 {hyperref}

\makeatletter

\makeatletter
\def\l@subsubsection#1#2{}

\@ifundefined{textcolor}{}
{%
 \definecolor{BLACK}{gray}{0}
 \definecolor{WHITE}{gray}{1}
 \definecolor{RED}{rgb}{1,0,0}
 \definecolor{GREEN}{rgb}{0,1,0}
 \definecolor{BLUE}{rgb}{0,0,1}
 \definecolor{CYAN}{cmyk}{1,0,0,0}
 \definecolor{MAGENTA}{cmyk}{0,1,0,0}
 \definecolor{YELLOW}{cmyk}{0,0,1,0}
}

\newcolumntype{L}[1]{>{\hsize=#1\hsize\raggedright\arraybackslash}X}%
\newcolumntype{R}[1]{>{\hsize=#1\hsize\raggedleft\arraybackslash}X}%
\newcolumntype{C}[1]{>{\hsize=#1\hsize\centering\arraybackslash}X}%

\newcommand{\nn}{\nonumber}
\newcommand{\sign}{\mathrm{sign}}

\newcommand{\chicrit}{\sqrt{2/(3\kappa^2)}}

\makeatother

\begin{document}

\title{Interacting 3-form dark energy models: distinguishing interactions and avoiding the Little Sibling of the Big Rip}

%%%%%%%%%%%%%%%%%%%%%%%%%
\author{Jo\~ao Morais}

\email{jviegas001@ikasle.ehu.eus}

\affiliation{Department of Theoretical Physics, University of the Basque Country UPV/EHU, P.O. Box 644, 48080 Bilbao, Spain}

%%%%%%%%%%%%%%%%%%%%%%%%%
\author{Mariam Bouhmadi-L\'opez}

\email{mbl@ubi.pt (On leave of absence from UPV/EHU and IKERBASQUE)}

\address{Departamento de F\'{i}sica, Universidade da Beira Interior, Rua Marqu\^es D'\'Avila e Bolama,
6201-001 Covilh\~a, Portugal}

\affiliation{Centro de Matem\'atica e Aplica\c{c}\~oes da Universidade da Beira Interior (CMA-UBI), Rua Marqu\^es D'\'Avila e Bolama,
6201-001 Covilh\~a, Portugal}

\affiliation{Department of Theoretical Physics, University of the Basque Country UPV/EHU, P.O. Box 644, 48080 Bilbao, Spain}

\affiliation{IKERBASQUE, Basque Foundation for Science, 48011, Bilbao, Spain}

%%%%%%%%%%%%%%%%%%%%%%%%%
\author{K. Sravan Kumar}

\email{sravan@ubi.pt}

\address{Departamento de F\'{i}sica, Universidade da Beira Interior, Rua Marqu\^es D'\'Avila e Bolama,
6201-001 Covilh\~a, Portugal}

\affiliation{Centro de Matem\'atica e Aplica\c{c}\~oes da Universidade da Beira Interior (CMA-UBI), Rua Marqu\^es D'\'Avila e Bolama,
6201-001 Covilh\~a, Portugal}

%%%%%%%%%%%%%%%%%%%%%%%%%
\author{Jo\~ao Marto}

\email{jmarto@ubi.pt}

\address{Departamento de F\'{i}sica, Universidade da Beira Interior, Rua Marqu\^es D'\'Avila e Bolama,
6201-001 Covilh\~a, Portugal}

\affiliation{Centro de Matem\'atica e Aplica\c{c}\~oes da Universidade da Beira Interior (CMA-UBI), Rua Marqu\^es D'\'Avila e Bolama,
6201-001 Covilh\~a, Portugal}

%%%%%%%%%%%%%%%%%%%%%%%%%

\author{Yaser Tavakoli}

\email{yaser.tavakoli@ut.ac.ir}

\affiliation{Department of Physics, University of Tehran, 14395-547 Tehran, Iran}

\affiliation{School of Physics, Institute for Research in Fundamental Sciences (IPM), 19395-5531 Tehran, Iran}

\affiliation{Departamento de F\'isica, Universidade Federal do Esp\'irito Santo, Av. Fernando Ferrari 514, 29075-910 Vit\'oria - ES, Brazil}

%%%%%%%%%%%%%%%%%%%%%%%%%

\begin{abstract}
In this paper we consider 3-form dark energy (DE) models with interactions in the dark sector. We aim to distinguish the phenomenological interactions that are defined through the dark matter (DM) and the DE energy densities. We do our analysis mainly in two stages. In the first stage, we identify the non-interacting 3-form DE model which generically leads to an abrupt late-time cosmological event which is known as the little sibling of the Big Rip (LSBR). We classify the interactions which can possibly avoid this late-time abrupt event. We also study the parameter space of the model that is consistent with the interaction between DM and DE energy densities at present as indicated by recent studies based on BAO and SDSS data. In the later stage, we observationally distinguish those interactions using the statefinder hierarchy parameters $\{ S_{3}^{(1)}\,,\, S_{4}^{(1)}\} \,,\,\{ S_{3}^{(1)}\,,\, S_{5}^{(1)}\} .$ We also compute the growth factor parameter $\epsilon(z)$ for the various interactions we consider herein and use the composite null diagnostic (CND) $\{ S_{3}^{(1)}\,,\,\epsilon(z)\} $ as a tool to characterise those interactions by measuring their departures from the concordance model. In addition, we make a preliminary analysis of our model in light of the recently released data by SDSS~III on the measurement of the linear growth rate of structure.
\end{abstract}



\date{\today}

\maketitle

\tableofcontents

%%%%%%%%%%%%%%%%%%%%%%%%%%%%%%%%%%%%%%%%
%
%	Introduction
%
%%%%%%%%%%%%%%%%%%%%%%%%%%%%%%%%%%%%%%%%

\section{Introduction}

Despite the huge advancement in cosmology, we still face enormous theoretical problems like finding a suitable explanation of the current dark energy (DE) era; i.e. we know that the Universe started speeding up recently as predicted by SnIeA observations more than a decade ago \cite{Riess:1998cb,Perlmutter:1998np}, and afterwards confirmed by several types of cosmological and astrophysical observations (cf. for example Ref.~\cite{Ade:2015rim} for a recent account on this issue), but we do not know from a well grounded theoretical framework what is causing this acceleration \cite{amendola2010dark,Capozziello:2010zz}. The simplest approach is to assume a cosmological constant that started recently dominating the late-time energy density budget of the Universe but then the issue {\it{of why is it so tiny?}} and {\it{why this cosmological constant has begun to be important only right now?}} have to be addressed as well (see for example: \cite{Sahni:1999gb,Carroll:2000fy,RevModPhys.61.1,Padmanabhan:2002ji}). This has led to a great interest on exploring other avenues to explain the late-time acceleration of the cosmos invoking either an additional matter component in the Universe, we name DE \cite{Tsujikawa:2010sc,Bamba:2012cp}, or modifying appropriately the laws of gravity (for a recent account on this issue see \cite{Morais:2015ooa} and the extensive list of references provided therein). 

In this paper, we will follow the first approach corresponding to invoke a (dynamical) DE component. Therefore, one of the simplest options is to consider a scalar field. So far in nature, we have detected a single scalar field, the Higgs, which is too heavy to describe the current acceleration of the Universe. Now, it turns out that the notion of a scalar field can be embedded within what is known as $p$-forms \cite{GK09,Koivisto:2009sd,Germani:2009gg}; i.e. differential forms which are natural ``inhabitants'' of any geometrical structure and theory \cite{Wald2010}. In addition, $p$-forms have been used in cosmology for quite a long time, for example in string cosmology \cite{Copeland:1994km,Lukas:1996iq} and in the pre-big-bang scenario \cite{Gasperini:1998bm}. In a Universe like ours, i.e. four dimensional, 4 types of $p$-forms can live in where%
\footnote{\label{footnote1}While 4-forms can exist in a 4-dimensional space-time they cannot have dynamics on the standard way as its strength tensor would be a 5-form which by definition vanishes in a 4-dimensional space-time. Still a 4-form can be endowed with dynamics as it mimics a scalar field, see for example \cite{DasGupta:2009jy}.}
$p=0,1,2,3$. The first two options correspond to scalars and vector fields. If in addition, we incorporate the condition of homogeneity and isotropy of our Universe on large scales, then we need to restrict to 0-forms and 3-forms if we want to invoke a single differential form%
\footnote{In fact, while a massive $p$-form with $p=0,3$ involves a single degree of freedom, a $p$-form with $p=1,2$ involves three degrees of freedom \cite{GK09}. Please see footnote~\ref{footnote1} for a discussion of the dynamics of 4-forms.}
(cf. for example: \cite{PhysRevD.40.967, Bento:1992wy, ArmendarizPicon:2004pm, Golovnev:2008cf, Golovnev:2008hv, GK09} to see how to accommodate 1-forms and 2-forms in a Friedmann-Lema\^itre-Robertson-Walker (FLRW) Universe). Therefore, and for the reasons explained at the beginning of this paragraph, from now on, we restrict our analysis to 3-form fields.

The 3-form fields have been proven to be very useful not only for the late-time Universe but also for the early Universe (cf. \cite{Koivisto:2009fb}). 
In what refers to inflation, a single 3-form field minimally coupled to gravity has been introduced and studied in Refs. \cite{Koivisto:2009fb, Koivisto:2009ew}. The issue of ghosts and Laplacian instabilities was addressed in Ref. \cite{DeFelice:2012jt}, where a suitable choice of self-interacting potential for the 3-form field has been described; in addition, the same authors provided evidences that potentials possessing a manifest quadratic dominance, in the small field limit, would allow the production of sufficient oscillations for reheating \cite{DeFelice:2012wy}, without the presence of ghost instabilities. The model observational predictions, in the single \cite{Mulryne:2012ax} and multiple 3-form fields cases \cite{Kumar:2014oka,Kumar:2016tdn}, have been obtained and adequately fitted within the available current bounds of Planck data. Consequently, the idea of a 3-form field generating inflation has been strengthening and this also sustains a relevant motivation to consider this field in different cosmological contexts. %
Brane-world inflation driven by a confined 3-form field on a brane hypersurface was also considered in \cite{Barros:2015evi}.

Likewise, 3-forms can play the role of DE. In fact, almost forty years ago, it was shown that they could give rise to a cosmological constant \cite{DUFF1980179}. More recently, it was shown in \cite{Koivisto:2009ew} and \cite{Koivisto:2009fb} that a minimally coupled 3-form with a properly self-interacting potential can drive the present acceleration of the Universe. In general for a FLRW Universe filled with a self-interacting 3-form on top of the DM component, it was shown that a DM era can be successfully followed by a de Sitter attractor phase as given in \cite{Koivisto:2009ew} and \cite{Koivisto:2009fb}. On this work, we will show that aside from the late-time de Sitter behaviour which is present in 3-forms cosmology, the asymptotic future behaviour of the Universe on this kind of setup can correspond to a Little Sibling of the Big Rip (LSBR) \cite{Bouhmadi-Lopez:2014cca,Albarran:2015cda}. We remind that the LSBR is an abrupt cosmological behaviour where the Hubble rate blows up at an infinite cosmic time and for a very large scale factor while its cosmic time derivative remains finite \cite{Bouhmadi-Lopez:2014cca}. Although the divergence of the Hubble parameter happens in the asymptotic future, at finite time a LSBR leads to serious consequences for local bounded structures in our Universe. As discussed in Ref.~\cite{Bouhmadi-Lopez:2014cca}, in such a scenario any initially bounded gravitational system will eventually become unstable and disappear. This effect of dissociation of structure was found to be scale dependent, with the dissociation time being smaller for large scale structures, e.g. $\sim10^{13}$ years for the Coma cluster, than for structures of the size of the Solar system, for which the dissociation time, $\sim10^{19}$ years, is several orders of magnitude larger. In addition, on this work, we will identify an infinite past fixed point inherent to most of these models which was not previously identified and that correspond to a DM era.

In order to remove the LSBR that might be present on this type of models, we invoke an interaction on the dark sector, i.e. an interaction between DM and DE which we will consider to be up to quadratic order on the energy density of DM and/or DE. Different types of interactions between DM and 3-forms have been previously considered in \cite{Ngampitipan:2011se, Boehmer:2011tp,Koivisto:2012xm}. In particular, it has been proven that for some appropriate choices of the interaction the coincidence problem can be alleviated \cite{Ngampitipan:2011se}. In addition, the centre manifold theorem of dynamical systems was employed in \cite{Boehmer:2011tp} for a correct analysis of some fixed points in this type of models. The cosmological perturbations, at first order and within the Newtonian limit were analysed in \cite{Koivisto:2012xm}. 

For those interactions between DM and the 3-form that are able to remove the LSBR, we use cosmography \cite{Visser:2004bf, Capozziello:2008qc,Vitagliano:2009et, BouhmadiLopez:2010pp, Capozziello:2011tj, Bouhmadi-Lopez:2014jfa, Morais:2015ooa} or the almost equivalent statefinder approach \cite{Sahni:2002fz, Alam:2003sc,Arabsalmani:2011fz, Li:2014mua, Hu:2015bpa, Yin:2015pqa} to constrain them observationally. Cosmography is a very simple approach which relies on the assumption that the Universe is homogeneous and isotropic on large scale and no dynamical theory is assumed a priori and it is based in Taylor expanding the scale factor \cite{Visser:2004bf, Capozziello:2008qc}. The cosmographic parameters can be redefined in such a way that they are equal to unity for a $\Lambda$CDM and this is what is known as the statefinder hierarchy. It has the advantage of graphically being able to distinguish easily a DE model from the $\Lambda$CDM (although physically both setups carry the same information). For this reason, we have applied this method on this paper instead of using the cosmographic parameters as we did in some of our previous works \cite{BouhmadiLopez:2010pp, Bouhmadi-Lopez:2014jfa, Morais:2015ooa}. On the last part of the work, we also present a preliminary analysis of the cosmological perturbations of the 3-forms models we have studied. For each model we calculate the growth rate of matter perturbations and compare it with the predictions of $\Lambda$CDM, through the growth factor and the composite null diagnosis (CND) \cite{Yin:2015pqa}, and with the recent observational data of the Sloan Digital Sky Survey III (SDSS III) Baryon Oscillation Spectroscopic Survey (BOSS) Data Release 12 (DR12) \cite{Satpathy:2016tct}.

The paper is outlined as follows, in Sect~\ref{3-form introduction} we review the cosmology of 3-forms in absence of interactions. In Sect~\ref{LSBR_rising}, we explain how does a LSBR appears on this kind of models. In Sect~\ref{interacting 3form}, we present the evolution equation for 3-form DE models with an arbitrary interaction with DM. Afterwards, and following some physical criterion, we fix the choice of the potentials and interactions that we analyse. Afterwards, in Sect~\ref{Dynsys} a thorough dynamical system analysis is carried where all the fixed points and their stability are analysed. In particular, we present for the first time the fixed points that exist at infinite values of the 3-form field, which we found to correspond to the asymptotic past of the system. In Sect~\ref{Statefinder Hierarchy}, the statefinder hierarchy (very similar in spirit to the cosmographic approach) is used to constrain the model in presence and absence of interaction on the dark sector. Afterwards, we present in Sect~\ref{perturbations} a preliminary analysis of the cosmological perturbations of this model where the growth rate of matter perturbations is presented and compared with the recent observational SDSS~III data. In Sect~\ref{Conclusions} we present our conclusions. We include also two appendices, on the first one the statefinders expression for the analysed model are presented and on the second one the Hurwitz criterion for cubic polynomials is briefly presented which can be extremely useful for studying the stability of some critical points of the system analysed in Sect~\ref{Dynsys}.

%%%%%%%%%%%%%%%%%%%%%%%%%%%%%%%%%%%%%%%%
%
%	Reviewing the 3-form 
%
%%%%%%%%%%%%%%%%%%%%%%%%%%%%%%%%%%%%%%%%

\section{Reviewing the 3-form Field}

\label{3-form introduction}

In this section, we briefly review the 3-form field model introduced in \cite{Koivisto:2009sd,Koivisto:2009ew,Koivisto:2009fb}. Then, we write the corresponding field equations in a suitable cosmological background.
%%%%%%%%%%%%%%%%%%%%%%%%%%%%%%%%%%%%%%%%
%
%	The 3-form Action
%
%%%%%%%%%%%%%%%%%%%%%%%%%%%%%%%%%%%%%%%%

\subsection{The 3-form action}

The general action for the 3-form field $A_{\mu\nu\rho}$ minimally coupled to gravity and with a potential $V$ can be written as%
\footnote{Throughout this paper, we will use Greek indices for 4-dimensional quantities and Latin indices for 3-dimensional space quantities.%
} 
\begin{align}
	\label{N3-form action}
	S
	 =
	\int \mathrm{d}^{4}\mathbf{x}\sqrt{-g}\mathcal{L}
	=\int \mathrm{d}^{4}\mathbf{x}\sqrt{-g} \left[
		-\frac{1}{48} F^{\mu\nu\rho\sigma} F_{\mu\nu\rho\sigma}
		-V\left(A^{\mu\nu\rho}A_{\mu\nu\rho}\right)
	\right]
	\,.
\end{align}
Here, $g$ is the determinant of the metric, and $F_{\mu\nu\rho\sigma}$ is the strength tensor of the 3-form, defined as \cite{GK09} 
\begin{align}
	\label{N3f-Maxw-1}
	F_{\mu\nu\rho\sigma}\equiv4\nabla_{[\mu}A_{\nu\rho\sigma]}
	=
	\nabla_{\mu}A_{\nu\rho\sigma}
	-\nabla_{\sigma}A_{\mu\nu\rho}
	+\nabla_{\rho}A_{\sigma\mu\nu}
	-\nabla_{\nu}A_{\rho\sigma\mu}
	\,,
\end{align}
where the square brackets denote anti-symmetrisation \cite{Wald2010}. 

Minimising the action \eqref{N3-form action} with respect to variations of the 3-form field we obtain the equations of motion of the 3-form \cite{Koivisto:2009sd}
\begin{align}
	\label{general_eq_motion}
	\nabla_{\sigma}{F^{\sigma}}_{\mu\nu\rho}
		-12\frac{\partial \,V}{\partial 	\left(A^2\right)} A_{\mu\nu\rho}=0
		\,,
\end{align}
where we use the notation $A^2=A^{\alpha\beta\gamma}A_{\alpha\beta\gamma}$ introduced in Refs.~\cite{Koivisto:2009sd,Koivisto:2009ew,Koivisto:2009fb}.
Finally, the energy-momentum tensor of the 3-form obtained from the action \eqref{N3-form action} reads \cite{Koivisto:2009sd,GK09}
\begin{align}
	\label{energy_momentum_3form}
	{T}_{\mu\nu}
	\equiv
	\frac{-2}{\sqrt{-g}}\frac{\partial \sqrt{-g}\mathcal{L}}{\partial g^{\mu\nu}}
	=&~
	\frac{1}{6}  {F}_{\mu}^{\phantom{\mu}\alpha\beta\gamma} F_{\nu\alpha\beta\gamma} 
	+6\frac{\partial \,V}{\partial \left(A^2\right)} {A}_{\mu}^{\phantom{\mu}\alpha\beta} A_{\nu\alpha\beta}
	-\left[
		\frac{1}{48}F^2
		+V \left(A^2\right)
	\right] g_{\mu\nu}
	\,.
\end{align}

In general, any $p$-form field in $d-$dimensions has a dual $\left(d-p\right)-$form \cite{GK09,Mulryne:2012ax,Wongjun:2016tva}. In our case, the 3-form field $A_{\mu\nu\rho}$ and its field tensor $F_{\mu\nu\rho\sigma}$, which is a 4-form, are dual to a vector and a scalar field as described in \cite{Mulryne:2012ax}. Therefore, the 3-form field is similar to a non-canonical scalar field. In other words, a 3-form model can come under a subclass of a more general $k$-essence model. This duality could be lost if a non-minimal coupling to gravity is assumed for the 3-form \cite{Koivisto:2009fb}.

%%%%%%%%%%%%%%%%%%%%%%%%%%%%%%%%%%%%%%%%
%
%	The 3-form Cosmology
%
%%%%%%%%%%%%%%%%%%%%%%%%%%%%%%%%%%%%%%%%

\subsection{3-form cosmology}

\label{3-form Cosmology}

We now consider a spatially flat FLRW cosmology, described by the metric
\begin{align}
	\label{FLRW_metric}
	ds^{2}=-dt{}^{2}+a^{2}(t)d\vec{x}^2
	\,,
\end{align}
where $t$ is the cosmic time and $a(t)$ is the scale factor. In such a Universe, the 3-form field depends only on the cosmic time, hence only the space-like components will be dynamical%
\footnote{We will set all the non-dynamical components $A_{0ij}$ to zero.
}%
, with its non-zero components given by \cite{Koivisto:2009sd,Koivisto:2009ew,Koivisto:2009fb,GK09}
\begin{align}
	\label{NNZ-comp-1-1}
	A_{ijk}=a^{3}(t)\,\chi(t)\,\epsilon_{ijk}
	\,,
\end{align}
while the non-zero components of the strength tensor are
\begin{align}
	\label{F_0ijk_comp}
	F_{0ijk} = a^{3}(t)\left[\dot\chi(t)+3H(t)\chi(t)\right]\epsilon_{ijk}
	\,.
\end{align}
Here, $\epsilon_{ijk}$ is the standard 3-dimensional Levi-Civita symbol and $\chi(t)$ is the comoving scalar quantity associated with the 3-form field. From Eqs.~\eqref{NNZ-comp-1-1} and \eqref{F_0ijk_comp} we find that $A^{\mu\nu\rho}A_{\mu\nu\rho}=6\chi^{2}$, which allows us to write the potential as $V(\chi^2)$, and $F^{\mu\nu\rho\sigma}F_{\mu\nu\rho\sigma}=-24\left(\dot\chi+3H\chi\right)^2$.
Substituting Eq.~\eqref{NNZ-comp-1-1} in the equation of motion of the 3-form field, Eq.~\eqref{general_eq_motion}, we find the equation of motion of the field $\chi(t)$ \cite{Koivisto:2009sd,Koivisto:2009ew,Koivisto:2009fb,GK09}:
\begin{align}
	\label{NDiff-syst-1-1}
	\ddot{\chi}+3H\dot{\chi}+3\dot{H}\chi+V_{,\chi}=0
	\,,
\end{align}
where a dot represents a derivative with respect to the cosmic time, $H\equiv \dot{a}/a$ is the Hubble parameter.
The energy density and pressure of the 3-form, obtained from the diagonal elements of the energy-momentum tensor \eqref{energy_momentum_3form}, can be written in terms of the field $\chi$ and its derivatives as \cite{Koivisto:2009sd,Koivisto:2009ew,Koivisto:2009fb}
\begin{align}
	\label{energy_3form}
	\rho_\chi =&~ -{T^0}_0 = \frac{1}{2}\left(\dot\chi+3H\chi\right)^2 + V
	\,,
	\\
	\label{pressure_3form}
	P_\chi =&~ \frac{1}{3}{T^i}_i = -\frac{1}{2}\left(\dot\chi+3H\chi\right)^2 - V + \chi V_{,\chi}
	\,.
\end{align}
The parameter of the equation of state (EoS) of the 3-form is therefore
\begin{align}
	\label{3form_w}
	w_\chi = \frac{P_\chi}{\rho_\chi} = -1 + \frac{\chi V_{,\chi}}{\frac{1}{2}\left(\dot\chi+3H\chi\right)^2 + V}
	\,.
\end{align}
Notice that whenever the derivative of the potential vanishes, the 3-form mimics a cosmological constant behaviour. If $\chi V_{,\chi}<0$, i.e., for potentials decreasing in $\chi^2$, the 3-form field takes a phantom-like behaviour with $w<-1$. On this paper, we will restrict our analysis to non-negative potentials, therefore, even in the case of a phantom-like behaviour the energy density of the 3-form always respects $\rho_{\chi}\geq0$.

In a homogeneous and isotropic Universe with the metric \eqref{FLRW_metric} and filled only by a 3-form, the Friedmann and Raychaudhuri equations read, respectively,
\begin{align}
	\label{NFriedm-1-1}
	H^{2}=\frac{\kappa^{2}}{3}\left[
		\frac{1}{2}\left(\dot{\chi}+3H\chi\right)^{2}
		+V
	\right]
	\,,
\end{align}
and
\begin{align}
	\label{NHdot-1-1}
	\dot{H}=-\frac{\kappa^{2}}{2}\chi V_{,\chi}
	\,,
\end{align}
where $\kappa^2=8\pi G$ and $G$ is the gravitational constant. Using the Raychaudhuri equation \eqref{NHdot-1-1} to eliminate $\dot H$ in Eq.~\eqref{NDiff-syst-1-1} we find
\begin{align}
	\label{chi_eq_motion}
	\ddot{\chi}+3H\dot{\chi}+\left(1 - \frac{3\kappa^2}{2}\chi^2\right)V_{,\chi}=0
	\,.
\end{align}
This equation shows that, in the absence of other kinds of matter, the field $\chi$ evolves as a scalar field with an effective potential such that $V^{\mathrm{eff}}_{,\chi}=(1- 3\kappa^2\chi^2/2 )V_{,\chi}$, as first noted in Refs.~\cite{Koivisto:2009ew,Koivisto:2009fb}. There is however a striking difference between the equation of motion of a minimally coupled scalar field and Eq.~\eqref{chi_eq_motion}: while the former only admits as static solutions such points where the derivative of the potential vanishes, the equation of motion of the homogeneous and isotropic 3-form admits as static solutions the stationary points of the effective potential, i.e. points where $V_{,\chi}=0$ as well as the points $\chi = \pm\chi_\mathrm{c}$, with $\chi_\mathrm{c}=\chicrit$.
The latter play a critical role in 3-form cosmology; as noted in Refs.~\cite{Koivisto:2009ew,Koivisto:2009fb} for positive-valued potentials the Friedmann equation \eqref{NFriedm-1-1} imposes the constraint $(\dot\chi+3H\chi)^2<(3H\chi_\mathrm{c})^2$ which for expanding Universes and values $|\chi|>\chi_\mathrm{c}$ implies that $\dot\chi\chi<0$, i.e., independently of the shape of potential, the field $\chi$ evolves monotonically towards the interval $[-\chi_\mathrm{c},\chi_\mathrm{c}]$ from which it does not escape. Since this is a compact interval, the critical points act as local maxima or minima of the effective potential, therefore corresponding to unstable and stable equilibrium points, respectively.
The stability of these two equilibrium points can also be understood by analysing the sign of the second derivative of the effective potential:
\begin{align}
	\label{second_derivative_Veff}
	V^\textrm{eff}_{,\chi\chi}\left(\pm\chi_ \textrm{c}\right) 
	= - 2\frac{V_{,\chi}\left(\pm\chi_ \textrm{c}\right)}{\pm\chi_{\textrm{c}}} 
	= -4 V_{,\chi^2}\left(\pm\chi_ \textrm{c}\right)
	\,.
\end{align}
Since $\chi=\pm\chi_\textrm{c}$ are stationary points of the effective potential, if $\partial V/\partial (\chi^2)>0$ at $\chi=\pm\chi_\textrm{c}$ then the right hand side of Eq.~\eqref{second_derivative_Veff} is negative and $V_\textrm{eff}$ has two maxima at the critical points which represent unstable equilibrium states. On the other hand, if $\partial V/\partial \chi^2<0$ at $\chi=\pm\chi_\textrm{c}$, the right hand side of Eq.~\eqref{second_derivative_Veff} becomes positive and the effective potential has two stable minima at $\chi=\pm\chi_\mathrm{c}$.

%%%%%%%%%%%%%%%%%%%%%%%%%%%%%%%%%%%%%%%%
%
%	The Little Sibling of the Big Rip from 3-forms
%
%%%%%%%%%%%%%%%%%%%%%%%%%%%%%%%%%%%%%%%%

\section{The rising of the LSBR: Absence of interaction on the dark sector}

\label{LSBR_rising}
\label{LSBR}

The LSBR is a late-time cosmological event which was first analysed in Ref.~\cite{Bouhmadi-Lopez:2014cca}. This event happens at an infinite cosmic time where the Hubble parameter diverges but the time derivative of the Hubble parameter remains constant. In addition, the presence of a LSBR event in the future of the Universe leads in a finite time to a dissociation of the local structure of the Universe, which begins by ripping apart the large scale structures, such as clusters of galaxies, and only later on affecting structures of the size of the Solar system. We now show how LSBR appears naturally in 3-form cosmological models with positive-valued potentials.

We begin by re-writing the Friedmann equation \eqref{NFriedm-1-1} as
\begin{align}
	H = \frac{1}{3}\frac{1}{\chi_\mathrm{c}^2 - \chi^2}\left[
		\chi\dot\chi
		\pm|\chi\dot\chi|\sqrt{
			1
			+ \left(\chi_\mathrm{c}^2 - \chi^2\right) 
			\frac{
				\dot\chi^2
				+2V
			}{(\chi\dot\chi)^2}
		}
	\right]
	\,.
\end{align}
As the 3-form field approaches the critical points $\pm\chi_\mathrm{c}$, the previous equation takes the limit
\begin{align}
	H_{(\chi\rightarrow\pm\chi_\mathrm{c})} = \frac{1}{6}\frac{\dot\chi^2+2V}{|\chi_\mathrm{c}\dot\chi|} 
	\,.
\end{align}
For positive valued potentials, the previous result shows that if the 3-form field evolves towards $\chi(t)=\chi_\mathrm{c}$, which was found to be a static solution of the evolution equation \eqref{chi_eq_motion}, then the Hubble parameter goes to positive infinity. This divergence, however, is not propagated to the first derivative of Hubble parameter, as we find from the Raychaudhuri equation \eqref{NHdot-1-1} that for a general potential the value of $\dot H$ at the critical point is given by the slope of the potential:
\begin{align}
	\dot H_{(\chi\rightarrow\pm\chi_\mathrm{c})}
	= \mp \frac{\kappa^2}{2} \chi_\mathrm{c} V_{,\chi}\left(\pm\chi_\mathrm{c}\right)
	=-\frac{2}{3} V_{,\chi^2}\left(\pm\chi_\mathrm{c}\right)
	\,.
\end{align}
If the Hubble parameter diverges to positive infinity (at an infinite cosmic time as will be shown below) then by consistency its derivative needs to be non-negative at the critical point. In particular, we reach the conclusion that if $\partial V/\partial (\chi^2)<0$ at $\chi=\pm\chi_\mathrm{c}$ the Universe can hit a LSBR event in the future%
\footnote{%
In this analysis, the conclusion that the Hubble parameter diverges when the static solution is reached depends strongly on the assumption that the potential is positive valued, and in particular that it does not have zeros at $\chi=\pm\chi_\mathrm{c}$. In addition, the presence of LSBR event requires the asymptotic value of $\dot H$ to be finite. In this work we will therefore not analyse models with either $V(\pm\chi_\mathrm{c})=0$ or with $V_{,\chi^2}(\pm\chi_\mathrm{c})=0$.
}%
. This result is in accordance with the analysis in Sect~\ref{3-form Cosmology}  where the same condition was obtained for the critical points to be stable minima of the potential (cf. Eq.~\eqref{second_derivative_Veff}). We point out that for potentials which have no other minimum in the interval $[-\chi_\textrm{c},\,\chi_\textrm{c}]$, the 3-form will always lead the Universe towards a LSBR event, independently of the initial value of the field $\chi$.  This includes, e.g., the exponential potential, the Gaussian potential (with $\xi>0$) and the Ginzburg-Landau potential (with $C>\chi_\textrm{c}$) discussed in Ref.~\cite{Koivisto:2009fb}.

So far we have shown that in a 3-form model with a negative slope at the critical points $\chi=\chi_\mathrm{c}$, if $\chi\rightarrow\pm\chi_\mathrm{c}$ and $\dot\chi\rightarrow0$, the Hubble parameter and the scale factor diverge whereas the derivative of the Hubble rate is asymptotically constant. Now we examine whether this event happens at a finite or infinite cosmic time. Let $t_f$ the final time at which the system reaches the state $\chi=\pm\chi_\mathrm{c}$ and $\dot\chi=0$, and $t_i$ be a moment in the evolution of the Universe sufficiently close to $t_f$ such that $\dot H$ does not change sign and is almost constant for $t>t_i$. Then, we can write
\begin{align}
	\label{lsbinfty}
	t_f - t_i = \int_{t_i}^{t_{f}}dt
	=\int_{H(t_i)}^{H(t_{f})}\frac{dH}{\dot{H}} \approx \frac{H(t_{f}) - H(t_i)}{\dot H(t_f)}
	\,.
\end{align}
Therefore, if $H(t_{f})\rightarrow\infty$ and $H(t_i)$ and $\dot H(t_f)$ are finite, we get $t_{f}\rightarrow+\infty$. Eq.~\eqref{lsbinfty} shows that in the late stage of the evolution of the Universe, at the leading order the Hubble parameter evolves linearly with the cosmic time. Under this approximation we find, after some algebra, that
\begin{align}
	\label{asymptotic_value}
	H^2 = 2\dot H(t_f)\log\left(\frac{a}{a_i}\right) + H_i^2
	\,.
\end{align}
Notice that this is precisely the behaviour of the DE fluid in Ref.~\cite{Bouhmadi-Lopez:2014cca} that gives rise to a LSBR. 
If we now plug back this solution in the Friedmann equation \eqref{NFriedm-1-1} we find that $\chi\rightarrow\pm\chi_\mathrm{c}$, thus confirming the validity of our analysis.

%%%%%%%%%%%%%%%%%%%%%%%%%%%%
%
%	Figure 1
%
%%%%%%%%%%%%%%%%%%%%%%%%%%%%

\begin{figure}[t]
\centering
\includegraphics[width=0.505\hsize]{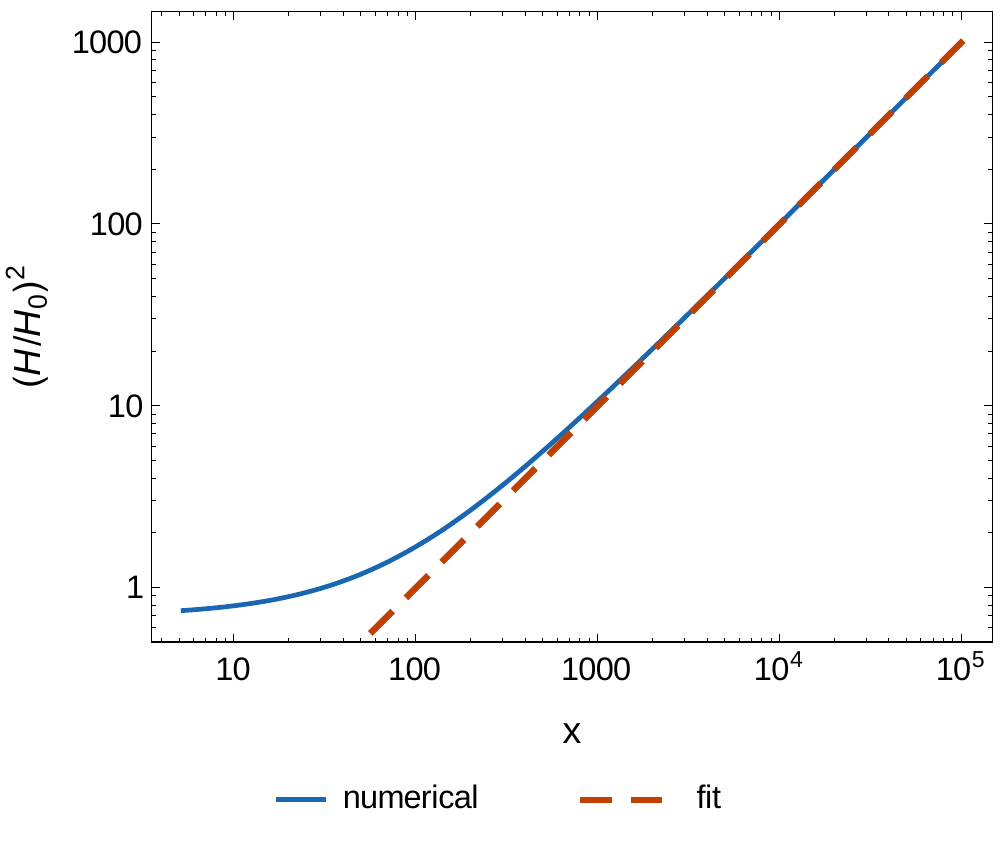}
\hfill
\includegraphics[width=0.45\hsize]{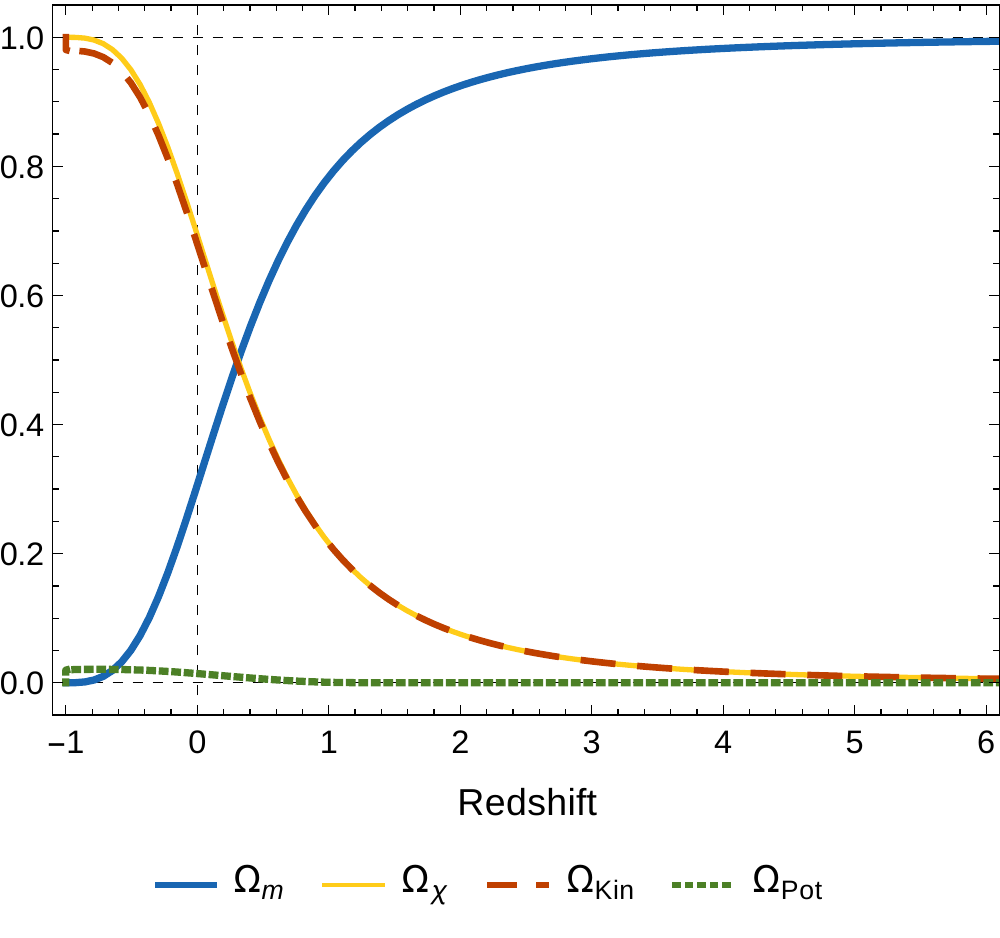}
\caption{\label{fig1}In the left panel we present the numerical solution of $(H/H_{0})^{2}$ (as a function of $x=\log\, a$) from Eq.
\eqref{NDiff-syst-1-1} and its late-time behaviour using the ansatz $(H/H_{0})^{2}=c_1\log(a/a_0)$, where $c_1\approx9.838\times10^{-3}$. We have taken $\Omega_{m,0}=0.3065$, $w_{\chi,{0}}=-1.006$, and $\chi_{0}\approx1.147\chi_\mathrm{c}$. A subscript $0$ stands for quantities evaluated at the present time. In the right panel we have the numerical evaluation, as a function of the redshift, of the energy density components in the model with non-interacting 3-form and DM. We have considered a Gaussian potential $V(\chi)=V_{0}\exp(-\xi\chi^{2}/6)$,
where (without loss of generality) $\xi=1$, which will also be used in all the numerical simulations to be presented in the remaining sections.}
\end{figure}
%%%%%%%%%%%%%%%%%%%%%%%%%%%%%

In the above analysis we have shown the possible existence of a LSBR event in the future evolution of a cosmological model where the Universe is filled uniquely by a 3-form field governed by the action \eqref{N3-form action}. Nevertheless, it is easy to show that the presence of non-interacting DM does not alter the fate of the Universe. If we include DM in our model then Eqs.~\eqref{NFriedm-1-1}, \eqref{NHdot-1-1}, and \eqref{chi_eq_motion} should be properly modified to
\begin{align}
	H^{2}=&~\frac{\kappa^{2}}{3}\left[
		\rho_m
		+
		\frac{1}{2}\left(\dot{\chi}+3H\chi\right)^{2}
		+V
	\right]
	\,,
	\\
	\dot{H}=&~-\frac{\kappa^{2}}{2}\left(\rho_m + \chi V_{,\chi}\right)
	\,,
\\
	\ddot{\chi}+&~3H\dot{\chi}+\left(1 - \frac{3\kappa^2}{2}\chi^2\right)V_{,\chi}= -\frac{3\kappa^2}{2}\chi\, \rho_m
	\,.
\end{align}
As the Universe expands, and $\rho_m$ decays with $1/a^{3}$, the contribution of DM becomes negligible and we recover the Eqs.~\eqref{NFriedm-1-1}, \eqref{NHdot-1-1}, and \eqref{chi_eq_motion}, and consequently LSBR event in the future.
In order to show that this is indeed the case we present in Fig.~\ref{fig1}, where for convenience we define the dimensionless
quantities
\begin{equation}
\Omega_{m}=\dfrac{\kappa^{2}\rho_{m}}{3H^{2}}
\,,
\qquad 
\Omega_{\chi}=\Omega_{Kin} + \Omega_{Pot}\,,
 \qquad
 \Omega_{Kin}=\dfrac{\kappa^{2}\left(\dot{\chi}+3H\chi\right)^{2}}{6H^{2}}
 \,,
 \qquad
 \Omega_{Pot}=\dfrac{\kappa^{2}V}{3H^{2}}\, ,
\end{equation}% 
the late-time behaviour of $H^{2}/H_{0}^{2}$ against its analytical fit for a cosmological model with non-interacting DM and a 3-form fluid with a Gaussian potential with negative exponent, where the relative energy density of DM at the present time satisfies the latest observational constraints \cite{Ade:2015xua,Adam:2015rua} and the parameter of EoS of the 3-form $w_{\chi,0}$ is slightly smaller than $-1$. The perfect agreement observed between the numerical result and the fit is an indication that asymptotically
\begin{align}
	\left(\frac{H}{H_{0}}\right)^{2} \propto \log\left(\frac{a}{a_0}\right)
	\Rightarrow H\sim\left(t-t_{0}\right)
	\,,
\end{align}
which is in agreement with the result of Eq.~\eqref{asymptotic_value}. We thus conclude that even in the presence of non-interacting DM, the Hubble parameter (and the scale factor) diverges at an infinite cosmic time.

In the next section we investigate whether a suitable interaction between DM and DE can avoid this LSBR event in the asymptotic future.

%%%%%%%%%%%%%%%%%%%%%%%%%%%%%%%%%%%%%%%%
%
%	Interacting 3-form DE models
%
%%%%%%%%%%%%%%%%%%%%%%%%%%%%%%%%%%%%%%%%

\section{Interacting 3-form DE models}

\label{interacting 3form}

In this section we introduce a cosmological model for the late-time evolution of the Universe. Our model consists of a spatially-flat FLRW Universe with the metric \eqref{FLRW_metric} and filled with DM and a 3-form field playing the role of DE. Additionally, we consider the possibility of an interaction between DM and the 3-form field, indicated by the presence of a term $Q$ in the equations (cf. Eq.~\eqref{conservation_equations}), whose form we will specify in a later section. Previous works on interacting DM and 3-form can be found in \cite{Boehmer:2011tp,Ngampitipan:2011se,Koivisto:2012xm}. In addition, in Ref.~\cite{DeFelice:2012wy} a model with interaction between a scalar field and a 3-form is considered within the context of reheating.

\subsection{Background equations}

As derived in the last section, cf. Eq.~\eqref{NNZ-comp-1-1}, we will encode the dynamics of the 3-form field in terms of the scalar field $\chi$ and its time derivatives.
The Friedmann equation for our model reads 
\begin{align}
	\label{hubble}
	H^{2}=\frac{\kappa^{2}}{3}\left(\rho_{m}+\rho_{\chi}\right)
	\,,
\end{align}
where $\rho_{m}$ and $\rho_{\chi}$ denote, respectively, the DM and 3-form energy densities that satisfy the conservation equations
\begin{align}
	\label{conservation_equations}
	\dot{\rho}_{m} =-3H\left(\rho_{m}+P_m\right)-Q
	\,,
	\qquad
	\dot{\rho}_{\chi} = -3H\left(\rho_{\chi}+P_{\chi}\right)+Q 
	\,.
\end{align}
Independently of the specific form of the interaction term $Q$ in the previous equations, its sign indicates the direction of the energy transfer between DM and the 3-form field: if $Q>0$ then energy is being transferred from DM to the 3-form field, and vice-versa.
Combining the individual conservation equations, we obtain the conservation of the total energy density $\dot\rho_\mathrm{tot}+3H(\rho_\mathrm{tot}+P_\mathrm{tot})=0$, where $\rho_\mathrm{tot}=\rho_m+\rho_\chi$ and $P_\mathrm{tot} = P_m + P_\chi$, in agreement with the Bianchi identity that must hold.

In this work, we assume DM to be cold (CDM) so that the pressure exerted by DM is zero ($P_{m}=0$). With the energy density and pressure of the 3-form field given by Eqs.~\eqref{energy_3form} and \eqref{pressure_3form} we can now re-write Eqs.~\eqref{conservation_equations} as
\begin{align}
	\label{int_rhom}
	&~\dot{\rho}_{m} + 3H\rho_{m} = -Q
	\,,
	\\
	&~
	\label{int_rhochi}
	 \ddot{\chi} + 3H\dot{\chi} + 3\dot{H}\chi + V_{,\chi} = \frac{Q}{\dot{\chi} + 3H\chi}
	 \,.
\end{align}
With respect to Eqs.~\eqref{conservation_equations}, we can also define the effective parameter of EoS of DM and the 3-form as follows
\begin{align}
	\label{eq:eff-eos}
	w_{m}^{\textrm{eff}} =&~
	 \frac{Q}{3H\rho_m}
	\,,
	\qquad
	w_{\chi}^{\textrm{eff}}=
	w_{\chi}-\frac{Q}{3H\rho_\chi}
	\,.
\end{align}

Finally, differentiating the Friedmann equation and combining it with the conservation equations \eqref{conservation_equations} and the expressions \eqref{energy_3form} and \eqref{pressure_3form} for the energy density and pressure of the 3-form, we obtain the Raychaudhury equation 
\begin{align}
	\label{raychaudhurieqn}
	\dot{H}=-\frac{\kappa^{2}}{2}\left(\rho_{m}+V_{,\chi}\chi\right)
	\,.
\end{align}

%%%%%%%%%%%%%%%%%%%%%%%%%%%%%%%%%%%%%%%%
%
%	Choice of the Potential
%
%%%%%%%%%%%%%%%%%%%%%%%%%%%%%%%%%%%%%%%%

\subsection{Choice of the potential}

\label{choicepot}

Several potentials have been studied in the context of a 3-form field playing the role of the inflaton or DE \cite{Koivisto:2009sd,GK09,Kobayashi:2009hj,Koivisto:2009ew,Koivisto:2009fb,Mulryne:2012ax,Koivisto:2012xm,Kumar:2014oka,DeFelice:2012jt}.
As we have previously shown, it is quite common to have a LSBR on 3-form DE models (in absence of interaction on the dark sector). This is the case whenever we have a phantom-like behaviour, $w<-1$, implying that the potential is a decreasing function of $\chi^2$ (cf. Eq,~\eqref{3form_w}), and in particular when this property is fulfilled at the critical points $\chi=\pm\chi_\mathrm{c}$. Our choice of potentials will guarantee that the above properties are fulfilled, given that one of the main goals of this paper is to remove the LSBR event for 3-form DE models by allowing an appropriate transfer of energy from DE to DM.

At the perturbative level, we want our model to be free of instabilities caused by superluminal and imaginary speeds of sound, $c_s$, of the 3-form \cite{Koivisto:2009fb,DeFelice:2012jt}. We will therefore look for potentials that can provide a squared speed of sound in the range $0<c_{s}^{2}<1$, taking into account that for a 3-form the squared speed of sound is determined by \cite{Koivisto:2009fb}
\begin{align}
	\label{soundspeed}
	c_{s}^{2}=\frac{\chi V_{,\chi\chi}}{V_{,\chi}}
	\,.
\end{align}
From this expression we can immediately conclude that any potential with stationary points other than the origin will lead to divergences in $c_s^2$. To avoid this, we will restrict our analysis to potentials that decrease monotonically with the absolute value of $\chi$. For non-negative valued potentials, this implies that for sufficiently large values of $\chi$ the second derivative of the potential must be positive and the right-hand-side of Eq.~\eqref{soundspeed} becomes negative. Therefore, at best we can choose a potential with a maximum at the origin so that in an interval around it we have $V_{\chi\chi}<0$ and a positive $c_s^2$.  If we make sure that this interval includes the critical points $\chi=\pm\chi_\mathrm{c}$ then at late-time the 3-form will be free of instabilities at the perturbative level%
\footnote{%
As stated in Sect \ref{3-form Cosmology}, the Friedmann equation constrains the evolution of the 3-form field, such that for large values of $\chi$ the field decays towards the interval $\chi\in[-\chi_\mathrm{c},\chi_\mathrm{c}]$. This result is independent of the shape of the potential in as much as $V\geq0$.
Even if the initial value of $\chi$ is  large enough to imply a negative squared speed of sound, $c_s^2$, the field will decay and reach rapidly the interval $[-\chi_\mathrm{c},\chi_\mathrm{c}]$, inside which the condition $c_s^2>0$ is met. Any instabilities at the linear level will therefore be present only during a finite interval of time.
}.
Additionally, we note that since we are looking for potentials which have no minima in the interval $(-\chi_\textrm{c},\,\chi_\textrm{c})$,  in the absence of interactions the 3-form always leads the Universe to a LSBR event in the asymptotic future.

With the previous considerations in mind we now scan the choices of 3-form potentials found in the literature for a suitable candidate:
\begin{enumerate}[(i)]
	\item Power law potentials: $V=V_{0}\chi^{2n}$ 
	
	This class of potentials includes the case of a 3-form field with a constant mass $m^2$ ($n=1$), which was the first model considered in the literature \cite{Koivisto:2009sd,Koivisto:2009ew,Koivisto:2009fb} and shown to be dual to a canonical scalar field, cf. Sect~\ref{3-form introduction} and references therein for more on the dualisation of a 3-form. In order to obtain phantom-like behaviour we would need to consider potentials with a negative exponent $n$. This, however, leads to a constant negative value of $c_s^2$ and therefore we will disregard this case.
	
	\item Exponential potentials: $V(\chi)=V_0e^{-\xi\kappa\chi/\sqrt6}$
	
	Although its mathematical simplicity makes the exponential potential an attractive model from a dynamical approach point of view and therefore is recurrently used in the literature, see e.g. Refs. \cite{Koivisto:2009fb,Ngampitipan:2011se,Boehmer:2011tp}, as stated in \cite{Koivisto:2009fb} the exponential potential is not compatible with the action \eqref{N3-form action} as it is not a function of $\chi^2$. We will therefore disregard this case altogether. Nevertheless, we point out that for an exponential potential the 3-form presents a phantom-like behaviour for $\xi\chi>0$. Since there are no minima in the potential, the end state of the Universe is always a LSBR event characterized by $\chi=\textrm{sgn}(\xi)\chi_\textrm{c}$.
	
	\item Ginzburg-Landau potentials: $V=V_{0}\left(\chi^{2}-C^{2}\right)^{2}$
	
	In the context of 3-forms this potential has been studied in detail in Refs.~\cite{Koivisto:2009fb,DeFelice:2012jt} and has the interesting characteristic that it can accommodate both quintessence and phantom-like behaviour in the same model. In particular, a LSBR event is present in this case whenever $C>\chi_\textrm{c}$. This however comes at a cost, as the speed of sound of the 3-form diverges at $\chi=\pm C$ and becomes superluminal in the large $\chi$ regime. Therefore, we also disregard this potential in our analysis.
	
	\item Gaussian potentials: $V=V_0 e^{-\xi\kappa^2\chi^2/6}$
	\label{gauss-pot}
	
	For positive values of the dimensionless parameter $\xi$, the Gaussian potential \cite{Koivisto:2009fb} presents a maximum value $V_0$ at $\chi=0$ and decays to zero as the value of $\chi^2$ goes to infinity. These are the characteristics that we are looking for, as this potential can provide phantom-like behaviour with the presence of a LSBR while maintaining positive values of $c_s^2$ near the origin. Substituting it in Eq.~\eqref{soundspeed}, we find that in this case
	\begin{align}
		c_{s}^{2}=1-\frac{\kappa^2}{3}\xi\chi^{2}
		\,,
	\end{align}
	meaning that the squared speed of sound satisfies $0\leq c_s^2\leq1$ in the interval $\chi\in[-\sqrt{9/(2\xi)}\chi_\mathrm{c},\sqrt{9/(2\xi)}\chi_\mathrm{c}]$. If we demand that the critical points $\pm\chi_\mathrm{c}$ are inside this interval, we arrive at the constraint $0<\xi<9/2$. For large values of $\chi^2$, however, $c_s^2$ becomes increasingly negative. Nevertheless, since in that regime the 3-form field behaves as a cosmological constant, we expect any instabilities arising from a negative value of $c_s^2$ to be suppressed.
\end{enumerate}

Based on these general qualitative features, we will consider a Gaussian potential to study some consequences of the interacting 3-form DE model in the rest of the paper. Despite the fact that we do not develop an extensive study of suitable potentials for the interacting 3-form DE model, it will be shown in the subsequent sections that the Gaussian potential is appropriate to define valid and general assumptions about what this model brings beyond the standard predictions of the $\Lambda$CDM model. This will become apparent in Sect.~\ref{Dynsys}, as the qualitative structure of the dynamical system employed does not depend on the specific shape of the potential as long as $V(\chi)$ has a maximum at $\chi=0$ and is a monotonically decreasing function of $|\chi|$, i.e., as long as the potential satisfies the conditions stated above for the existence of a LSBR event and a positive squared speed of sound at late-time.

%%%%%%%%%%%%%%%%%%%%%%%%%%%%%%%%%%%%%%%%
%
%	Choice of the interaction
%
%%%%%%%%%%%%%%%%%%%%%%%%%%%%%%%%%%%%%%%%

\subsection{Choice of the interaction}

\label{Choice of interaction}

In this work we consider a model for the late-time evolution of the Universe with interactions between DM and DE and where a 3-form field plays the role of DE. 
We expect that a DM/DE interaction will remove LSBR as long as there is a suitable energy transfer from DE to DM. 
Despite the extensive literature on the subject of interacting DM and DE, see e.g. Refs.~\cite{Bolotin:2013jpa,Wang:2016lxa} and references within, to the best of our knowledge there are up to date only four published works on interacting 3-form models: Refs.~\cite{Boehmer:2011tp,Ngampitipan:2011se,Koivisto:2012xm} consider models of interacting DM and DE in which the 3-form field plays the role of DE, while Ref.~\cite{DeFelice:2012wy} considers a coupling between a 3-form inflaton and a scalar field as a mean to describe the reheating period that ends the 3-form fuelled inflation.

Starting from a phenomenological point of view, we introduce the class of interactions
\begin{align}
	\label{Int}
	Q & =3H\left(\rho_m+\rho_\chi\right) \sum_{i,j=0} \lambda_{ij} \left(\frac{\rho_{m}}{\rho_m+\rho_\chi}\right)^{i} \left(\frac{\rho_{\chi}}{\rho_m+\rho_\chi}\right)^{j}
	\,,
	\nonumber \\
	 &=3H\left(\rho_m+\rho_\chi\right) \left[ 
	 	\left(\lambda_{00}+\lambda_{10}+\lambda_{20}+\dots\right)
	 	+\left(\lambda_{01}-\lambda_{10}+\lambda_{11}-2\lambda_{20}+\dots\right) \left(\frac{\rho_{\chi}}{\rho_m+\rho_\chi}\right)
	\right.
	\nn\\
	&\qquad\qquad\qquad\qquad
	\left.
	 	+\left(\lambda_{20}+\lambda_{02}-\lambda_{11}+\dots\right)\left(\frac{\rho_{\chi}}{\rho_m+\rho_\chi}\right)^{2}
	 	+...
	\right]
	\,,
\end{align}
where $i,j$ are non-negative integers and $\lambda_{ij}$ are dimensionless couplings that determine the strength of the interaction. This is a natural generalisation to higher orders of the frequently considered linear interaction $Q=3H\,(\lambda_{m}\rho_{m}+\lambda_{DE}\rho_{DE})$, \cite{Bolotin:2013jpa}, and include as well the case of $Q=3H\lambda\, \rho_{m}\rho_{DE}/(\rho_{m}+\rho_{DE})$, inspired from two-body chemical reactions \cite{Arevalo:2011hh,Bouhmadi-Lopez:2016dcs}. This class of interactions has the advantage of leaving the evolution equations with a rather simple mathematical structure, which in some particular cases even allows for the finding of analytical solutions for the energy densities of the individual components. Furthermore, as we will see in the next section, with interaction terms of the class \eqref{Int} we can employ a dynamical system approach to our model without introducing new variables, therefore maintaining the dimensionality of the problem. Despite these mathematical advantages, this class of interactions is phenomenological in nature and as such is not derived from any considerations coming from particle physics or field theory, as those introduced in \cite{Bolotin:2013jpa,Boehmer:2009tk}.

We can divide the interactions introduced in Eq.~\eqref{Int} in three different categories: the cases where $Q$ depends only on the relative energy density of DM, therefore $\lambda_{i\,0}$ for $i\neq0$; the cases where $Q$ depends only on the relative energy density of DE, and therefore $\lambda_{0\,j}$ for $j\neq0$; and the ``mixed'' interactions where both the relative energy densities of DM and DE appear. These three types of interaction will be considered in the following sections, and we truncate Eq.~\eqref{Int} at the quadratic order, thus allowing us to write
\begin{align}
	\label{quadratic_int}
	Q=3H\left(\rho_m+\rho_\chi\right)\sum_{i=0}^2\alpha_{i}\left(\frac{\rho_{\chi}}{\rho_m+\rho_\chi}\right)^{i}
	\,.
\end{align}
Here the coefficients $\alpha_i$, linear combinations of the coefficients $\lambda_{ij}$, are preferred as they eliminate the degeneracy of the coefficients in the original formulation \eqref{Int}. 

Many of the recent studies based on CMB, BAO and Large Scale Structure (LSS) data indicate possible, although weak, interactions between DE and DM \cite{Salvatelli:2014zta,Abdalla:2014cla,Valiviita:2015dfa,Costa:2013sva,Bolotin:2013jpa,Bolotin:2015fea,Zhang:2013zyn,Xia:2013nua, Nunes:2016dlj, Costa:2016tpb, Sola:2016ecz}. This feature alleviates the discrepancies between CMB and LSS data at lower multipoles $l<40$ \cite{Valiviita:2015dfa,Costa:2013sva,Wang:2014xca,Zhang:2013zyn}. In Ref.~\cite{Feng:2008fx}, the authors study the observational constraints on the linear interaction $Q=3H\lambda_{\chi}\rho_{\chi}$ and obtained $\lambda_{\chi}\sim0.04$. A similar value was recently obtained for the same interaction in Ref.~\cite{Costa:2016tpb}. In addition, the recent observational constraints \cite{Valiviita:2015dfa} for wCDM model with interaction $Q=\Gamma\rho_{m}$, suggest $-0.14<\Gamma/H_{0}<0.02$ at $95\%$ CL for non-phantom models and $-0.46<\Gamma/H_{0}<-0.01$ at $95\%$ CL for phantom models. Moreover, in few recent works \cite{Costa:2013sva,Bolotin:2013jpa,Bolotin:2015fea,Xia:2013nua,Zhang:2013zyn}, the dimensionless coupling constants of various interactions in the dark sector were argued to be small at present. Taking into consideration all these recent developments, we consider small values of the dimensionless coupling constants for all the interactions we are considering in this work.

%%%%%%%%%%%%%%%%%%%%%%%%%%%%%%%%%%%%%%%%
%
%	Dynamical System
%
%%%%%%%%%%%%%%%%%%%%%%%%%%%%%%%%%%%%%%%%

\section{Dynamical system analysis}

\label{Dynsys}
In this section, we apply a dynamical system analysis to the interacting 3-form DE model. We aim to identify its fixed points, study their stability and characterise the physical scenarios to which they correspond.

\subsection{Dynamical System}
We choose to employ the same set of variables introduced in Ref.~\cite{Boehmer:2011tp},
\begin{align}
	\label{eq:Var1}
	\begin{cases}
		u\equiv\dfrac{2}{\pi}\arctan\left( \sqrt{\dfrac{3\kappa^2}{2}}\chi\right)
		\,,\\
		y\equiv\sqrt{\dfrac{\kappa^2}{6}}\,\dfrac{\dot{\chi}+3H\chi}{H}
		\,,\\
		z\equiv\sqrt{\dfrac{\kappa^2V}{3H^{2}}}
		\,,\\
		s\equiv\sqrt{\dfrac{\kappa^2\rho_{m}}{3H^{2}}}
		\,,
	\end{cases}
\end{align}
which represent a compactification of the variables used in the early 3-form paper~\cite{Koivisto:2009fb} and in the ensuing works \cite{Ngampitipan:2011se,Koivisto:2012xm,DeFelice:2012jt,Kumar:2014oka}. From the definition in Eq.~\eqref{eq:Var1}, it is immediate to verify that for non-negative potentials the dynamical variables are defined within the intervals%
\footnote{ Notice that in order to be able to capture the asymptotic behaviour of the system here we are considering the original variables in the extended intervals $-\infty\leq\chi\leq+\infty$, $0\leq\rho_m\leq+\infty$ and $-\infty\leq\dot\chi\leq+\infty$, which in turn imply $0\leq H\leq+\infty$. Here we do not include negative values of the Hubble parameter as we are considering only expanding cosmologies.}
\begin{align}
	\label{Vlimits}
	-1\leq u\leq1
	\,,
	\qquad
	0\leq s\leq1
	\,,
	\qquad
	0\leq z\leq1
	\,,
	\qquad
	-1\leq y\leq1
	\,.
\end{align}

Using these variables, the Friedmann constraint reads 
\begin{align}
	\label{friefconstrant}
	y^{2}+z^{2}+s^{2}=1
	\,,
\end{align}
while the Raychaudhury equation can be recast as 
\begin{align}
	\label{h1h}
	\frac{\dot{H}}{H^{2}}=-\frac{3}{2}\left[s^{2}-\frac{1}{3}\lambda(u)\, z^{2}\tan\left(\frac{\pi}{2}u\right)\right]
	\,,
	\qquad
	\lambda(u)\equiv -\sqrt{\frac{6}{\kappa^2}}\frac{V_{,\chi}}{V}
	=-\frac{6}{\pi}\cos^2\left(\frac{\pi}{2}u\right)\frac{V_{,u}}{V}
	\,.
\end{align}
In addition, the EoS parameters for the 3-form and for the total energy density of the Universe can be written in terms of the dimensionless variables defined in \eqref{eq:Var1} as 
\begin{align}
	\label{state_param}
	w_{\chi}=-1-\frac{1}{3}\,\lambda(u)\,\frac{z^{2}}{y^{2}+z^{2}}\tan\left(\frac{\pi}{2}u\right)
	\,,
	\qquad
	w_\mathrm{tot}=-\left(y^{2}+z^{2}\right)-\frac{1}{3}\,\lambda(u)\, z^{2}\tan\left(\frac{\pi}{2}u\right)
	\,,
\end{align}
while the effective parameters of EoS in Eq.~\eqref{eq:eff-eos} can be recast as
\begin{align}
	\label{eq:Eff-eos}
	w_{m}^{\textrm{eff}} =&~
	 \frac{\kappa^2Q}{9H^3s^2}
	\,,
	\qquad
	w_{\chi}^{\textrm{eff}}=
	-1-\frac{1}{3}\,\lambda(u)\,\frac{z^{2}}{y^{2}+z^{2}}\tan\left(\frac{\pi}{2}u\right)
	-\frac{\kappa^2Q}{9H^3\left(y^2+z^2\right)}
	\,.
\end{align}

In order to obtain the set of evolution equations for our dimensionless variables, we combine the definitions in \eqref{eq:Var1} with Eqs.~\eqref{int_rhom}, \eqref{int_rhochi}, and \eqref{h1h}, and obtain the following autonomous dynamical system:
\begin{align}
	\label{eq_u}
	u' & =\frac{6}{\pi}\cos^{2}\left(\frac{\pi u}{2}\right)\left[y-\tan\left(\frac{\pi u}{2}\right)\right]
	\,,
	\\
	\label{eq_y}
	y' & =\frac{1}{2}\left\{ 3s^{2}y+\lambda\left(u\right)z^{2}\left[1-\tan\left(\frac{\pi}{2}u\right)y\right]\right\} +\frac{\kappa^{2}}{6yH^{3}}\, Q
	\,,
	\\
	\label{eq_z}
	z' & =\frac{1}{2}z\left\{ 3s^{2}-\lambda\left(u\right)\left[y-\tan\left(\frac{\pi}{2}u\right)\left(1-z^{2}\right)\right]\right\} 
	\,,
	\\
	\label{eq_s}
	s' & =-\frac{3}{2}s\left[1-s^{2}+\frac{\lambda\left(u\right)}{3}\tan\left(\frac{\pi}{2}u\right)z^{2}\right]-\frac{\kappa^{2}}{6sH^{3}}\, Q
	\,.
\end{align}
Here, we have used $x=\log (a/a_0)$ as our time variable and a prime indicates a derivative with respect to $x$. The set of dynamical equations \eqref{eq_u}, \eqref{eq_y}, \eqref{eq_z}, \eqref{eq_s}, is complemented by the Friedmann constraint \eqref{friefconstrant}, which means that our system has only three degrees of freedom. In fact, we can use the Friedmann constraint \eqref{friefconstrant} to eliminate the variable $s$, obtaining
\begin{align}
	\label{eq_u_1}
	u' & =\frac{6}{\pi}\cos^{2}\left(\frac{\pi u}{2}\right)\left[y-\tan\left(\frac{\pi u}{2}\right)\right]
	\,,
	 \\
	\label{eq_y_1}
	y' & =\frac{1}{2}\left\{ 3\left[1-y^{2}-z^{2}\right]y+\lambda\left(u\right)z^{2}\left[1-\tan\left(\frac{\pi}{2}u\right)y\right]\right\} +\frac{\kappa^{2}}{6yH^{3}}\, Q
	\,,
	 \\
	\label{eq_z_1}
	z' & =\frac{1}{2}z\left\{3\left[1-y^{2}-z^{2}\right]
	-\lambda\left(u\right)\left[y-\tan\left(\frac{\pi}{2}u\right)
	\left(1-z^{2}\right)\right]\right\} 
	\,.
\end{align}

The set of Eqs.~\eqref{eq_u_1}, \eqref{eq_y_1}, and \eqref{eq_z_1} constitutes a three-dimensional autonomous system that defines the evolution of the dimensionless variables $(u,\, y,\, z)$ which encode the evolution of our model of interacting DM and 3-form. Although we could have used the Friedmann constraint \eqref{friefconstrant} to eliminate either the variable $y$ or $z$ instead of $s$, there are some advantages in eliminating the variable $s$ in models with interacting DM and DE. Due to the fact that we make a decomposition of $\rho_m/H^2$ in terms of a squared variable $s^2$, when we derive the evolution equation for $s$ from Eq.~\eqref{int_rhom} the interaction term appears with a $1/s$ factor. This can introduce an artificial divergence%
\footnote{Here we employ the term ``artificial'' to point out the fact that the divergences in the dynamical system appear from the decomposition employed and not as a generic feature of the model.
}
in the equations at $s=0$, unless the interaction is at least proportional to $s$. Notice that interactions proportional to the energy density of DM, which appear often in the literature, eliminate this divergence as in those cases $Q\propto s^2$. However, since we will consider a more general class of interactions, we choose to eliminate the variable $s$ from our dynamical system.
Another feature of these equations is that, for potentials in which $\lambda(u)$ is well defined for all values\footnote{From the definition in Eq.~\eqref{h1h} we find that $\lambda(u)$ is well defined for all $u$ whenever the potential does not have any isolated zeros.} of
$u$, the plane $z=0$, in which the 3-form behaves like a cosmological constant \cite{Koivisto:2012xm}, is an invariant subset of the dynamical system. In the absence of interaction the system is also invariant with regards to the transformation $(u,y)\rightarrow(-u,-y)$, a consequence of the considering symmetric potentials in Eq.~\eqref{N3-form action}, and the plane $s=0$ constitutes an invariant subset of the system. As we turn on the interaction, however, these symmetries can be broken unless $Q$ vanishes sufficiently fast as $s\rightarrow0$ and is symmetric on $\chi$.

%%%%%%%%%%%%%%%%%%%%%%%%%%%%%%%%%%%%%%%%
%
%	FIxed Points
%
%%%%%%%%%%%%%%%%%%%%%%%%%%%%%%%%%%%%%%%%

\subsection{Fixed Points}
\label{subsec_Fixed_Points}

We now determine the conditions for the existence of fixed points $(u_{fp},y_{fp},z_{fp})$, defined by $u'=y'=z'=0$, of the dynamical system presented in Eqs.~\eqref{eq_u_1}, \eqref{eq_y_1}, and \eqref{eq_z_1}. The behaviour of the system in the vicinity of these points, i.e., their stability, will be studied in a subsequent section for the particular case of the Gaussian potential and the class of interactions introduced in Eq.~\eqref{quadratic_int}. In the following analysis, we separate the fixed points in three categories:
\begin{itemize}

%%%%%%%%%%%%%%%%%%%%%%%%%%%%%%%%%%%%%%%%
%
%	FIxed Points: Type I
%
%%%%%%%%%%%%%%%%%%%%%%%%%%%%%%%%%%%%%%%%
\item \textbf{Type~I:} these fixed points verify $z=0$ for finite values of the 3-form field $\chi$, i.e., $u_{fp}\neq\pm1$. The fixed points corresponding to a LSBR event, { when present, verify} $(u_{fp},y_{fp},z_{fp}) = (\pm1/2,\,\pm1,\,0)$ { and are therefore} included in this category. Solving Eqs.~\eqref{eq_u_1} and \eqref{eq_y_1} for $u$ and $y$ we find that, for choices of potentials such that $\lambda(u)$ is well defined for all values of $u$, any fixed point $(u_{fp},y_{fp},z_{fp})$ of this category must verify%
\footnote{On the second condition of \eqref{TypeIfix} we leave $y_{fp}$ in the denominator of the right hand side in conformity with Eq.~\eqref{eq_y_1}}%
\begin{align}
	\label{TypeIfix}
	\left(u_{fp}, \, y_{fp}, \, z_{fp}\right):
	\begin{cases}
		u_{fp} & =\dfrac{2}{\pi}\arctan\left(y_{fp}\right)
		\,,
		\\
		 y_{fp}\left(1-y_{fp}^{2}\right) & =-\dfrac{\kappa^2}{9H^{3}y_{fp}}Q
		\,,
		\\
		z_{fp} & =0
		\,.
	\end{cases}
\end{align}
Since the quotient $\lambda(u)$ is absent from these conditions it is possible to conclude that the existence of fixed points of Type~I depends on the choice of the interaction but not on the potential. Nevertheless, in general their stability will depend on the shape of the potential.

From Eq.~\eqref{state_param}, we calculate the total parameter of EoS at the fixed points of Type~I and obtain $w_\mathrm{tot}=-y_{fp}^2$. In the absence of interaction, we have only two solutions: a matter era with $w_\mathrm{tot}=0$ for $y_{fp}=0$ and two fixed points with $y_{fp}^2=1$ that corresponds to an asymptotic scenario where $\chi=\pm\chicrit$ and $\dot\chi=0$. As discussed in Sect~\ref{LSBR_rising}, these fixed points correspond to LSBR events. When we turn on the interaction we find that there is a possibility of obtaining new solutions with $-1<y_{fp}<1$, meaning that the interaction between DM and the 3-form field leads to a scaling behaviour of the two components. Nevertheless, from Eq.~\eqref{TypeIfix} we find that LSBR is not removed from the system of interacting DM and 3-form if there are roots of $Q=0$ with $y^2=1$.

%%%%%%%%%%%%%%%%%%%%%%%%%%%%%%%%%%%%%%%%
%
%	FIxed Points: Type II
%
%%%%%%%%%%%%%%%%%%%%%%%%%%%%%%%%%%%%%%%%
\item \textbf{Type~II:} in this category we consider all the fixed points that, for finite values of $\chi$, have non-zero values of the variable $z$. Solving the system of Eqs.~\eqref{eq_u_1}, \eqref{eq_y_1}, and \eqref{eq_z_1} for $u'=y'=z'=0$, we thus find that any solution $(u_{fp},y_{fp},z_{fp})$ in this category must verify
\begin{align}
	\label{TypeIfix-2}
	\left(u_{fp}, \, y_{fp}, \, z_{fp}\right):
	\begin{cases}
		u_{fp} & =\dfrac{2}{\pi}\arctan\left(y_{fp}\right)
		\,,\\
		\lambda(u_{fp})z_{fp}^{2} & =-\dfrac{\kappa^2}{3y_{fp}H^{3}}Q
		\,,\\
		1-y_{fp}^{2}-z_{fp}^{2} & =\frac{1}{3}\lambda\left(u_{fp}\right)y_{fp}z_{fp}^{2}
		\,.
	\end{cases}
\end{align}
Here we observe that, contrary to what was found in \eqref{TypeIfix}, the position of the fixed points of Type~II will depend both on the choice of the interaction and of the 3-form potential. However, when combining the first and third conditions in \eqref{TypeIfix-2} with the expression for the total parameter of EoS, Eq.~\eqref{state_param}, we find that for these fixed points $w_\textrm{tot}=-1$, independently of the potential of the 3-form and of the interaction between DM and DE.
%%%%%%%%%%%%%%%%%%%%%%%%%%%%%%%%%%%%%%%%
%
%	FIxed Points: Type 0
%
%%%%%%%%%%%%%%%%%%%%%%%%%%%%%%%%%%%%%%%%
\item \textbf{Type~III:} characterised by lying on the planes $u=\pm1$, Type~III fixed points correspond to infinite values of the 3-form field $\chi$.
As discussed in Sect~\ref{3-form Cosmology} the constraint in the Friedmann equation forces the 3-form field to decay to the interval $[-\chi_\mathrm{c},\chi_\mathrm{c}]$, where $\chi_\mathrm{c}=\chicrit$, meaning that at late-time the variable $u$ is constrained to the interval $[-1/2,1/2]$. As such, any Type~III fixed point present in the system will necessarily be unstable%
\footnote{The existence of Type~III fixed points can only be identified once the variable $\chi$ is compactified. Therefore this was not noticed in previous works where a dynamical system analysis is employed with the variable $\chi$ or a linear rescaling of it. In fact, even though the compactification employed here was first introduced in Ref.~\cite{Boehmer:2011tp}, the existence of fixed points with $u=\pm1$ was not identified and to the best of our knowledge is being discussed for the first time in this work. In a different paper, we supply suitable mathematical tools to study the stability of fixed points at infinity \cite{ Bouhmadi:2016}}. Due to the appearance of divergent terms in the $y'$ and $z'$ equations (e.g. $\tan(\pi u/2)\rightarrow\pm\infty$ as $u\rightarrow\pm1$), extra care is needed when identifying the position of fixed points of Type~III and a general analysis for any type of potential is not possible. For example, in the case of power law potentials we find that $\lambda(u)\propto1/\tan(\pi u/2)$ and all divergences in the equations are automatically cancelled, while in the case of a Gaussian potential with positive $\xi$ the variable $z$ must vanish sufficiently fast as $u\rightarrow\pm1$ in order to cancel the divergences of $\lambda(u)\propto\tan(\pi u/2)$ and $\tan(\pi u/2)$.
 \end{itemize} 

As stated above, the stability of the fixed points found depends on (i) the kind of interaction and (ii) the potential considered. In the following subsections we study this issue for the specific class of interactions presented in \ref{Choice of interaction}, while considering a Gaussian potential $V=V_0\exp(-\kappa^2\chi^2/6)$ with a positive parameter $\xi$. For completeness and as a mean of comparison, we begin by reproducing the results  when the interaction is switched off, which were first obtained in Ref.~\cite{Koivisto:2009fb}.

%%%%%%%%%%%%%%%%%%%%%%%%%%%%%%%%%%%%%%%%
%
%	Stability Analysis: No interaction
%
%%%%%%%%%%%%%%%%%%%%%%%%%%%%%%%%%%%%%%%%

\subsection{Stability in the non-interacting case}

\label{No Interaction}

In the absence of interaction, the term $Q$ vanishes in Eq.~\eqref{eq_y_1} and in the conditions \eqref{TypeIfix} and \eqref{TypeIfix-2} for the existence of fixed points. Solving \eqref{TypeIfix} for $u$ and $y$, we find three fixed points corresponding to the Type~I category:
\begin{align}
	\begin{array}{c}
		\\
		\textbf{Type~I fixed points:}\\
		\textrm{(no interaction)}
	\end{array}
	\qquad
	\begin{cases}
	p_0:
	\quad&
	\left(u_{fp}, \, y_{fp}, \, z_{fp}\right) = \left(0,\,0,\,0\right)
	\,,
	\\
	p_1^\pm :
	\quad&
	\left(u_{fp}, \, y_{fp}, \, z_{fp}\right) = \left(\pm\dfrac{1}{2},\,\pm1,\,0\right)
	\,.
	\end{cases}
\end{align}
The point $p_0$ corresponds to a matter dominated scenario and the eigenvalues of the Jacobian are $\{-3,3/2,3/2\}$, leading to the conclusion that this point is saddle and does not correspond to a late-time attractor. On the other hand, the points $p_1^\pm$ correspond to a LSBR event and the eigenvalues of the Jacobian of the system are $\{-3,-3,0\}$. The existence of a null eigenvalue indicates that the linear approximation in this case is not valid for the stability analysis. To decide on the stability of these points, we resort to the method described in \cite{Rendall:2001it,Boehmer:2011tp} based on Centre Manifold Theory \cite{Carr}. We find that $p_1^+$ ($p_1^-$) is an attractor if $\lambda>0$ ($\lambda<0$) and a saddle-node if $\lambda<0$ ($\lambda>0$) \cite{Koivisto:2009fb}. For the case of a Gaussian potential this means that both $p_1^\pm$ are attractive and lead to the existence of a LSBR event in the future evolution of the Universe if $\xi>0$. We remind the reader that for the Gaussian potential defined in Sect.~\ref{choicepot} we have $\lambda = 2\xi/3$.

In the absence of interaction on the dark sector, we find from the condition \eqref{TypeIfix-2} that the fixed points of Type~II are characterised by $\lambda(u_{fp})=0$, i.e., they must correspond to extrema of the potential \cite{Koivisto:2009fb}. In the case of the Gaussian potential, which only has a minimum for $\chi=0$, i.e., $u=0$, we find only one such solution:
\begin{align}
	\begin{array}{c}
	\\
		\textbf{Type~II fixed points:}\\
		\textrm{(no interaction)}
	\end{array}
	\qquad
	p_2:
	\quad
	\left(u_{fp}, \, y_{fp}, \, z_{fp}\right) =~ \left(0,\,0,\,1\right)
	\,.
\end{align}
This corresponds to a scenario where the content of the Universe is completely dominated by the 3-form potential and enter a de Sitter inflationary era. The eigenvalues of the Jacobian in this case are
\begin{align}
	\left\{-3, -\frac{3}{2}\left(1+\sqrt{1+\frac{4\xi}{9}}\right), -\frac{3}{2}\left(1-\sqrt{1+\frac{4\xi}{9}}\right)\right\}
	\,,
\end{align}
which leads to the conclusion that $p_2$ is an attractive focus node for $\xi<0$ and a saddle for $\xi>0$ \cite{Koivisto:2009fb}.

\begin{figure}[t]

\includegraphics[width=.45\hsize]{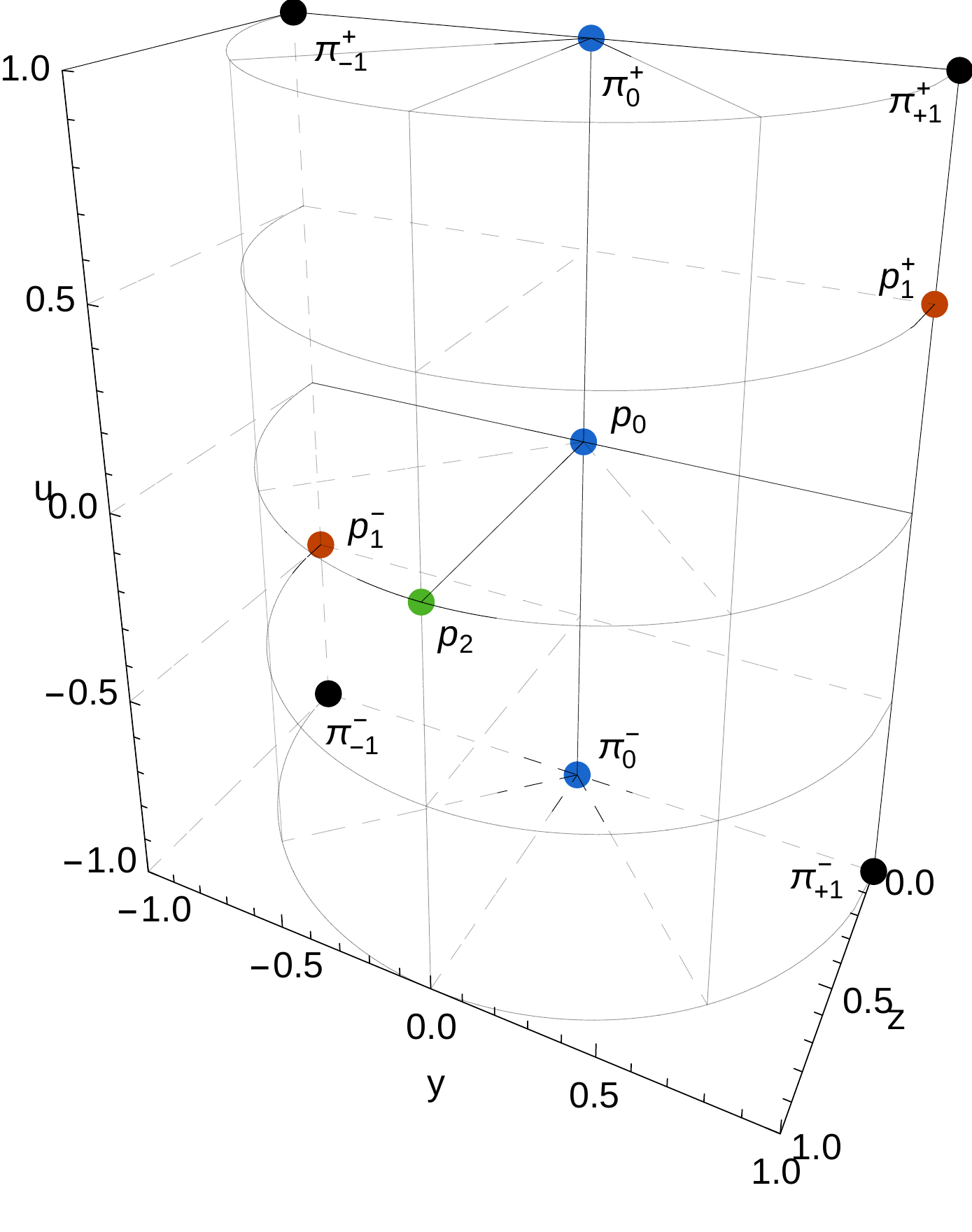}
\hfill
\includegraphics[width=.45\hsize]{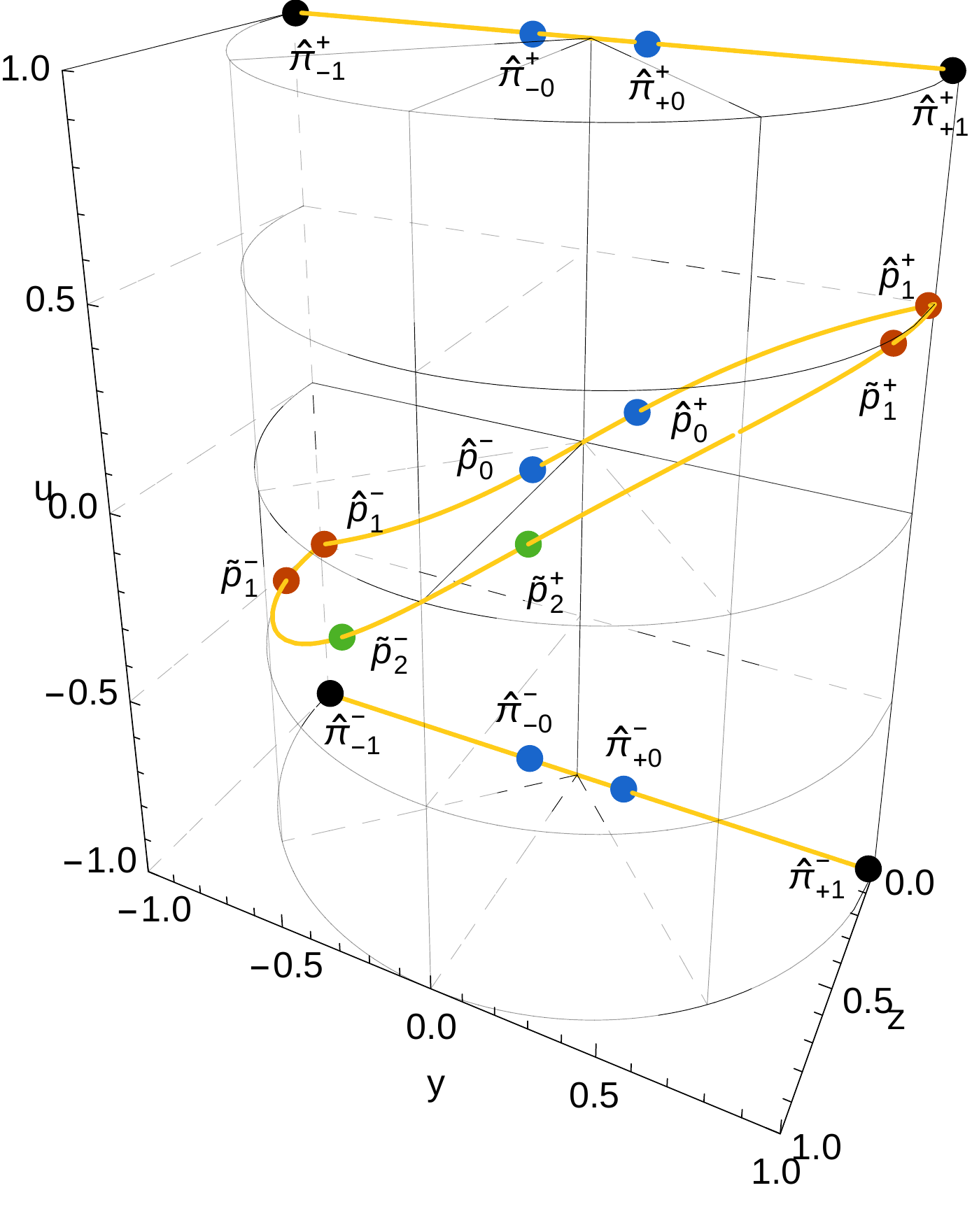}

\caption{\label{FixedPoints}%
On the left panel, we present the position of the fixed points of the dynamical system $(u,y,z)$ with a Gaussian potential and when the interaction between DM and DE is switched off. On the right panel, we show how the position of the fixed points changes when, for the same potential, we turn on a quadratic interaction of the kind described in Eq.~\eqref{quadratic_int_dynamical} with small coefficients $\alpha_i$. It is possible to observe how each fixed point in the non-interacting case, except for $\hat\pi_{\pm1}^\pm$, splits in two points once the interaction is turned on. As the strength of the interaction coefficients changes, the new fixed points move along the yellow curves. For sufficiently flat potentials, the pairs $\tilde p_1^\pm$ and $\tilde p_2^\pm$ may coalesce and give rise to the fixed points $\tilde p_\Delta^\pm$, cf. Eq.~\eqref{Delta0_solutions}. To obtain this figure we used the values $\xi=1$, $\alpha_0=\alpha_2=-0.03$, and $\alpha_1=+0.03$.
}
\end{figure}

In addition to the fixed points of Type~I and Type~II, which were first identified in Ref~\cite{Koivisto:2009fb}, in the case of the Gaussian potential with a positive $\xi$ we find three additional fixed points corresponding to the Type~III category:
\begin{align}
	\begin{array}{c}
	\\
		\textbf{Type~III fixed points:}\\
		\textrm{(no interaction)}
	\end{array}
	\qquad
	\begin{cases}
	\pi_0^\pm:
	\quad&
	\left(u_{fp}, \, y_{fp}, \, z_{fp}\right) = \left(\pm1,\,0,\,0\right)
	\,,
	\\
	\pi_{+1}^\pm :
	\quad&
	\left(u_{fp}, \, y_{fp}, \, z_{fp}\right) = \left(\pm1,\,1,\,0\right)
	\,,
	\\
	\pi_{-1}^\pm :
	\quad&
	\left(u_{fp}, \, y_{fp}, \, z_{fp}\right) = \left(\pm1,\,-1,\,0\right)
	\,.
	\end{cases}
\end{align}
The eigenvalues of the Jacobian at these points are%
\footnote{%
The infinite value of the eigenvalue was obtained by computing the characteristic polynomial of the Jacobian near the fixed point and then taking an appropriate limit of the formulas obtained. In particular, we first took the limit of $z\rightarrow0$ and only afterwards the limit $u\rightarrow\pm1$. This is in accordance with the previous statement that the variable $z$ vanishes faster than $\tan(\pi u/2)$ divergence as $u\rightarrow \pm1$.
}
\begin{align}
	\left\{3, \frac{3}{2}\left(1-3y_{fp}^2\right), +\infty\right\}
	\,,
\end{align}
while the total parameter of EoS is $w_\mathrm{tot}=-y_{fp}^2$, leading to the conclusion that $\pi_0^\pm$ are repulsive, matter dominated, fixed points while the four fixed points $\pi_{\pm1}^\pm$ are all saddles. Since $\pi_0^\pm$ are the only repulsive fixed points in the system, we expect that all trajectories converge to one of these points as we go sufficiently into the past.
This has been corroborated by our numerical analysis (see the left hand side panel in Fig.~\ref{Trajectories})
The results obtained are summarised in Tables~\ref{Table_fixedpoints} and \ref{Table_fixedpoints_II} and represented on the left panel of Fig.~\ref{FixedPoints}.

%%%%%%%%%%%%%%%%%%%%%%%%%%%%%%%%%%%%%%%%
%
%	Interactions
%
%%%%%%%%%%%%%%%%%%%%%%%%%%%%%%%%%%%%%%%%

\subsection{Stability for a General Quadratic Interaction}

\label{General Quadratic Interaction}

%%%%%%%%%%%%%%%%%%%%%%%%%%%%%%%%%%%%%%%%
%
%	Figure No Interaction vs Quadratic Interaction
%
%%%%%%%%%%%%%%%%%%%%%%%%%%%%%%%%%%%%%%%%

%%%%%%%%%%%%%%%%%%%%%%%

In this section we perform the stability analysis of the dynamical system \eqref{eq_u_1}, \eqref{eq_y_1}, and \eqref{eq_z_1} taking into account an interaction of the form \eqref{quadratic_int}. For the 3-form potential we choose a Gaussian potential $V=V_0\exp(-\kappa^2\chi^2/6)$ with a positive parameter $\xi$. Using the definitions \eqref{eq:Var1} and the Friedmann constraint \eqref{friefconstrant} we can write the interaction in \eqref{quadratic_int} as a function of $y$ and $z$ as
\begin{align}
	\label{quadratic_int_dynamical}
	Q = \frac{9H^3}{\kappa^2}\sum_{i=0}^2\alpha_{i}\left(y^2+z^2\right)^{i}
	\,.
\end{align}
The coefficients $\alpha_i$ are dimensionless constants that module the strength of the interaction and in this work we will consider them to be small, i.e., $|\alpha_i|\ll1$.
Previously we have found that a fixed point corresponding to a LSBR event exists in our dynamical system if and only if $y_{fp}^2=1$ is a root of $Q=0$. By using Eq.~\eqref{quadratic_int_dynamical}, we find that this leads to the relation
\begin{align}
	\label{no_LSBR_removal}
	\alpha_0+\alpha_1+\alpha_2=0
	\,,
\end{align}
between the interaction coefficients. If we impose this condition back on Eq.~\eqref{quadratic_int_dynamical} we obtain the general form of the interactions belonging to the class \eqref{quadratic_int} that do not remove LSBR event:
\begin{align}
	\label{Q=0_y^2=1}
	Q = \frac{9H^3}{\kappa^2}
	\left[\left(y^2+z^2\right)-1\right]
	\left[\alpha_1+\alpha_2 +\alpha_2\left(y^2+z^2\right)\right]
	\,.
\end{align}
Since we can use the Friedmann constraint \eqref{friefconstrant} to write $(y^2+z^2)-1=-s^2$, we conclude that the interactions that verify \eqref{no_LSBR_removal}, and therefore do not remove LSBR event, are the ones proportional to the energy density of DM. In order to understand this result we can express the interaction term in Eq.~\eqref{Q=0_y^2=1} as $Q=3H\rho_m\, g(\alpha_1,\,\alpha_2,\,\Omega_\chi)$, with $g$ a linear function of $\Omega_\chi$. Upon substitution in Eq.~\eqref{int_rhom},  we obtain
\begin{align}
	\label{interaction_rhom_eq1}
	\dot\rho_m + 3H\left(1+g\right)\rho_m &~=0
	\,.
\end{align}
Within the assumption of small interaction coefficients, i.e., considering a weak interaction between DM and the 3-form, the function $g$ will satisfy $|g|\ll1$ at all times during the evolution of the universe. Therefore, from Eq.~\eqref{interaction_rhom_eq1} we find that the presence of the interaction is not able to prevent the complete decay of the energy density of DM in the asymptotic future. This means that, just as in the non-interacting case, at very late-time the 3-form field is not affected by the presence of DM and drives the universe towards a LSBR event. Finally, we point out that this condition is completely independent of the shape of the potential, therefore, it is not limited to the case studied with more detail in this work of a 3-form with a Gaussian potential.

\subsubsection{Type I fixed points}

Substituting the expression \eqref{quadratic_int_dynamical} in the condition \eqref{TypeIfix}, we find that the fixed points of Type~I are given by
\begin{align}
	\label{TypeI_fixedpoints_interactoin}
	\begin{array}{c}
		\\
		\textbf{Type~I fixed points:}\\
		\textrm{(Quadratic interaction)}
	\end{array}
	\qquad
	\begin{cases}
	\hat{p}_0^\pm:
	\quad&
	\left(u_{fp}, \, y_{fp}, \, z_{fp}\right) = \left(\pm\dfrac{2}{\pi}\arctan\left(\hat{y}_{-}\right),\,\pm \hat{y}_-,\,0\right)
	\,,
	\\
	\hat{p}_1^\pm :
	\quad&
	\left(u_{fp}, \, y_{fp}, \, z_{fp}\right) = \left(\pm\dfrac{2}{\pi}\arctan\left(\hat{y}_{+}\right),\,\pm \hat{y}_+,\,0\right)
	\,,
	\end{cases}
\end{align}
where $y_\pm$ is defined in terms of the interaction parameters $\alpha_i$ as
\begin{align}
	\label{typ1Q1y}
	\hat{y}_{\pm}^2=
		\frac{1 +\alpha_{1}}{2\left(1-\alpha_{2}\right)}\left[
			1
			\pm\sqrt{
				1
				+4\alpha_0\frac{1-\alpha_{2}}{\left(1 + \alpha_{1}\right)^2}
			}
		\right]
	\,.
\end{align}
As we switch off the interaction the pairs of fixed points $\hat{p}_0^\pm$ and $\hat{p}_1^\pm$ converge, respectively, to the points $p_0$ and $p_1^\pm$ obtained above. In the regime of small coefficients $\alpha_i$ we recover $\hat{y}_-^2=0$ if $\alpha_0=0$, while $\hat{y}_+^2=1$ if the condition \eqref{no_LSBR_removal} is verified. If we expand \eqref{typ1Q1y} at first order in the $\alpha_i$'s we find
\begin{align}
	\hat{y}_{-} ^2 \simeq -\alpha_0
	\,,
	\qquad
	\hat{y}_{+} ^2 \simeq 1 + \left(\alpha_0 + \alpha_1 + \alpha_2\right)
	\,.
\end{align}
Thus, for small interaction coefficients, the existence of $\hat{p}_0^\pm$ is conditioned to $\alpha_0\leq0$ while the existence of $\hat{p}_1^\pm$ is verified only if $\alpha_0+\alpha_1+\alpha_2\leq0$.
As previously stated, for the Type~I fixed points we have $w_\mathrm{tot}=-y_{fp}^2$. Therefore, $\hat{p}_0^\pm$ correspond either to a matter era, in the case of $\alpha_0=0$, or to a scaling behaviour between DM and DE with DM dominance, while the pair $\hat{p}_0^\pm$ corresponds to LSBR events if $\alpha_0+\alpha_1+\alpha_2=0$, or to scaling behaviour with 3-form dominance if otherwise.

The eigenvalues of the Jacobian at $\hat{p}_0^\pm$ and $\hat{p}_1^\pm$ can be written in terms of $\hat{y}_\pm^2$, $\alpha_1$, and $\alpha_2$ as
\begin{align}
	\label{type1_eigenvalues}
	\left\{
		-3,
		\frac{3}{2}(1-\hat{y}_{\pm}^{2}),
		-3\left[
			2\left(1-\alpha_{2}\right)\hat{y}_{\pm}^{2}
			-\left(1+\alpha_1\right)
		\right]
	\right\}
	\,,
\end{align}
where $\hat{y}_-$ ($\hat{y}_+$) on the above formula corresponds to the case $\hat{p}^\pm_0$ ($\hat{p}^\pm_1$).
We find that the Jacobian has two negative and one positive eigenvalues if $\hat{y}_\pm^2\neq1$ and two negative and one null eigenvalues when $\hat{y}_\pm^2=1$. Therefore the pair of fixed points $\hat{p}_0^\pm$ always correspond to two saddle points, while the pair $\hat{p}_1^\pm$ also corresponds to two saddles if $\hat{y}_+^2\neq1$, i.e., when LSBR event is removed. In the case of $\hat{y}_+^2=1$, we employ once more the method described in \cite{Rendall:2001it,Boehmer:2011tp}, based on Centre Manifold Theory, to decide on the stability of the system. { The results obtained are analogous to the ones found in subsection~\ref{No Interaction} for the case of no interaction: $\hat{p}_1^+$ ($\hat{p}_1^-$) is an attractor if $\lambda(u_{fp})>0$ ($\lambda(u_{fp})<0$) and a saddle node if $\lambda(u_{fp})<0$ ($\lambda(u_{fp})>0$). For the types of potentials considered in this work, Gaussian potentials with $\xi>0$, both $\hat{p}_1^\pm$ are late-time attractors.}

\subsubsection{Type II fixed points}

When we take into account the interactions given by Eq.~\eqref{quadratic_int_dynamical} with a Gaussian potential, we find that the condition \eqref{TypeIfix-2} that defines the Type~II fixed points can be re-written as
\begin{align}
	\label{C1type2}
	\alpha_2 \beta^2 + \left(\alpha_1-1\right)\beta + \left(1+\alpha_0\right)=&~0
	\,,
	\\
	\label{C2type2}
	\left(\beta - y_{fp}^{2}\right)y_{fp}^2 - \frac{9}{2\xi}\left(1-\beta\right) =&~ 0
	\,,
\end{align}
with
\begin{align}
	\beta\equiv y_{fp}^2+z_{fp}^2
	\,,
\end{align}
 and $u_{fp}=(2/\pi)\arctan(y_{fp})$. Hence, $0<\beta\leq1$ corresponds to the relative energy density of the 3-form at the fixed point. An immediate conclusion from these solutions is that, for the class of interactions considered, the value of $\beta$ is determined solely by the interaction coefficients $\alpha_i$ and does not depend on the choice of the potential. { A similar result was found in subsection \ref{subsec_Fixed_Points} for the total parameter of EoS as in the case of Type~II fixed points we have $w_\mathrm{tot}=-1$ independently of the potential considered.}

The solutions to Eq.~\eqref{C1type2} are
\begin{align}
	\label{beta_solutions}
	\beta_\pm = \frac{1-\alpha_1}{2\alpha_2}\left[
		1
		\pm\sqrt{1 - 4\alpha_2\frac{1+\alpha_0}{\left(1-\alpha_1\right)^2}}
	\right]
	\,,
\end{align}
which, at first order in the coefficients $\alpha_i$ can be written as
\begin{align}
	\beta_+ \simeq \frac{1-\alpha_1}{\alpha_2} - 1 - \left(\alpha_0 + \alpha_1 + \alpha_2\right)
	\,,
	\qquad
	\beta_- \simeq 1 + \left(\alpha_0 + \alpha_1 + \alpha_2\right)
	\,.
\end{align}
In the regime of small coefficients, $|\alpha_i|\ll1$, the solution $\beta_+$ is outside the interval $(0,1)$ and can be disregarded, while the solution $\beta_-$ lies within the interval $(0,1)$ and is close to unity ($\beta_-\lesssim1$) for $\alpha_0+\alpha_1+\alpha_2\lesssim0$, meaning that the the 3-form dominates over DM near the fixed points ($\rho_\chi\gg\rho_m$). Notice that this is the same condition found for the existence of the Type~I fixed points $\hat{p}_1^\pm$.

Setting $\beta=\beta_-$ in Eq.~\eqref{C2type2} and solving for $y_{fp}$ we find that the dynamical system has at most two pairs of Type~II fixed points:
\begin{align}
	\label{TypeII_fixedpoints_interactoin}
	\begin{array}{c}
		\\
		\textbf{Type~II fixed points:}\\
		\textrm{(Quadratic interaction)}
	\end{array}
	\qquad
	\begin{cases}
	\tilde{p}_1^\pm :
	\quad&
	\left(u_{fp}, \, y_{fp}, \, z_{fp}\right) = \left(\pm\dfrac{2}{\pi}\arctan\left(\tilde{y}_{+}\right),\,\pm \tilde{y}_+,\,\sqrt{\beta_- - \tilde{y}_+^2}\right)
	\,,
	\\
	\tilde{p}_2^\pm:
	\quad&
	\left(u_{fp}, \, y_{fp}, \, z_{fp}\right) = \left(\pm\dfrac{2}{\pi}\arctan\left(\tilde{y}_{-}\right),\,\pm \tilde{y}_-,\, \sqrt{\beta_- - \tilde{y}_-^2}\right)
	\,,
	\end{cases}
\end{align}
where
\begin{align}
	\label{typeII_y_sol}
	\tilde{y}_{\pm}^2 = \frac{\beta_-}{2}\left(
		1
		\pm\Delta
	\right)
	\,,
	\qquad
	\Delta\left(\beta,\xi\right)\equiv\sqrt{1-\frac{18}{\xi}\frac{1-\beta_-}{\beta_-^2}}
	\,.
\end{align}
For $\xi>0$, where $\xi$ is the slope of the Gaussian potential, we find that $0\leq\Delta<1$, thus ensuring the existence of Type~II fixed points, as long as
\begin{align}
	\label{cs2_constraint}
	\frac{9}{\xi}\left(\sqrt{1+\frac{2\xi}{9}}-1\right)\leq\beta_-\leq1
	\,,
	\qquad
	0<\xi<\frac{9}{2}
	\,.
\end{align}

{ As we switch off the interaction, we find from Eq.~\eqref{C1type2} that $\beta_-=1$. This in turn implies that $\Delta=1$, cf. Eq.~\eqref{typeII_y_sol}, and that the pairs $\tilde{p}_1^\pm$ and $\tilde{p}_2^\pm$ converge, respectively, to the points $p_1^\pm$ and $p_2$ obtained above.
In fact, this result is obtained whenever $\beta_-=1$, for which the no-interaction case is only a particular example, as Eq.~\eqref{C1type2} with $\beta_-=1$ implies the condition \eqref{no_LSBR_removal} for the existence of the fixed points corresponding to the LSBR event.
Since the points $p_1^\pm$ belong to the Type~I category, when $\beta_-=1$ the only fixed point of Type~II is $\tilde{p}_2=p_2$. 

Another limiting case occurs when $\beta_-$ approaches the lower bound in Eq.~\eqref{cs2_constraint}, i.e. $\Delta$ vanishes and $\tilde{y}_-^2=\tilde{y}_+^2$. In this situation $\tilde{p}_1^\pm$ and $\tilde{p}_2^\pm$ merge and we obtain just two Type~II fixed points:
\begin{align}
	\label{Delta0_solutions}
	\begin{array}{c}
		\\
		\textbf{Type~II fixed points for $\Delta=0$:}\\
		\textrm{(Quadratic interaction)}
	\end{array}
	\qquad
	\tilde{p}_{\Delta}^\pm :
	\quad&
	\left(u_{fp},\,y_{fp},\,z_{fp}\right) = \left(
		\frac{2}{\pi}\arctan\left(\sqrt{\frac{\beta_-}{2}}\right),\,
		\pm\sqrt{\frac{\beta_-}{2}},\,
		\sqrt{\frac{\beta_-}{2}}
	\right)
	\,.
\end{align}
We thus reach the conclusion that the solutions presented in Eq.~\eqref{TypeII_fixedpoints_interactoin} represent four different fixed points as long as $\Delta\neq0$ and $\beta_-\neq1$. Since we expect the structure of the dynamical system around the fixed points to change when $\beta_-=1$ or $\Delta=0$, we will first present the mechanism to study the stability of the fixed points in the general case and then treat each of the specific cases individually.
}

For general values of the parameters of the model $(\xi,\alpha_0,\alpha_0,\alpha_2)$, the expressions for the eigenvalues of the Jacobian for the Type~II fixed points are too cumbersome to be analysed analytically. A much more fruitful approach is then to use the Hurwitz criterion {(cf. Ref.~\cite{Olver2010a} for the definition and Appendix~\ref{AppendixB} for its application to the cubic case)}. If we define the characteristic polynomial of the Jacobian, $J$, as $\det \left(J-\gamma \mathbb{I}_3\right) = a_0 + a_1 \gamma + a_2 \gamma^2- \gamma^3$, where $\mathbb{I}_3$ is the $3\times3$ identity matrix and $a_i$, $i=0,1,2$, are constant coefficients, then using Hurwitz criterion we can state that all three eigenvalues $\gamma$ of the Jacobian have negative real part if and only if
\begin{align}
	\label{Hurwitz_degree3}
	a_0<0
	\,,
	\qquad
	\sign\,a_1=\sign\,a_0
	\,,
	\qquad
	a_1a_2+a_0 >0
	\,.
\end{align}
As a corollary, when $a_0>0$ there is at least one positive eigenvalue and the fixed point is unstable. For interactions that only have one free parameter and $\beta_-\neq1$, we could obtain the regions of stability of $\tilde{p}_1^\pm$ and $\tilde{p}_2^\pm$ in terms of $\beta_-$ and $\Delta$. Although we did not obtain a general result for the stability of these points, from analogy with the non-interaction case it might be expected that late-time attractors are provided by the pair $\tilde{p}_1^\pm$, while the pair $\tilde{p}_2^\pm$ gives unstable points. 

{
Next, we look at the particular case of $\beta_-=\Delta=1$. In this case only one fixed point of Type~II exists: $\tilde{p}_2=(0,\,0,\,1)$. The analysis of the system around this point presents serious challenges as the interaction term in Eq.~\eqref{eq_y_1} diverges at the fixed point. However, as stated above $\beta_-=1$ occurs only for those interactions for which the LSBR event is not removed. As such, all the trajectories of interest converge to fixed point corresponding to the LSBR and are always far away from $\tilde{p}_2$. We will not look into the stability of this point with further detail.

Finally, for $\Delta=0$, which occurs when $\xi=18(1-\beta_-)/\beta_-^2$, the only two fixed points of Type~II are the ones in Eq.~\eqref{Delta0_solutions}. Notice, however, that for values of $\beta_-\lesssim1$ this implies very small values of $\xi$, i.e., potentials that are almost flat. In this case we find that one eigenvalue is always zero, therefore the analysis of the linearised system is not valid to decide on the stability of the system near the fixed points and we need to employ methods based on Centre Manifold Theory, see e.g. Ref.~\cite{Rendall:2001it,Boehmer:2011tp}.
The reader may notice that the solutions in \eqref{TypeII_fixedpoints_interactoin} and \eqref{Delta0_solutions} do not have the same limit when we take $\beta_-\rightarrow1$. This difference comes from the fact that in \eqref{Delta0_solutions} we are imposing the equality $\xi=18(1-\beta_-)/\beta_-^2$. In the limit of $\beta_-=1$, $\xi=0$ and we obtain a 3-form with a constant potential which, as described above, is a special case where the 3-form behaves identically as a cosmological constant.
}

\subsubsection{Type III fixed points}

As mentioned in the previous section, the Type~III fixed points in 3-form models with Gaussian potential with positive $\xi$ exist only along the lines $u_{fp}=\pm1\,\wedge\,z_{fp}=0$. The fact that $z_{fp}=0$ suggests an analogy with the results found for the Type~I. In fact, we find that as we turn on the interaction, for each set of coefficients $(\alpha_0,\,\alpha_1,\,\alpha_2)$ the value of $y_{fp}^2$ is given by Eq.~\eqref{typ1Q1y}, and
\begin{align}
	\begin{array}{c}
	\\
		\textbf{Type~III fixed points:}\\
		\textrm{(Quadratic Interaction)}
	\end{array}
	\qquad
	\begin{cases}
	\hat{\pi}_{-0}^\pm:
	\quad&
	\left(u_{fp}, \, y_{fp}, \, z_{fp}\right) = \left(\pm1,\,-\hat y_-,\,0\right)
	\,,
	\\
	\hat{\pi}_{+0}^\pm:
	\quad&
	\left(u_{fp}, \, y_{fp}, \, z_{fp}\right) = \left(\pm1,\,+\hat y_-,\,0\right)
	\,,
	\\
	\hat\pi_{-1}^\pm :
	\quad&
	\left(u_{fp}, \, y_{fp}, \, z_{fp}\right) = \left(\pm1,\,-\hat y_+,\,0\right)
	\,,
	\\
	\hat\pi_{+1}^\pm :
	\quad&
	\left(u_{fp}, \, y_{fp}, \, z_{fp}\right) = \left(\pm1,\,+\hat y_+,\,0\right)
	\,.
	\end{cases}
\end{align}
The eigenvalues of the Jacobian for these fixed points are
\begin{align}
	\label{eigenvalues_III}
	\left\{3, 3\left[1+\alpha_1 - 2\left(1 - \alpha_2\right)y_{fp}^2\right], +\infty\right\}
	\,.
\end{align} 
Therefore, the fixed points $\pi_{\pm0}^\pm$, with $y_{fp}^2=y_-^2\sim0$ correspond to highly repulsive points representing the asymptotic past, while $\pi_{\pm1}^\pm$, with $y_{fp}^2=y_+^2\sim1$ correspond to saddle points%
\footnote{Notice that here we are assuming the interpretation that an positive infinity eigenvalue means that the system is highly repulsive in the direction of the corresponding eigenvector. We explore in more details the mathematical tools to study the stability of fixed points at infinity in a different work \cite{ Bouhmadi:2016}.}.

In the next section we analyse in more detail some particular cases of the interaction class \eqref{quadratic_int_dynamical}.

%%%%%%%%%%%%%%%%%%%%%%%%%%%%%%%%%%%%%%%%
%
%	Specific Examples
%
%%%%%%%%%%%%%%%%%%%%%%%%%%%%%%%%%%%%%%%%

\subsection{Specific Examples}

\label{Examples}

%%%%%%%%%%%%%%%%%%%%%%%%%%%%%%%%%%%%%%%%
%
%	Fixed Points Table
%
%%%%%%%%%%%%%%%%%%%%%%%%%%%%%%%%%%%%%%%%

\begin{table}[t]
\footnotesize
\centering
\begin{tabularx}{\textwidth}{ C{.35} C{.35} C{1.25} C{2.35} C{.6} C{1.1}}
\hline
	{\textbf{Inter.}}
	&
	{\textbf{Fixed Points}}
	& 
	{\textbf{Constraints}}
	& 
	{\bf{$ (u_{fp},\,y_{fp},\,z_{fp})$}}
	& 
	{\textbf{Stability}}
	& 
	{\textbf{Description}}
\\ \hline
%%%%%%%%%%%%%
	\multirow{ 6}{\hsize}{\centering$Q=0$}
	&
	$p_0$
	&
	
	&
	$\left(0,\,0,\,0\right)$
	&
	Saddle
	&
	Matter Domination
 \\ 
 %%%%%%%%%%%%%
	
	&
	$p_1^\pm$
	&
	
	&
	$\left(\pm1/2,\,\pm1,\,0\right)$
	&
	Focus Node
	&
	LSBR
 \\ 
 %%%%%%%%%%%%%
	
	&
	$p_2$
	&
	
	&
	$\left(0,\,0,\,1\right)$
	&
	Saddle Node
	&
	Potential Domination
 \\ 
%%%%%%%%%%%%%
	 
	&
	$\pi_0^\pm$
	&
	
	&
	$\left(\pm1,\,0,\,0\right)$
	&
	Repulsive
	&
	Matter Domination
 \\ 
 %%%%%%%%%%%%%
	
	&
	$\pi_{+1}^\pm$
	&
	
	&
	$\left(\pm1,\,1,\,0\right)$
	&
	Saddle
	&
	Kinetic Domination
 \\ 
 %%%%%%%%%%%%%
	
	&
	$\pi_{-1}^\pm$
	&
	
	&
	$\left(\pm1,\,-1,\,0\right)$
	&
	Saddle
	&
	Kinetic Domination
 \\ \hline
%%%%%%%%%%%%%
	\multirow{9}{\hsize}{\centering I}
	&
	$\hat{p}_0^\pm$
	&
	$-1<\alpha_m<0$
	&
	$\left(\pm\frac{2}{\pi}\arctan\sqrt{-\alpha_m},\, \pm\sqrt{-\alpha_m},\,0\right)$
	&
	Saddle
	&
	Scal. sol. ($w_\mathrm{tot}\lesssim0$)\quad Matter Domination
 \\ 
%%%%%%%%%%%%%
	
	&
	$\hat{p}_1^\pm$
	&
	
	&
	$\left(\pm1/2,\,\pm1,\,0\right)$
	&
	Focus Node
	&
	LSBR
 \\ 
%%%%%%%%%%%%%
	
	&
	$\tilde{p}_2$
	&
	
	&
	$\left(0,\,0,\,1\right)$
	&
	Saddle
	&
	Potential Domination
\\ 
 %%%%%%%%%%%%%
	
	&
	$\hat\pi_{+0}^\pm$
	&
	$-1<\alpha_m<0$
	&
	$\left(\pm1,\,\sqrt{-\alpha_m},\,0\right)$
	&
	Repulsive
	&
	Scal. sol. ($w_\mathrm{tot}\lesssim0$)\quad Matter Domination
 \\ 
 %%%%%%%%%%%%%
	
	&
	$\hat\pi_{-0}^\pm$
	&
	$-1<\alpha_m<0$
	&
	$\left(\pm1,\,-\sqrt{-\alpha_m},\,0\right)$
	&
	Repulsive
	&
	Scal. sol. ($w_\mathrm{tot}\lesssim0$)\quad Matter Domination
 \\ 
 %%%%%%%%%%%%%
	
	&
	$\hat\pi_{+1}^\pm$
	&
	
	&
	$\left(\pm1,\,1,\,0\right)$
	&
	Saddle
	&
	Kinetic Domination
 \\ 
 %%%%%%%%%%%%%
	
	&
	$\hat\pi_{-1}^\pm$
	&
	
	&
	$\left(\pm1,\,-1,\,0\right)$
	&
	Saddle
	&
	Kinetic Domination
 \\ \hline
%%%%%%%%%%%%%
	\multirow{ 9}{\hsize}{\centering II}
	&
	$\hat{p}_0^\pm$
	&
	$\alpha_{mm}<0$
	&
	$\left(\pm\frac{2}{\pi}\arctan\sqrt{-\frac{\alpha_{mm}}{1-\alpha_{mm}}},\, \pm\sqrt{-\frac{\alpha_{mm}}{1-\alpha_{mm}}},\,0\right)$
	&
	Saddle
	&
	Scal. sol. ($w_\mathrm{tot}\lesssim0$)\quad Matter Domination
 \\ 
%%%%%%%%%%%%%
	
	&
	$\hat{p}_1^\pm$
	&
	
	&
	$\left(\pm1/2,\,\pm1,\,0\right)$
	&
	Focus Node
	&
	LSBR
 \\ 
%%%%%%%%%%%%%
	
	&
	$\tilde{p}_2$
	&
	
	&
	$\left(0,\,0,\,1\right)$
	&
	Undecided
	&
	Potential Domination
\\ 
 %%%%%%%%%%%%%
	
	&
	$\hat\pi_{+0}^\pm$
	&
	$\alpha_{mm}<0$
	&
	$\left(\pm1,\,\sqrt{-\frac{\alpha_{mm}}{1-\alpha_{mm}}},\,0\right)$
	&
	Repulsive
	&
	Scal. sol. ($w_\mathrm{tot}\lesssim0$)\quad Matter Domination
 \\ 
 %%%%%%%%%%%%%
	
	&
	$\hat\pi_{-0}^\pm$
	&
	$\alpha_{mm}<0$
	&
	$\left(\pm1,\,-\sqrt{-\frac{\alpha_{mm}}{1-\alpha_{mm}}},\,0\right)$
	&
	Repulsive
	&
	Scal. sol. ($w_\mathrm{tot}\lesssim0$)\quad Matter Domination
\\
 %%%%%%%%%%%%%
	
	&
	$\hat\pi_{+1}^\pm$
	&
	
	&
	$\left(\pm1,\,1,\,0\right)$
	&
	Saddle
	&
	Kinetic Domination
 \\ 
 %%%%%%%%%%%%%
	
	&
	$\hat\pi_{-1}^\pm$
	&
	
	&
	$\left(\pm1,\,-1,\,0\right)$
	&
	Saddle
	&
	Kinetic Domination
 \\ \hline
%%%%%%%%%%%%%
	\multirow{ 6}{\hsize}{\centering III}
	&
	$\hat{p}_0$
	&
	
	&
	$\left(0,\,0,\,0\right)$
	&
	Saddle
	&
	Matter Domination
 \\ 
 %%%%%%%%%%%%%
	
	&
	$\hat{p}_1^\pm$
	&
	
	&
	$\left(\pm1/2,\,\pm1,\,0\right)$
	&
	Focus Node
	&
	LSBR
 \\ 
%%%%%%%%%%%%%
	
	&
	$\tilde{p}_2$
	&
	
	&
	$\left(0,\,0,\,1\right)$
	&
	Undecided
	&
	Potential Domination
 \\ 
%%%%%%%%%%%%%
	 
	&
	$\hat\pi_0^\pm$
	&
	
	&
	$\left(\pm1,\,0,\,0\right)$
	&
	Repulsive
	&
	Matter Domination
 \\ 
 %%%%%%%%%%%%%
	
	&
	$\hat\pi_{+1}^\pm$
	&
	
	&
	$\left(\pm1,\,1,\,0\right)$
	&
	Saddle
	&
	Kinetic Domination
 \\ 
 %%%%%%%%%%%%%
	
	&
	$\hat\pi_{-1}^\pm$
	&
	
	&
	$\left(\pm1,\,-1,\,0\right)$
	&
	Saddle
	&
	Kinetic Domination
 \\ \hline
 %%%%%%%%%%%%%

\end{tabularx}
\caption{\label{Table_fixedpoints}%
The fixed points found for the 3-form DE model with DM in the case of no interaction and for the interactions I, II, and III studied in Sect~\ref{Examples}. Notice that none of the interactions presented in this Table remove the points $(\pm1/2,\,\pm1,\,0)$ corresponding to the LSBR event.
}
\end{table}

%%%%%%%%%%%%%%%%%%%%%%%%%%%%%%%%%%%%%%%%

%%%%%%%%%%%%%%%%%%%%%%%%%%%%%%%%%%%%%%%%
%
%	Fixed Points Table
%
%%%%%%%%%%%%%%%%%%%%%%%%%%%%%%%%%%%%%%%%

\begin{table}[t]
\footnotesize
\centering
\begin{tabularx}{\textwidth}{ C{.35} C{.35} C{1.25} C{2.35} C{.6} C{1.1}}
\hline
	{\textbf{Inter.}}
	&
	{\textbf{Fixed Points}}
	& 
	{\textbf{Constraints}}
	& 
	{\bf{$ (u_{fp},\,y_{fp},\,z_{fp})$}}
	& 
	{\textbf{Stability}}
	& 
	{\textbf{Description}}
\\ \hline
%%%%%%%%%%%%%
	\multirow{ 6}{\hsize}{\centering$Q=0$}
	&
	$p_0$
	&
	
	&
	$\left(0,\,0,\,0\right)$
	&
	Saddle
	&
	Matter Domination
 \\ 
 %%%%%%%%%%%%%
	
	&
	$p_1^\pm$
	&
	
	&
	$\left(\pm1/2,\,\pm1,\,0\right)$
	&
	Focus Node
	&
	LSBR
 \\ 
 %%%%%%%%%%%%%
	
	&
	$p_2$
	&
	
	&
	$\left(0,\,0,\,1\right)$
	&
	Saddle Node
	&
	Potential Domination
 \\ 
%%%%%%%%%%%%%
	 
	&
	$\pi_0^\pm$
	&
	
	&
	$\left(\pm1,\,0,\,0\right)$
	&
	Repulsive
	&
	Matter Domination
 \\ 
 %%%%%%%%%%%%%
	
	&
	$\pi_{+1}^\pm$
	&
	
	&
	$\left(\pm1,\,1,\,0\right)$
	&
	Saddle
	&
	Kinetic Domination
 \\ 
 %%%%%%%%%%%%%
	
	&
	$\pi_{-1}^\pm$
	&
	
	&
	$\left(\pm1,\,-1,\,0\right)$
	&
	Saddle
	&
	Kinetic Domination
 \\ \hline
%%%%%%%%%%%%%
	\multirow{ 11}{\hsize}{\centering IV}
	&
	$\hat{p}_0$
	&
	
	&
	$\left(0,\,0,\,0\right)$
	&
	Saddle
	&
	Matter Domination
 \\ 
 %%%%%%%%%%%%%
	
	&
	$\hat{p}_1^\pm$
	&
	$-1<\alpha_\chi<0$
	&
	$\left(\pm\frac{2}{\pi}\arctan\sqrt{1+\alpha_\chi},\, \pm\sqrt{1+\alpha_\chi},\,0\right)$
	&
	Saddle
	&
	Scal. sol. ($w_\mathrm{tot}\gtrsim-1$)\quad Kinetic Domination
\\
 %%%%%%%%%%%%%
	
	&
	$\tilde{p}_1^\pm$
	&
	%\multirow{2}{\hsize}{
	%\centering
	$-\frac{\sqrt{1+\frac{2\xi}{9}}-1}{2} < \alpha_\chi <0$
	%}
	&
	$\left(\pm\frac{2}{\pi}\arctan\sqrt{\frac{1+\Delta}{2}\beta_-},\, \pm\sqrt{\frac{1+\Delta}{2}\beta_-},\,\sqrt{\frac{1-\Delta}{2}\beta_-}\right)$
	&
	Focus Node
	&
	Scal. sol. (de Sitter)\quad Kinetic Domination
 \\ 
%%%%%%%%%%%%%
	
	&
	$\tilde{p}_2^\pm$
	&
	$-\frac{\sqrt{1+\frac{2\xi}{9}}-1}{2} < \alpha_\chi <0$
	&
	$\left(\pm\frac{2}{\pi}\arctan\sqrt{\frac{1-\Delta}{2}\beta_-},\, \pm\sqrt{\frac{1-\Delta}{2}\beta_-},\,\sqrt{\frac{1+\Delta}{2}\beta_-}\right)$
	&
	Unstable
	&
	Scal. sol. (de Sitter)\quad Potential Domination
 \\ 
%%%%%%%%%%%%%
	
	&
	$\tilde{p}_\Delta^\pm$
	&
	$\alpha_\chi = -\frac{\sqrt{1+\frac{2\xi}{9}}-1}{2}$
	&
	$\left(\pm\frac{2}{\pi}\arctan\sqrt{\frac{\beta_-}{2}},\, \pm\sqrt{\frac{\beta_-}{2}},\,\sqrt{\frac{\beta_-}{2}}\right)$
	&
	Saddle Node
	&
	Scal. sol. (de Sitter)\quad 3-form Domination
 \\ 
%%%%%%%%%%%%%
	 
	&
	$\pi_0^\pm$
	&
	
	&
	$\left(\pm1,\,0,\,0\right)$
	&
	Repulsive
	&
	Matter Domination
 \\ 
 %%%%%%%%%%%%%
	
	&
	$\pi_{+1}^\pm$
	&
	
	&
	$\left(\pm1,\,+\sqrt{1+\alpha_\chi},\,0\right)$
	&
	Saddle
	&
	Kinetic Domination
 \\ 
 %%%%%%%%%%%%%
	
	&
	$\pi_{-1}^\pm$
	&
	
	&
	$\left(\pm1,\,-\sqrt{1+\alpha_\chi},\,0\right)$
	&
	Saddle
	&
	Kinetic Domination
 \\ \hline
 %%%%%%%%%%%%%
	\multirow{ 12}{\hsize}{\centering V}
	&
	$\hat{p}_0$
	&
	
	&
	$\left(0,\,0,\,0\right)$
	&
	Saddle
	&
	Matter Domination
 \\ 
 %%%%%%%%%%%%%
	
	&
	$\hat{p}_1^\pm$
	&
	$\alpha_{\chi\chi}<0$
	&
	$\left(\pm\frac{2}{\pi}\arctan\frac{1}{\sqrt{1-\alpha_{\chi\chi}}},\,\pm\frac{1}{\sqrt{1-\alpha_{\chi\chi}}},\,0\right)$
	&
	Saddle
	&
	Scal. sol. ($w_\mathrm{tot}>-1$)\quad Kinetic Domination
\\
 %%%%%%%%%%%%%
	
	&
	$\tilde{p}_1^\pm$
	&
	%\multirow{2}{\hsize}{
	%\centering
	$-\frac{\xi}{18}<\alpha_{\chi\chi}<0$
	%}
	&
	$\left(\pm\frac{2}{\pi}\arctan\sqrt{\frac{1+\Delta}{2}\beta_-},\, \pm\sqrt{\frac{1+\Delta}{2}\beta_-},\,\sqrt{\frac{1-\Delta}{2}\beta_-}\right)$
	&
	Focus Node
	&
	Scal. sol. (de Sitter)\quad Kinetic Domination
 \\ 
%%%%%%%%%%%%%
	
	&
	$\tilde{p}_2^\pm$
	&
	$-\frac{\xi}{18}<\alpha_{\chi\chi}<0$
	&
	$\left(\pm\frac{2}{\pi}\arctan\sqrt{\frac{1-\Delta}{2}\beta_-},\, \pm\sqrt{\frac{1-\Delta}{2}\beta_-},\,\sqrt{\frac{1+\Delta}{2}\beta_-}\right)$
	&
	Unstable
	&
	Scal. sol. (de Sitter)\quad Potential Domination
 \\ 
%%%%%%%%%%%%%
	
	&
	$\tilde{p}_\Delta^\pm$
	&
	$\alpha_{\chi\chi} = -\frac{\xi}{18}$
	&
	$\left(\pm\frac{2}{\pi}\arctan\sqrt{\frac{\beta_-}{2}},\, \pm\sqrt{\frac{\beta_-}{2}},\,\sqrt{\frac{\beta_-}{2}}\right)$
	&
	Saddle Node
	&
	Scal. sol. (de Sitter)\quad 3-form Domination
 \\ 
%%%%%%%%%%%%%
	 
	&
	$\pi_0^\pm$
	&
	
	&
	$\left(\pm1,\,0,\,0\right)$
	&
	Repulsive
	&
	Matter Domination
 \\ 
 %%%%%%%%%%%%%
	
	&
	$\pi_{+1}^\pm$
	&
	
	&
	$\left(\pm1,\,+\frac{1}{\sqrt{1-\alpha_{\chi\chi}}},\,0\right)$
	&
	Saddle
	&
	Kinetic Domination
 \\ 
 %%%%%%%%%%%%%
	
	&
	$\pi_{-1}^\pm$
	&
	
	&
	$\left(\pm1,\,-\frac{1}{\sqrt{1-\alpha_{\chi\chi}}},\,0\right)$
	&
	Saddle
	&
	Kinetic Domination
 \\ \hline

\end{tabularx}
\caption{\label{Table_fixedpoints_II}%
The fixed points found for the 3-form DE model with DM in the case of no interaction an for the interactions IV and V studied in Sect~\ref{Examples}. The values of $\beta_-$ and $\Delta$ are defined in Eqs.~\eqref{Case_IV_beta_sol}, \eqref{Delta_interaction_IV}, \eqref{Case_V_beta_sol}, and \eqref{Delta_interaction_V}. For non-vanishing values of the interaction parameters, the two interactions presented in this Table remove the points $(\pm1/2,\,\pm1,\,0)$ corresponding to the LSBR event.
}
\end{table}

%%%%%%%%%%%%%%%%%%%%%%%%%%%%%%%%%%%%%%%%
%
%	Q propto rho_m
%
%%%%%%%%%%%%%%%%%%%%%%%%%%%%%%%%%%%%%%%%

\subsubsection{Interaction I: $Q\propto \rho_m$}

The first interaction we analyse is of the form
\begin{align}
	\label{interaction_I}
	Q = 3H \alpha_m \rho_m = \frac{9H^3}{\kappa^2} \alpha_m \left[1-\left(y^2+z^2\right)\right]
	\,.
\end{align}
In the context of 3-forms this interaction has been studied before in Refs.~\cite{Boehmer:2011tp,Ngampitipan:2011se}.
Comparison with Eq.~\eqref{quadratic_int_dynamical} immediately shows that this corresponds to setting $\alpha_0=\alpha_m$, $\alpha_1=-\alpha_m$, and $\alpha_2=0$. Since this interaction is proportional to the energy density of DM the fixed points corresponding to LSBR are not removed. In fact, it can be easily verified that this choice of coefficients satisfies the condition \eqref{no_LSBR_removal}.

{ With an interaction of the kind of \eqref{interaction_I} we find the following pairs of fixed points of Type~I:
\begin{align}
	\label{Interaction_I_TypeI}
	\begin{array}{c}
		\\
		\textbf{Type~I fixed points:}\\
		\textrm{(Interaction I)}
	\end{array}
	\qquad
	\begin{cases}
	\hat{p}_0^\pm:
	\quad&
	\left(u_{fp}, \, y_{fp}, \, z_{fp}\right) = \left(\pm\dfrac{2}{\pi}\arctan\sqrt{-\alpha_m},\,\pm \sqrt{-\alpha_m},\,0\right)
	\,,
	\\
	\hat{p}_1^\pm:
	\quad&
	\left(u_{fp}, \, y_{fp}, \, z_{fp}\right) = \left(\pm\dfrac{1}{2},\,\pm 1,\,0\right)
	\,.
	\end{cases}
\end{align}
Notice that the pair $\hat{p}_0^\pm$ verifies $0<y_{fp}^2<1$ as long as $-1<\alpha_m<0$. From the results found previously in Eq.~\eqref{type1_eigenvalues} we deduce that the pair $\hat{p}_0^\pm$ correspond to saddle points while $\hat{p}_1^\pm$ are late-time attractors that lead the system to a LSBR event in the future.}
In this case, the only solution to Eq.~\eqref{beta_solutions} is $\beta_-=1$ which implies $\Delta=1$ in Eq.~\eqref{typeII_y_sol}. Therefore, $\tilde{p}_1^\pm$ correspond to $\hat{p}_1^\pm$ which belong to the Type~I category and the only new fixed point of Type~II is 
\begin{align}
	\label{Interaction_I_TypeII}
	\begin{array}{c}
		\\
		\textbf{Type~II fixed points:}\\
		\textrm{(Interaction I)}
	\end{array}
	\qquad
	\tilde{p}_2:
	\quad&
	\left(u_{fp}, \, y_{fp}, \, z_{fp}\right) = \left(0,\,0,\,1\right)
	\,.
\end{align}
Since there are terms in the Jacobian that diverge at this fixed point, we were unable to calculate its eigenvalues. Nevertheless, from the analogy with the non-interacting case, we expect the point to be unstable.

{ Finally, the fixed points of Type~III are
\begin{align}
	\label{Interaction_I_TypeIII}
	\begin{array}{c}
		\\
		\textbf{Type~III fixed points:}\\
		\textrm{(Interaction I)}
	\end{array}
	\qquad
	\begin{cases}
	\hat{\pi}_{-0}^\pm:
	\quad&
	\left(u_{fp}, \, y_{fp}, \, z_{fp}\right) = \left(\pm1,\,-\sqrt{-\alpha_m},\,0\right)
	\,,
	\\
	\hat{\pi}_{+0}^\pm:
	\quad&
	\left(u_{fp}, \, y_{fp}, \, z_{fp}\right) = \left(\pm1,\,\sqrt{-\alpha_m},\,0\right)
	\,,
	\\
	\hat\pi_{-1}^\pm :
	\quad&
	\left(u_{fp}, \, y_{fp}, \, z_{fp}\right) = \left(\pm1,\,-1,\,0\right)
	\,,
	\\
	\hat\pi_{+1}^\pm :
	\quad&
	\left(u_{fp}, \, y_{fp}, \, z_{fp}\right) = \left(\pm1,\,+1,\,0\right)
	\,.
	\end{cases}
\end{align}
 As reported above, \eqref{eigenvalues_III}, the first two pairs correspond to highly repulsive fixed points representing the past of the system, while the last two pairs are saddles and, therefore, are unstable.}

%%%%%%%%%%%%%%%%%%%%%%%%%%%%%%%%%%%%%%%%
%
%	Q propto rho_m^2
%
%%%%%%%%%%%%%%%%%%%%%%%%%%%%%%%%%%%%%%%%

\subsubsection{Interaction II: $Q\propto \rho_m^2$}

The second interaction we analyse is of the form
\begin{align}
	\label{interaction_II}
	Q = 3H \alpha_{mm} \frac{\rho_m^2}{\rho_m+\rho_\chi}= \frac{9H^3}{\kappa^2} \alpha_{mm} \left[1-\left(y^2+z^2\right)\right]^2
	\,.
\end{align}
A comparison with Eq.~\eqref{quadratic_int_dynamical} immediately shows that this corresponds to setting $\alpha_0=\alpha_{mm}$, $\alpha_1=-2\alpha_{mm}$, and $\alpha_2=\alpha_{mm}$. As in the case of the previous linear interaction, the fixed points corresponding to LSBR are not removed, because the interaction coefficients satisfy the condition \eqref{no_LSBR_removal}.

{
In the case of interaction II we find two pairs of Type~I fixed points 
\begin{align}
	\label{Interaction_II_TypeI}
	\begin{array}{c}
		\\
		\textbf{Type~I fixed points:}\\
		\textrm{(Interaction II)}
	\end{array}
	\qquad
	\begin{cases}
	\hat{p}_0^\pm:
	\quad&
	\left(u_{fp}, \, y_{fp}, \, z_{fp}\right) = \left(\pm\dfrac{2}{\pi}\arctan\sqrt{-\frac{\alpha_{mm}}{1-\alpha_{mm}}},\,\pm \sqrt{-\frac{\alpha_{mm}}{1-\alpha_{mm}}},\,0\right)
	\,,
	\\
	\hat{p}_1^\pm:
	\quad&
	\left(u_{fp}, \, y_{fp}, \, z_{fp}\right) = \left(\pm\dfrac{1}{2},\,\pm 1,\,0\right)
	\,.
	\end{cases}
\end{align}
While $\hat{p}_0^\pm$ exists within the domain of the system only for $\alpha_{mm}<0$ and correspond to saddle points, cf. Eq.~\eqref{type1_eigenvalues}, the two points $\hat{p}_1^\pm$ are late-time attractors that lead the system to a LSBR event in the future.}
In this case, the only solution to Eq.~\eqref{beta_solutions} that satisfies $0<\beta\leq1$ is $\beta_-=1$ which leads to $\Delta=1$ in Eq.~\eqref{typeII_y_sol}. Therefore $\tilde{p}_1^\pm$ correspond to $\hat{p}_1^\pm$, which belong to the Type~I category, and the only new fixed point of Type~II is 
\begin{align}
	\label{Interaction_I_TypeII}
	\begin{array}{c}
		\\
		\textbf{Type~II fixed points:}\\
		\textrm{(Interaction II)}
	\end{array}
	\qquad
	\tilde{p}_2:
	\quad&
	\left(u_{fp}, \, y_{fp}, \, z_{fp}\right) = \left(0,\,0,\,1\right)
	\,.
\end{align}
Since there are terms in the Jacobian that diverge at this fixed point, we were unable to calculate its eigenvalues. Nevertheless, from the analogy with the non-interacting case, we expect the point to be unstable.

{
Finally, the fixed points of Type~III are
\begin{align}
	\label{Interaction_I_TypeIII}
	\begin{array}{c}
		\\
		\textbf{Type~III fixed points:}\\
		\textrm{(Interaction I)}
	\end{array}
	\qquad
	\begin{cases}
	\hat{\pi}_{-0}^\pm:
	\quad&
	\left(u_{fp}, \, y_{fp}, \, z_{fp}\right) = \left(\pm1,\,-\sqrt{-\frac{\alpha_{mm}}{1-\alpha_{mm}}},\,0\right)
	\,,
	\\
	\hat{\pi}_{+0}^\pm:
	\quad&
	\left(u_{fp}, \, y_{fp}, \, z_{fp}\right) = \left(\pm1,\,\sqrt{-\frac{\alpha_{mm}}{1-\alpha_{mm}}},\,0\right)
	\,,
	\\
	\hat\pi_{-1}^\pm :
	\quad&
	\left(u_{fp}, \, y_{fp}, \, z_{fp}\right) = \left(\pm1,\,-1,\,0\right)
	\,,
	\\
	\hat\pi_{+1}^\pm :
	\quad&
	\left(u_{fp}, \, y_{fp}, \, z_{fp}\right) = \left(\pm1,\,+1,\,0\right)
	\,.
	\end{cases}
\end{align}
 As reported above, \eqref{eigenvalues_III}, the first two pairs correspond to highly repulsive fixed points representing the past of the system, while the last two pairs are saddles and therefore are unstable. We note that these types of quadratic interactions seem pretty much similar to their linear analogue.
}

%%%%%%%%%%%%%%%%%%%%%%%%%%%%%%%%%%%%%%%%
%
%	Q propto rho_m rho_chi
%
%%%%%%%%%%%%%%%%%%%%%%%%%%%%%%%%%%%%%%%%

\subsubsection{Interaction III: $Q\propto \rho_m\rho_\chi$}

The second interaction we analyse is of the form
\begin{align}
	\label{interaction_II}
	Q = 3H \alpha_{m\chi} \frac{\rho_m\rho_\chi}{\rho_m+\rho_\chi}= \frac{9H^3}{\kappa^2} \alpha_{m\chi} \left[1-\left(y^2+z^2\right)\right]\left(y^2+z^2\right)
	\,.
\end{align}
Comparison with Eq.~\eqref{quadratic_int_dynamical} immediately shows that this corresponds to setting $\alpha_0=0$, $\alpha_1=\alpha_{m\chi}$, and $\alpha_2=-\alpha_{m\chi}$. As in the previous two cases, the interaction coefficients satisfy the condition \eqref{no_LSBR_removal} and, therefore, the fixed points corresponding to LSBR are not removed.

With this interaction we obtain three fixed points of Type~I
\begin{align}
	\label{Interaction_II_TypeI}
	\begin{array}{c}
		\\
		\textbf{Type~I fixed points:}\\
		\textrm{(Interaction III)}
	\end{array}
	\qquad
	\begin{cases}
	\hat{p}_0:
	\quad&
	\left(u_{fp}, \, y_{fp}, \, z_{fp}\right) = \left(0,\,0,\,0\right)
	\,,
	\\
	\hat{p}_1^\pm :
	\quad&
	\left(u_{fp}, \, y_{fp}, \, z_{fp}\right) = \left(\pm\dfrac{1}{2},\,\pm1,\,0\right)
	\,,
	\end{cases}
\end{align}
where $\hat{p}_0$ corresponds to a matter dominated epoch and $\hat{p}_1^\pm$ to late-time attractors that lead the system to a LSBR event in the future.
For the point $\hat p_0$ we find from Eq.~\eqref{type1_eigenvalues} that the Jacobian has two negative and one positive eigenvalues and, therefore, $\hat{p}_0$ correspond to saddle fixed points. 

In this case, the only solution to Eq.~\eqref{beta_solutions} that satisfies $0<\beta\leq1$ is $\beta_-=1$ which leads to $\Delta=1$ in Eq.~\eqref{typeII_y_sol}. Therefore $\tilde{p}_1^\pm$ correspond to $\hat{p}_1^\pm$ and the only new fixed point of Type~II is 
\begin{align}
	\label{Interaction_I_TypeII}
	\begin{array}{c}
		\\
		\textbf{Type~II fixed points:}\\
		\textrm{(Interaction III)}
	\end{array}
	\qquad
	\tilde{p}_2:
	\quad&
	\left(u_{fp}, \, y_{fp}, \, z_{fp}\right) = \left(0,\,0,\,1\right)
	\,.
\end{align}
For this point there are terms in the Jacobian that diverge at this fixed point, we were unable to calculate its eigenvalues. Nevertheless, from the analogy with the non-interacting case, we expect the point to be unstable.

Finally, the fixed points of Type~III
\begin{align}
	\label{Interaction_I_TypeIII}
	\begin{array}{c}
		\\
		\textbf{Type~III fixed points:}\\
		\textrm{(Interaction I)}
	\end{array}
	\qquad
	\begin{cases}
	\hat\pi_{0}^\pm :
	\quad&
	\left(u_{fp}, \, y_{fp}, \, z_{fp}\right) = \left(\pm1,\,0,\,0\right)
	\,,
	\\
	\hat\pi_{-1}^\pm :
	\quad&
	\left(u_{fp}, \, y_{fp}, \, z_{fp}\right) = \left(\pm1,\,-1,\,0\right)
	\,,
	\\
	\hat\pi_{+1}^\pm :
	\quad&
	\left(u_{fp}, \, y_{fp}, \, z_{fp}\right) = \left(\pm1,\,+1,\,0\right)
	\,.
	\end{cases}
\end{align}
As reported above, \eqref{eigenvalues_III}, the points $\hat{\pi}_0$ are extremely repulsive and correspond to the asymptotic past matter era, while the points $\hat \pi_{\pm1}^{\pm}$ are saddles and therefore unstable.
%%%%%%%%%%%%%%%%%%%%%%%%%%%%%%%%%%%%%%%%
%
%	Q propto rho_chi
%
%%%%%%%%%%%%%%%%%%%%%%%%%%%%%%%%%%%%%%%%

\subsubsection{Interaction IV: $Q\propto \rho_\chi$}
\label{FixedPoints_InteractionIV}

The second interaction we analyse is of the form
\begin{align}
	\label{interaction_II}
	Q = 3H \alpha_{\chi} \rho_\chi = \frac{9H^3}{\kappa^2} \alpha_{\chi} \left(y^2+z^2\right)
	\,.
\end{align}
Comparison with Eq.~\eqref{quadratic_int_dynamical} immediately shows that this corresponds to setting $\alpha_0=0$, $\alpha_1=\alpha_{\chi}$, and $\alpha_2=0$. In this case the interaction coefficients do not satisfy the condition \eqref{no_LSBR_removal} and therefore the fixed points corresponding to LSBR will be removed.
{
From Eq.~\eqref{conservation_equations} we find that for an interaction of the type of Eq.~\eqref{interaction_II} the energy transfer between DM and DE is not stopped when the energy density of DM vanishes. As such, in order to guarantee that the energy density of DM does not evolve to negative values we impose the constraint $\alpha_{\chi}\leq0$.
}

With this interaction we obtain three fixed points of Type~I
\begin{align}
	\label{Interaction_IV_TypeI}
	\begin{array}{c}
		\\
		\textbf{Type~I fixed points:}\\
		\textrm{(Interaction IV)}
	\end{array}
	\qquad
	\begin{cases}
	\hat{p}_0:
	\quad&
	\left(u_{fp}, \, y_{fp}, \, z_{fp}\right) = \left(0,\,0,\,0\right)
	\,,
	\\
	\hat{p}_1^\pm :
	\quad&
	\left(u_{fp}, \, y_{fp}, \, z_{fp}\right) = \left(\pm\dfrac{2}{\pi}\arctan\sqrt{1+\alpha_\chi},\,\pm\sqrt{1+\alpha_\chi},\,0\right)
	\,.
	\end{cases}
\end{align}
where $\hat{p}_0$ corresponds to a matter dominated epoch and $\hat{p}_1^\pm$ to scaling solutions with 3-form dominance. Notice that the pair $\hat{p}_1^\pm$ satisfies the constraints on the variable $y$ as long as the parameter coefficient respects $-1<\alpha_\chi\leq0$, which is consistent with the assumption of small coefficients, i.e., $|\alpha_\chi|\ll1$. As found in \eqref{type1_eigenvalues} the Jacobian at these points has two negative and one positive eigenvalues and therefore $\hat{p}_1^\pm$ correspond to saddle fixed points. The same result is found for the matter point $\hat p_0$.

In this case, the only solution to Eq.~\eqref{beta_solutions} is $\beta_-$ given by
\begin{align}
	\label{Case_IV_beta_sol}
	\beta_- = \frac{1}{1-\alpha_\chi}
	\,.
\end{align}
which respects $0<\beta_-<1$ for all $\alpha_\chi<0$. Substituting the solution \eqref{Case_IV_beta_sol} in the condition \eqref{TypeII_fixedpoints_interactoin}, we obtain two pairs of fixed points of Type~II:
\begin{align}
	\begin{array}{c}
		\\
		\textbf{Type~II fixed points:}\\
		\textrm{(Interaction IV)}
	\end{array}
	\qquad
	\begin{cases}
	\tilde{p}_1^\pm :
	\quad&
	\left(u_{fp}, \, y_{fp}, \, z_{fp}\right) = \left(\pm\dfrac{2}{\pi}\arctan\sqrt{\frac{1+\Delta}{2}\beta_-},\,\pm \sqrt{\frac{1+\Delta}{2}\beta_-},\,\sqrt{\frac{1-\Delta}{2}\beta_-}\right)
	\,,
	\\
	\tilde{p}_2^\pm:
	\quad&
	\left(u_{fp}, \, y_{fp}, \, z_{fp}\right) = \left(\pm\dfrac{2}{\pi}\arctan\sqrt{\frac{1-\Delta}{2}\beta_-},\,\pm \sqrt{\frac{1-\Delta}{2}\beta_-},\, \sqrt{\frac{1+\Delta}{2}\beta_-}\right)
	\,,
	\end{cases}
\end{align}
where, as derived from Eqs.~\eqref{typeII_y_sol} and \eqref{Case_IV_beta_sol}, $\Delta$ is given by
\begin{align}
	\label{Delta_interaction_IV}
	\Delta\left(\alpha_\chi,\xi\right)\equiv\sqrt{1+\frac{18}{\xi}\left(1-\alpha_\chi\right)\alpha_\chi}
	\,.
\end{align}
From Eq.~\eqref{cs2_constraint}, we find that in order for the system to have fixed points of Type~II the interaction coefficient $\alpha_\chi$ and the potential parameter $\xi$ must respect
\begin{align}
	\label{InteractionIV_constraints}
	-\frac{\sqrt{1+2\xi/9}-1}{2} \leq \alpha_\chi \leq 0
	\,,
	\qquad
	0<\xi<9/2
	\,.
\end{align}

For general values of $0<\beta_-<1$ and $0<\Delta<1$ the expressions for the eigenvalues of the Jacobian are too cumbersome to be analysed with analytical methods. Instead, we will analyse the stability of the characteristic polynomial using the condition \eqref{Hurwitz_degree3} to decide on the stability of the Type~II fixed points. The coefficients of the characteristic polynomial of the Jacobian at the fixed points $\tilde{p}_1^\pm$ are
\begin{align}
	a_0 =&~ -54 \frac{\Delta}{1+\Delta}\frac{1-\beta_-}{\beta_-}
	\,,
	\\
	a_1 =&~ -9\frac{1+\Delta - 2\left(2+\Delta^2\right)\beta_- + 6\beta_-^2 - \left(2+\Delta-\Delta^2\right)\beta_-^3}{\left(1-\Delta^2\right)\beta_-^2}
	\,,
	\\
	a_2 =&~ -3\left[1 + \frac{\beta_-+\Delta}{\left(1+\Delta\right)\beta_-}\right]
	\,,
	\\
	a_3 =&~-1
	\,.
\end{align}
With these values, the three conditions in \eqref{Hurwitz_degree3} are satisfied for any $\beta$ and $\Delta$ such that $0<\beta_-<1$ and $0<\Delta<1$ and therefore the fixed points $\tilde{p}_1^\pm$ correspond to late-time attractors. When we calculate the coefficients of the characteristic polynomial of the Jacobian for the pair $\tilde{p}_2^\pm$, we find that $a_0<0$ for $0<\beta_-<1$ and $0<\Delta<1$, therefore the conditions \eqref{Hurwitz_degree3} are not satisfied and the fixed points $\tilde{p}_2^\pm$ are unstable.

In the limiting case of $\Delta=0$, where we only have one pair of Type~II fixed points given by Eq.~\eqref{Delta0_solutions}, we obtain the eigenvalues:
\begin{align}
	\left\{
		0,\,
		-3\left[
			1 + \frac{1-\beta_-}{\beta_-}\sqrt{-1+2\beta_-}
		\right],\,
		-3\left[
			1 - \frac{1-\beta_-}{\beta_-}\sqrt{-1+2\beta_-}
		\right]
	\right\}
	\,.
\end{align}
{ The real parts of the two non-null eigenvalues are always negative for $0<\beta_-<1$. However, due to the existence of a null eigenvalue, the linear theory fails to determine the stability of the fixed point and other methods are required. By employing the Centre Manifold Theory \cite{Rendall:2001it,Boehmer:2011tp}, we find that the points $\tilde{p}_{\Delta}^\pm$ correspond to saddle nodes and, therefore, are unstable fixed points.}

Finally, the fixed points of Type~III are
\begin{align}
	\label{Interaction_I_TypeIII}
	\begin{array}{c}
		\\
		\textbf{Type~III fixed points:}\\
		\textrm{(Interaction I)}
	\end{array}
	\qquad
	\begin{cases}
	\hat\pi_{0}^\pm :
	\quad&
	\left(u_{fp}, \, y_{fp}, \, z_{fp}\right) = \left(\pm1,\,0,\,0\right)
	\,,
	\\
	\hat\pi_{-1}^\pm :
	\quad&
	\left(u_{fp}, \, y_{fp}, \, z_{fp}\right) = \left(\pm1,\,-\sqrt{1+\alpha_\chi},\,0\right)
	\,,
	\\
	\hat\pi_{+1}^\pm :
	\quad&
	\left(u_{fp}, \, y_{fp}, \, z_{fp}\right) = \left(\pm1,\,+\sqrt{1+\alpha_\chi},\,0\right)
	\,.
	\end{cases}
\end{align}
As reported above, cf. Eq.~\eqref{eigenvalues_III}, the points $\hat{\pi}_0^\pm$ are extremely repulsive and correspond to the asymptotic past matter era, while the points $\hat \pi_{\pm1}^{\pm}$ are saddles and therefore unstable.

%%%%%%%%%%%%%%%%%%%%%%%%%%%%%%%%%%%%%%%%
%
%	Q propto rho_chi^2
%
%%%%%%%%%%%%%%%%%%%%%%%%%%%%%%%%%%%%%%%%

\subsubsection{Interaction V: $Q\propto \rho_\chi^2$}

The second interaction we analyse is of the form
\begin{align}
	\label{interaction_V}
	Q = 3H \alpha_{\chi\chi} \frac{\rho_\chi^2}{\rho_m+\rho_\chi} = \frac{9H^3}{\kappa^2} \alpha_{\chi\chi} \left(y^2+z^2\right)^2
	\,.
\end{align}
Comparison with Eq.~\eqref{quadratic_int_dynamical} immediately shows that this corresponds to setting $\alpha_0=0$, $\alpha_1=0$, and $\alpha_2=\alpha_{\chi\chi}$. In this case the interaction coefficients do not satisfy the condition \eqref{no_LSBR_removal} and therefore the fixed points corresponding to LSBR will again be removed.
{
As in the previous case of the linear interaction IV, the energy transfer between DM and DE is not stopped when the energy density of DM vanishes. Therefore, in order to guarantee that the system does not evolve to negative values of the DM energy density, see Eq.~\eqref{conservation_equations}, we can impose the constraint $\alpha_{\chi\chi}\leq0$.
}
With this interaction we obtain three fixed points of Type~I
\begin{align}
	\label{Interaction_V_TypeI}
	\begin{array}{c}
		\\
		\textbf{Type~I fixed points:}\\
		\textrm{(Interaction V)}
	\end{array}
	\qquad
	\begin{cases}
	\hat{p}_0:
	\quad&
	\left(u_{fp}, \, y_{fp}, \, z_{fp}\right) = \left(0,\,0,\,0\right)
	\,,
	\\
	\hat{p}_1^\pm :
	\quad&
	\left(u_{fp}, \, y_{fp}, \, z_{fp}\right) = \left(\pm\dfrac{2}{\pi}\arctan\frac{1}{\sqrt{1-\alpha_{\chi\chi}}},\,\pm\frac{1}{\sqrt{1-\alpha_{\chi\chi}}},\,0\right)
	\,,
	\end{cases}
\end{align}
where $\hat{p}_0$ corresponds to a matter dominated epoch and $\hat{p}_1^\pm$ to scaling solutions with 3-form dominance. Notice that for the pair $\hat{p}_1^\pm$, we find that $0<y_{fp}^2<1$ as long as the parameter coefficient respects $\alpha_{\chi\chi}<0$, which is consistent with the assumption of small coefficients, i.e., $|\alpha_{\chi\chi}|\ll1$. As found in \eqref{type1_eigenvalues}, the Jacobian at these points has two negative and one positive eigenvalues and therefore $\hat{p}_1^\pm$ correspond to saddles. The same result is found for the matter point $\hat p_0$.

In this case, the only solution to Eq.~\eqref{beta_solutions} that satisfies $0<\beta_-<1$ is $\beta_-$ given by
\begin{align}
	\label{Case_V_beta_sol}
	\beta_- = \frac{1}{2\alpha_{\chi\chi}}\left(1-\sqrt{1-4\alpha_{\chi\chi}}\right)
	\,.
\end{align}
From this expression, we find that the above constraints on $\beta_-$ are satisfied for all $\alpha_{\chi\chi}<0$. If we also enforce the constraint derived in \eqref{cs2_constraint}, we find the more restrictive interval range $-1/4<\alpha_{\chi\chi}<0$ for the interaction parameter. Substituting the solution \eqref{Case_V_beta_sol} in the condition \eqref{TypeII_fixedpoints_interactoin}, we obtain two pairs of fixed points of Type~II
\begin{align}
	\begin{array}{c}
		\\
		\textbf{Type~II fixed points:}\\
		\textrm{(Interaction V)}
	\end{array}
	\qquad
	\begin{cases}
	\tilde{p}_1^\pm :
	\quad&
	\left(u_{fp}, \, y_{fp}, \, z_{fp}\right) = \left(\pm\dfrac{2}{\pi}\arctan\sqrt{\frac{1+\Delta}{2}\beta_-},\,\pm \sqrt{\frac{1+\Delta}{2}\beta_-},\,\sqrt{\frac{1-\Delta}{2}\beta_-}\right)
	\,,
	\\
	\tilde{p}_2^\pm:
	\quad&
	\left(u_{fp}, \, y_{fp}, \, z_{fp}\right) = \left(\pm\dfrac{2}{\pi}\arctan\sqrt{\frac{1-\Delta}{2}\beta_-},\,\pm \sqrt{\frac{1-\Delta}{2}\beta_-},\, \sqrt{\frac{1+\Delta}{2}\beta_-}\right)
	\,,
	\end{cases}
\end{align}
where, as derived from Eqs.~\eqref{typeII_y_sol} and \eqref{Case_V_beta_sol}, $\Delta$ is given by
\begin{align}
	\label{Delta_interaction_V}
	\Delta\left(\alpha_\chi,\xi\right)\equiv\sqrt{1+\frac{18}{\xi}\alpha_{\chi\chi}}
	\,.
\end{align}
From Eq.~\eqref{cs2_constraint}, we find that in order for the system to have fixed points of Type~II the interaction coefficient $\alpha_\chi$ and the potential parameter $\xi$ must respect
\begin{align}
	-\frac{\xi}{18} \leq \alpha_{\chi\chi} \leq 0
	\,,
	\qquad
	0<\xi<9/2
	\,.
\end{align}

As was the case in the previous interaction, for general values of $0<\beta_-<1$ and $0<\Delta<1$ the expressions for the eigenvalues of the Jacobian are too cumbersome to be analysed with analytical methods. Instead, we will analyse the stability of the characteristic polynomial using the condition \eqref{Hurwitz_degree3} to decide on the stability of the Type~II fixed points. The coefficients of the characteristic polynomial of the Jacobian at the fixed points $\tilde{p}_1^\pm$ are
\begin{align}
	a_0 =&~ -54 \frac{\Delta}{1+\Delta}\frac{2-3\beta_-+\beta_-^2}{\beta_-}
	\,,
	\\
	a_1 =&~ -9\frac{
		2\Delta
		-2\Delta\left(1+2\Delta\right)\beta_-
		+\left(2+\Delta+3\Delta^2\right)\beta_-^2
		-\left(1+\Delta\right)\beta_-^3
	}{\left(1-\Delta^2\right)\beta_-^2}
	\,,
	\\
	a_2 =&~ -3\frac{1+2\Delta+\beta_-}{\left(1+\Delta\right)\beta_-}
	\,,
	\\
	a_3 =&~-1
	\,.
\end{align}
With these values the three conditions in \eqref{Hurwitz_degree3} are satisfied for all $0<\beta_-<1$ and $0<\Delta<1$ and the fixed points $\tilde{p}_1^\pm$ correspond to late-time attractors. When we calculate the coefficients of the characteristic polynomial of the Jacobian for the pair $\tilde{p}_2^\pm$, we find $a_0>0$ for all values $0<\beta_-<1$ and $0<\Delta<1$, therefore the conditions \eqref{Hurwitz_degree3} are not satisfied and the fixed points $\tilde{p}_2^\pm$ correspond at most to saddles.

In the limiting case of $\Delta=0$, where we only have one pair of Type~II fixed points given by Eq.~\eqref{Delta0_solutions}, we obtain one null and two negative eigenvalues:
\begin{align}
	\left\{
		0,\,
		-\frac{3}{2\beta_-}\left[\left(1+\beta_-\right) + \left(1-\beta_-\right)\sqrt{1+4\beta_-}\right],\,
		-\frac{3}{2\beta_-}\left[\left(1+\beta_-\right) - \left(1-\beta_-\right)\sqrt{1+4\beta_-}\right]
	\right\}
	\,.
\end{align}
By employing the Centre Manifold Theory \cite{Rendall:2001it,Boehmer:2011tp}, we find that just like in the case of interaction IV, $\tilde{p}_{\Delta}^\pm$ correspond to saddle nodes and, therefore, are unstable fixed points.

Finally, the fixed points of Type~III are
\begin{align}
	\label{Interaction_I_TypeIII}
	\begin{array}{c}
		\\
		\textbf{Type~III fixed points:}\\
		\textrm{(Interaction I)}
	\end{array}
	\qquad
	\begin{cases}
	\hat\pi_{0}^\pm :
	\quad&
	\left(u_{fp}, \, y_{fp}, \, z_{fp}\right) = \left(\pm1,\,0,\,0\right)
	\,,
	\\
	\hat\pi_{-1}^\pm :
	\quad&
	\left(u_{fp}, \, y_{fp}, \, z_{fp}\right) = \left(\pm1,\,-\frac{1}{\sqrt{1-\alpha_{\chi\chi}}},\,0\right)
	\,,
	\\
	\hat\pi_{+1}^\pm :
	\quad&
	\left(u_{fp}, \, y_{fp}, \, z_{fp}\right) = \left(\pm1,\,+\frac{1}{\sqrt{1-\alpha_{\chi\chi}}},\,0\right)
	\,.
	\end{cases}
\end{align}
As reported above, \eqref{eigenvalues_III}, the points $\hat{\pi}_0^\pm$ are extremely repulsive and correspond to the asymptotic past matter era, while the points $\hat \pi_{\pm1}^{\pm}$ are saddles and therefore unstable.

\subsection{Overview}

Throughout this section, we apply a dynamical system approach to study a cosmological model with an interaction between DM and a massive 3-form playing the role of DE. 
We have classified the fixed points of the system in three different categories (cf. Sect.~\ref{subsec_Fixed_Points}), depending on their coordinates $(u_{fp},\,y_{fp},\,z_{fp})$: Type~I fixed points characterised by $z_{fp}=0$ and $|u_{fp}|\neq1$; Type~II fixed points characterised by $z_{fp}\neq0$ and $|u_{fp}|\neq1$; Type~III fixed points characterised by $|u_{fp}|=1$.
In addition, we studied the conditions for the existence of such fixed points in our dynamical system, their coordinates and stability, for an interaction with the general form given in Eq.~\eqref{quadratic_int_dynamical}.
An interesting feature of the solutions found in \eqref{TypeI_fixedpoints_interactoin} and \eqref{TypeII_fixedpoints_interactoin} is the possibility to obtain fixed points where $y_{fp}^2+z_{fp}^2$ is neither $0$ nor $1$ and therefore there is no complete DM or DE dominance. These cases correspond to scaling behaviours near the fixed points with a constant non-vanishing ratio $\rho_m/\rho_\chi$:
\begin{align}
	\frac{\Omega_m}{\Omega_\chi} 
	= \frac{1 - \left(y_{fp}^2+z_{fp}^2\right)}{y_{fp}^2+z_{fp}^2}
	\neq0
	\,,
\end{align}
At the fixed points the parameter of EoS assumes the value $w_\mathrm{tot}=-y_{fp}^2$, in the case of the Type~I solutions, and $w_\mathrm{tot}=-1$, in the case of Type~II solutions.

In particular, we find that when $\alpha_0+\alpha_1+\alpha_2\neq0$ the interaction \eqref{quadratic_int_dynamical} can shift the fixed points away from $(u_{fp},y_{fp},z_{fp})=(\pm1/2,\pm1,0)$ and therefore avoid LSBR event completely. This condition implies that LSBR is removed if and only if the interaction is not proportional to a power of the DM energy density. Since in this case the Type~I solutions are always saddle points, we conclude that when LSBR is removed any existent late-time attractor will be of Type~II and consequently correspond to a future de Sitter inflationary era. If, however, the fixed points corresponding to a LSBR event are not removed, they will be the only late-time attractors of the system.

In Sect.~\ref{Examples} we present the results obtained for some examples of interactions commonly present in the literature, which we compile in Tables~\ref{Table_fixedpoints} and \ref{Table_fixedpoints_II}. Although the analysis performed is focused on the model of a 3-form field with a Gaussian potential, the classification and methods employed are general and can be applied to other models. In addition, as long as the potential chosen is similar to the Gaussian one, in the sense that it is a potential that has a maximum at  $\chi=0$ and decreases monotonically with $|\chi|$, the qualitative features of the results, such as the existence of scaling solutions due to the interaction and the condition of the removal of the LSBR event, will not be affected.

%%%%%%%%%%%%%%%%%%%%%%%%%%%%%%%%%%%%%%%%
%
%	 statefinder Hierarchy
%
%%%%%%%%%%%%%%%%%%%%%%%%%%%%%%%%%%%%%%%%

\section{Statefinder Hierarchy}

\label{Statefinder Hierarchy}

\subsection{Introducing the statefinders}
The statefinder hierarchy was shown to be useful to distinguish different DE models \cite{Sahni:2002fz,Alam:2003sc,Arabsalmani:2011fz,Li:2014mua}. Recent studies based on BAO data reinforce the status of the statefinder hierarchy as a suitable tool to distinguish DE models \cite{Hu:2015bpa}. In Ref.~\cite{Yin:2015pqa}, statefinders were studied in the context of DE and DM interaction. In this paper, we aim to study the statefinder hierarchy for different kinds of interactions between DE, modelled by a 3-form field, and DM.

The scale factor of the Universe, $a(t)/a(t_0)=1/(1+z)$, can be Taylor expanded around the present time $t_{0}$ as follows \cite{Visser:2004bf,Vitagliano:2009et},
\begin{align}
	\label{scale factor expansion}
	\frac{a\left(t\right)}{a\left(t_{0}\right)}=1+\overset{\infty}{\underset{n=1}{\sum}}\frac{A_{n}\left(t_{0}\right)}{n!}\left[H_{0}\left(t-t_{0}\right)\right]^{n}
	\,,
\end{align}
where 
\begin{align}
	\label{An}
	A_{n}=\frac{a^{\left(n\right)}}{aH^{n}},\quad n\in \mathbb{N}
	\,,
\end{align}
and $a^{\left(n\right)}$ is the $n^{th}$-derivative of the scale factor with respect to the cosmic time. Historically, the parameters $A_{2},\, A_{3},\, A_{4},\, A_{5}$ were termed as the deceleration parameter $q=-A_{2}$, the jerk%
\footnote{The terminology jerk, as far as we know, was introduced in \cite{Visser:2004bf}. The
same parameter was defined as ``$r$'' and named statefinder $r$
in \cite{Alam:2003sc}.%
}
$j=A_{3}$, the snap $s=A_{4}$ and the lerk $l=A_{5}$ respectively (See for example \cite{Visser:2004bf,Capozziello:2008qc,Bouhmadi-Lopez:2014jfa} and the extensive list of references in \cite{Morais:2015ooa}. Using these definitions the statefinders were defined as \cite{Arabsalmani:2011fz,Yin:2015pqa}
\begin{align}
	\label{sn1}
	S_{3}^{\left(1\right)}= & A_{3}
	\,,
	\\
	S_{4}^{\left(1\right)}= & A_{4}+3\left(1+q\right)
	\,,
	\\
	S_{5}^{\left(1\right)}= & A_{5}-2\left(4+3q\right)\left(1+q\right)
	\,.
\end{align}
%where the deceleration parameter for any FLRW Universe reads
%\begin{align}
%	\label{q_definition}
%	q=-A_{2}= & -\left(1+\frac{\dot{H}}{H^{2}}\right)
%	\,.
%\end{align}
By construction, the statefinder hierarchy defines a null diagnostic for the $\Lambda$CDM model as \cite{Arabsalmani:2011fz}
\begin{align}
	\label{statestandard}
	S_{n}^{(1)}\vert_{\Lambda\textrm{CDM}}=1
	\,.
\end{align}

In order to re-write the statefinders $S_3^{(1)}$, $S_4^{(1)}$, and $S_5^{(1)}$, in terms of our dynamical variables $u$, $y$, and $z$, we begin by combining Eqs.~\eqref{friefconstrant}, \eqref{h1h}, and \eqref{An}, to obtain $A_2(u,y,z)$:
\begin{align}
	\label{q_of_uyz}
	A_2= & 1 
	-\frac{3}{2}\left[
		1 
		- \left(y^{2}+z^2\right) 
		-\frac{2}{9}\xi\, z^{2}\tan^2\left(\frac{\pi}{2}u\right)
	\right]
	\,.
\end{align}
We can now make use of the following recursive relations for the $A_{n}$ parameters%
\footnote{This relation can be derived by differentiating Eq.~\eqref{An} with regards to $x=\log(a/a_0)$ and isolating the term in $a^{(n)}$.
}
\begin{align}
	A_{n+1} = A_{n}' + A_{n}\left[1-n(1+q)\right]
	\,,
\end{align}
and the set of evolution equations \eqref{eq_u_1}, \eqref{eq_y_1}, and \eqref{eq_z_1}, to derive the $A_n$ and $S_n^{(1)}$, $n\geq3$, as functions of the dynamical variables. Due to their cumbersome size the expressions found are presented only in the Appendix~\ref{ statefinders_uyz}.

%%%%%%%%%%%%%%%%%%%%%%%%%%%%%%%%%%%%%%%%
%
%	 statefinders: non-interacting model
%
%%%%%%%%%%%%%%%%%%%%%%%%%%%%%%%%%%%%%%%%

\subsection{Statefinder diagnosis for non-interacting 3-form DE model}

\label{statefinder_nointeraction}

%%%%%%%%%%%%%%%%%%%%%%%%%%%%
\begin{figure}[t]

{\includegraphics[width=.325\hsize]{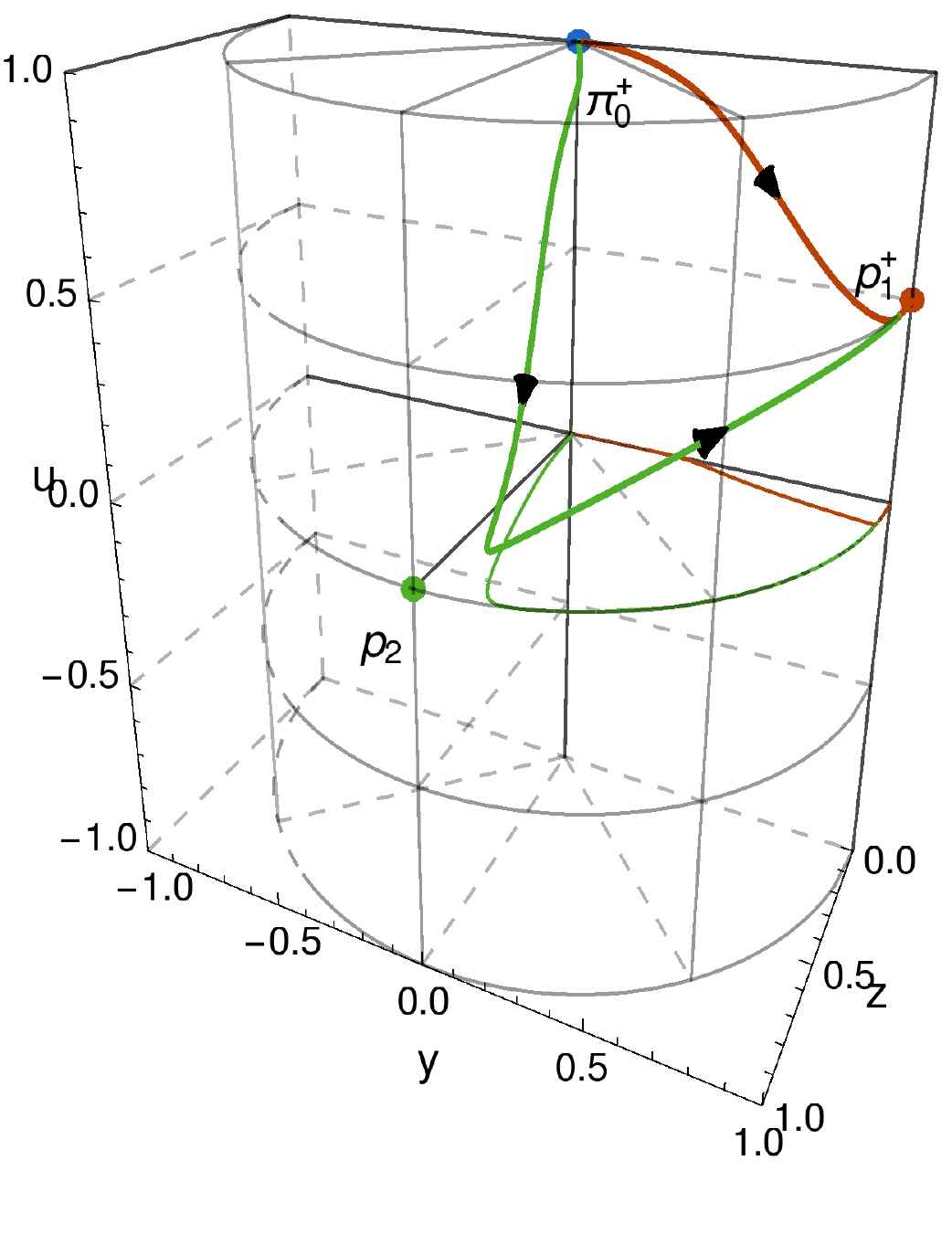}}
\hfill
{\includegraphics[width=.325\hsize]{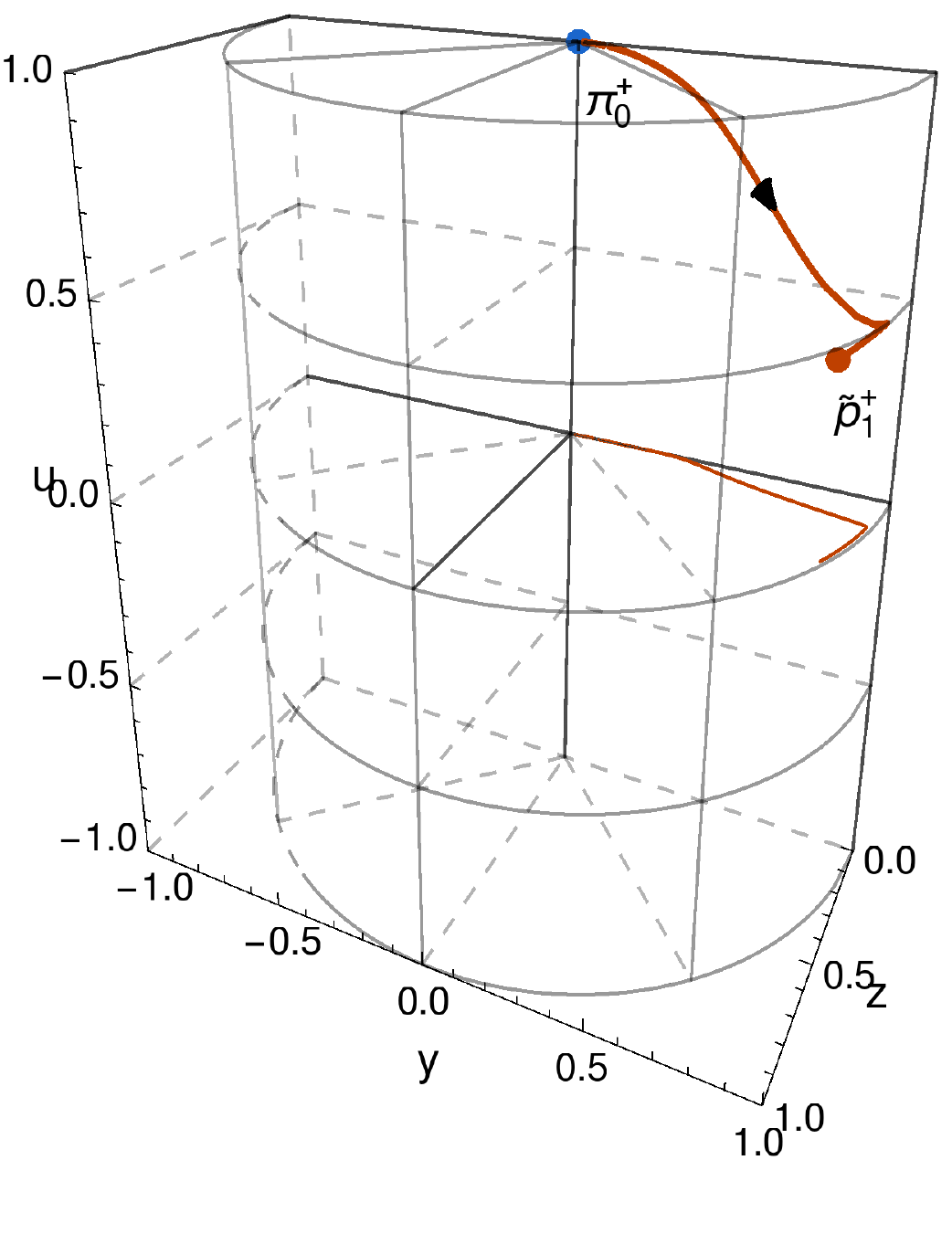}}
\hfill
{\includegraphics[width=.325\hsize]{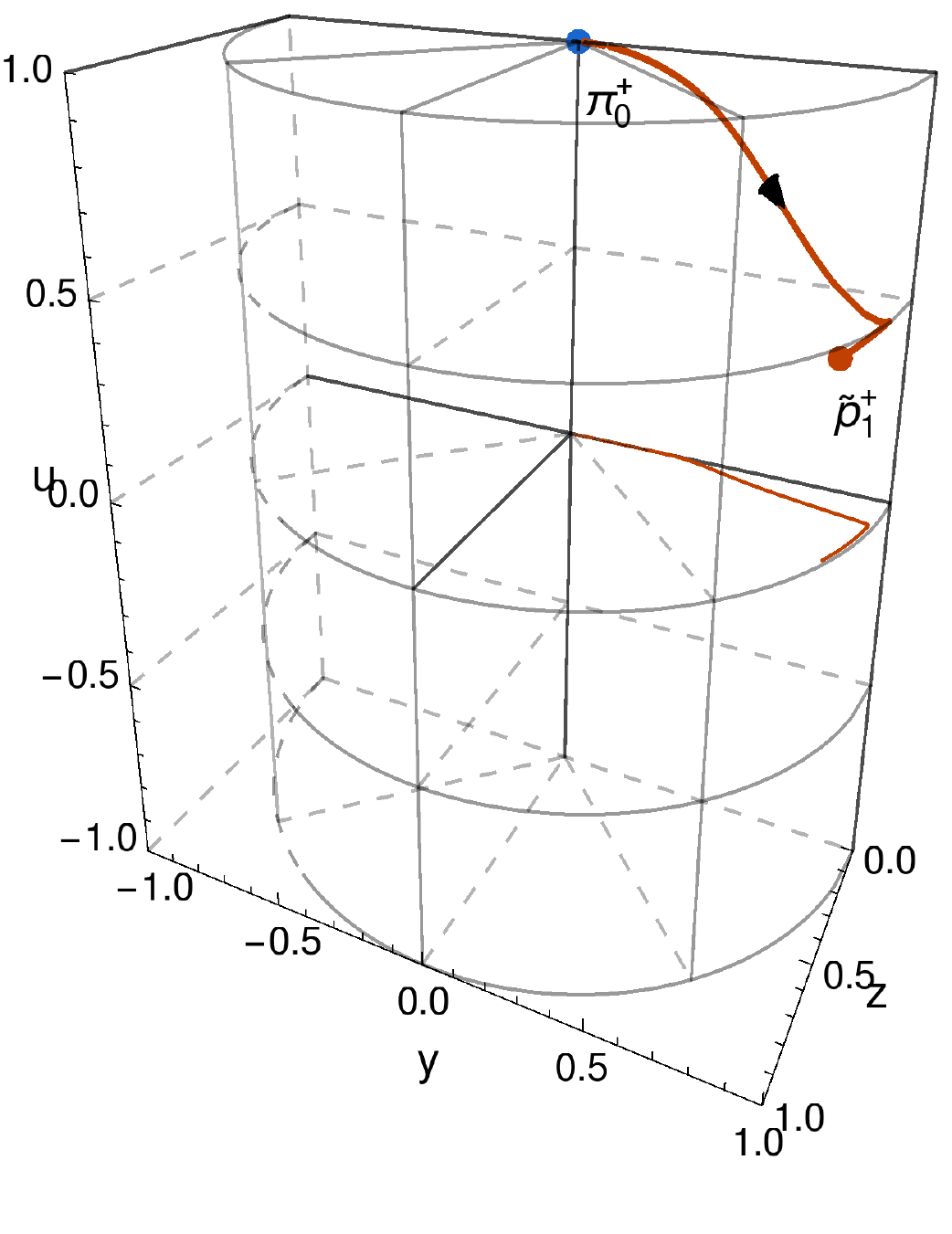}}

\caption{\label{Trajectories}%
On the left panel of this figure we plot the trajectories, in the 3-dimensional space $(u,\,y,\,z)$, for the two solutions $\mathcal{P}_{z}^{I}$ (red thick line) and $\mathcal{P}_{z}^{II}$ (green thick line) that were found to minimise the deviation from $\Lambda$CDM in the case of no interaction. On the middle panel we plot the trajectory obtained when interaction IV is considered with $\alpha_\chi=-0.03$. On the right panel we plot the trajectory obtained when interaction V is considered with $\alpha_{\chi\chi}=-0.03$. On both the middle and right panels, the initial conditions are set in the past, at redshift 6, using the values of the dynamical variables for the trajectory $\mathcal{P}_{z}^{I}$. The thin lines indicate the projection, onto the plane yOz, of the trajectory of the same colour. The labelled points indicate the fixed points, see Table~\ref{Table_fixedpoints_II} for the classification, near which the trajectories pass.
}

\end{figure}

%%%%%%%%%%%%%%%%%%%%%%%%%%%%

%%%%%%%%%%%%%%%%%%%%%%%%%%%

\begin{table}[t]
\centering
\begin{tabularx}{\textwidth}{C{.4} C{.7} C{1.3} C{.7} C{1.3} C{.7} C{1.3} C{.7} C{1.3} C{1.3} C{1.3}}
\hline
	{\footnotesize{$\mathbf{\#}$}}
	&
	{$\mathbf{\mathcal{P}_z^\#}$}
	&
	{$\mathbf{ X}$}
	& 
	{$\mathbf{ S_3^{(1)} }$}
	& 
	{$\mathbf{\dfrac{\delta \log S_3^{(1)}}{\delta \log X} }$}
	& 
	{$\mathbf{ S_4^{(1)} }$}
	& 
	{$\mathbf{\dfrac{\delta \log S_4^{(1)}}{\delta \log X} }$}
	& 
	{$\mathbf{ S_5^{(1)} }$}
	& 
	{$\mathbf{\dfrac{\delta \log S_5^{(1)}}{\delta \log X} }$}
	& 
	{$\mathbf{\dfrac{\delta \log \Omega_{m,0}}{\delta \log X} }$}
	& 
	{{$\mathbf{\dfrac{\delta \log w_{\chi,0}}{\delta \log X} }$}}
\\ \hline
\footnotesize
%%%%%%%%%%%%%
	\multirow{ 3}{\hsize}{\centering I}
	&
	\multirow{ 3}{\hsize}{\centering$0.02051$}
	&
	$u_0=0.5436$
	&
	\multirow{ 3}{\hsize}{\centering$1.009$}
	&
	$-7.554\times10^{-4}$
	&
	\multirow{ 3}{\hsize}{\centering$1.019$}
	&
	$2.522\times10^{-2}$
	&
	\multirow{ 3}{\hsize}{\centering$1.014$}
	&
	$1.182\times10^{-1}$
	&
	
	&
	$2.057\times10^{-2}$
\\ 
%%%%%%%%%%%%%
	&
	&
	$y_0=0.8242$
	&
	
	&
	$2.275\times10^{-2}$
	&
	
	&
	$-1.786\times10^{-2}$
	&
	
	&
	$1.972\times10^{-1}$
	&
	$1.959$
	&
	$-1.168\times10^{-2}$
 \\ 
 %%%%%%%%%%%%%
	&
	&
	$z_0=0.1193$
	&
	
	&
	$1.926\times10^{-2}$
	&
	
	&
	$3.621\times10^{-2}$
	&
	
	&
	$3.095\times10^{-2}$
	&
	$4.103\times10^{-2}$
	&
	$1.168\times10^{-2}$
\\ \hline
%%%%%%%%%%%%%
%		Pz2
%%%%%%%%%%%%%
	\multirow{ 3}{\hsize}{\centering II}
	&
	\multirow{ 3}{\hsize}{\centering$0.9721$}
	&
	$u_0=0.1052$
	&
	\multirow{ 3}{\hsize}{\centering$1.013$}
	&
	$-6.063\times10^{-3}$
	&
	\multirow{ 3}{\hsize}{\centering$1.032$}
	&
	$6.291\times10^{-2}$
	&
	\multirow{ 3}{\hsize}{\centering$0.9991$}
	&
	$-4.478\times10^{-1}$
	&
	
	&
	$1.215\times10^{-2}$
\\ 
%%%%%%%%%%%%%
	&
	&
	$y_0=0.1394$
	&
	
	&
	$3.076\times10^{-2}$
	&
	
	&
	$-4.786\times10^{-4}$
	&
	
	&
	$4.454\times10^{-1}$
	&
	$5.577\times10^{-2}$
	&
	$-3.326\times10^{-4}$
 \\ 
 %%%%%%%%%%%%%
	&
	&
	$z_0=0.8158$
	&
	
	&
	$2.477\times10^{-2}$
	&
	
	&
	$4.235\times10^{-2}$
	&
	
	&
	$1.737\times10^{-1}$
	&
	$1.944$
	&
	$3.326\times10^{-4}$
 \\ \hline
\end{tabularx}
\caption{\label{NoInteraction_2minima}%
For the two solutions $\mathcal{P}_z^I$ and $\mathcal{P}_z^{II}$, we present the current values of the dynamical variables and the statefinder parameters. For each statefinder parameter, and for the cosmological parameters $\Omega_m$ and $w_\chi$, we present the deviation of the present day values for small perturbations of the values of the dynamical variables. The present day values of $\Omega_{m}$ and $w_\chi$ are $\Omega_{m,0}=0.3065$ and $w_{\chi,0}=-1.006$ \cite{Ade:2015xua,Adam:2015rua}.}
\end{table}
%%%%%%%%%%%%%%%%%%%%%%%%%%%%

%%%%%%%%%%%%%%%%%%%%%%%%%%%%
\begin{figure}[t]
{\includegraphics[width=.48\hsize]{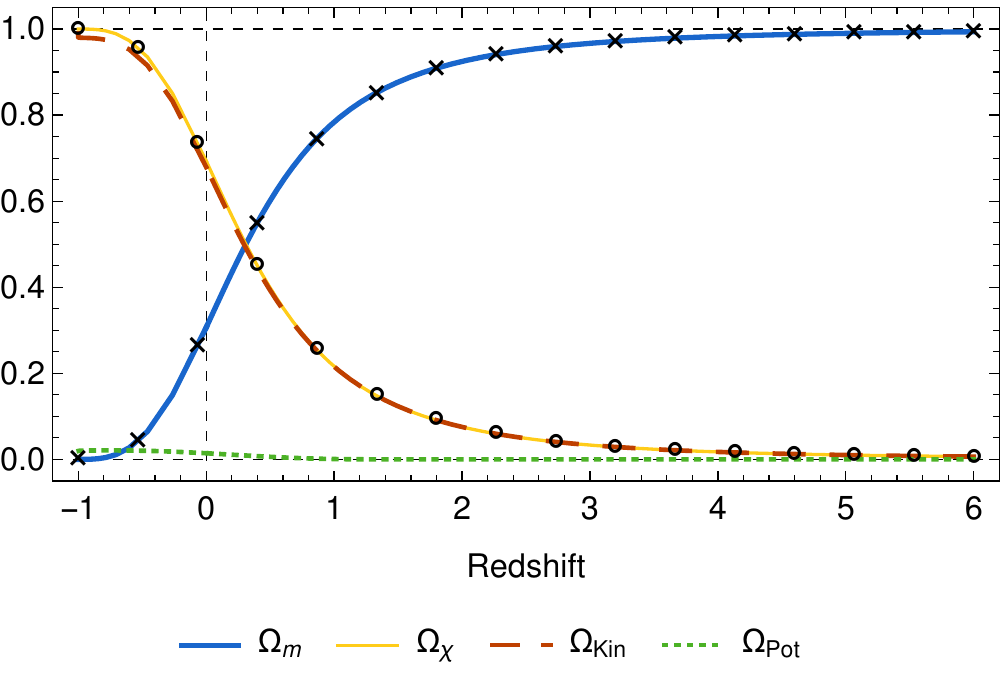}}
\hfill
{\includegraphics[width=.48\hsize]{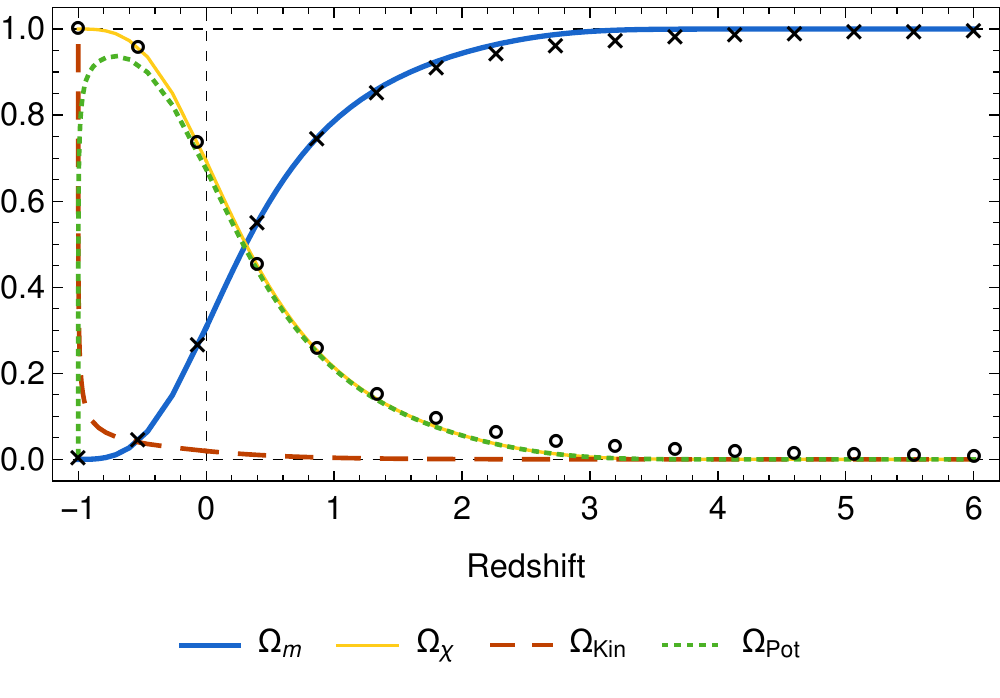}}

\caption{\label{Cosmology_NoInteraction}%
The evolution of the relative energy densities of DM (blue thick line) and DE (yellow thin line), and of the individual components that contribute to the energy density of DE: kinetic energy (red dashed line) and potential energy (green dotted line), in terms of the redshift. The plot on the left represents the evolution of the system for the case $\mathcal{P}_{z}^{I}$, while the plot on the right represents the evolution for the case $\mathcal{P}_{z}^{II}$. The circles and crosses indicate, respectively, the values of $\Omega_\Lambda$ and $\Omega_m$ for the $\Lambda$CDM model.
}

\end{figure}

%%%%%%%%%%%%%%%%%%%%%%%%%%%%
\begin{figure}[t]
{\includegraphics[width=.48\hsize]{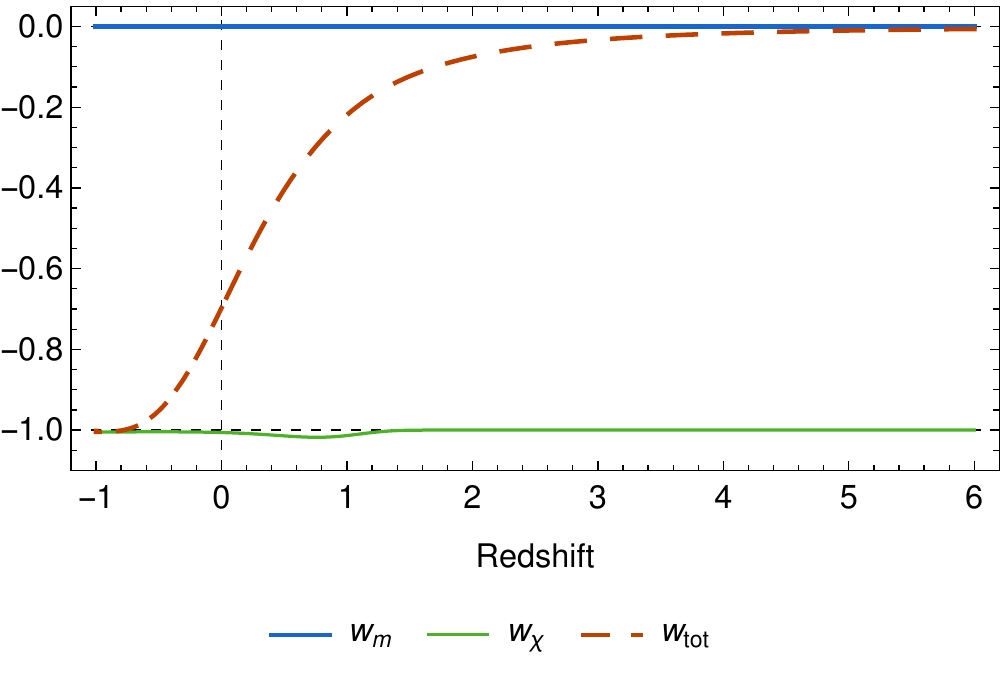}}
\hfill
{\includegraphics[width=.48\hsize]{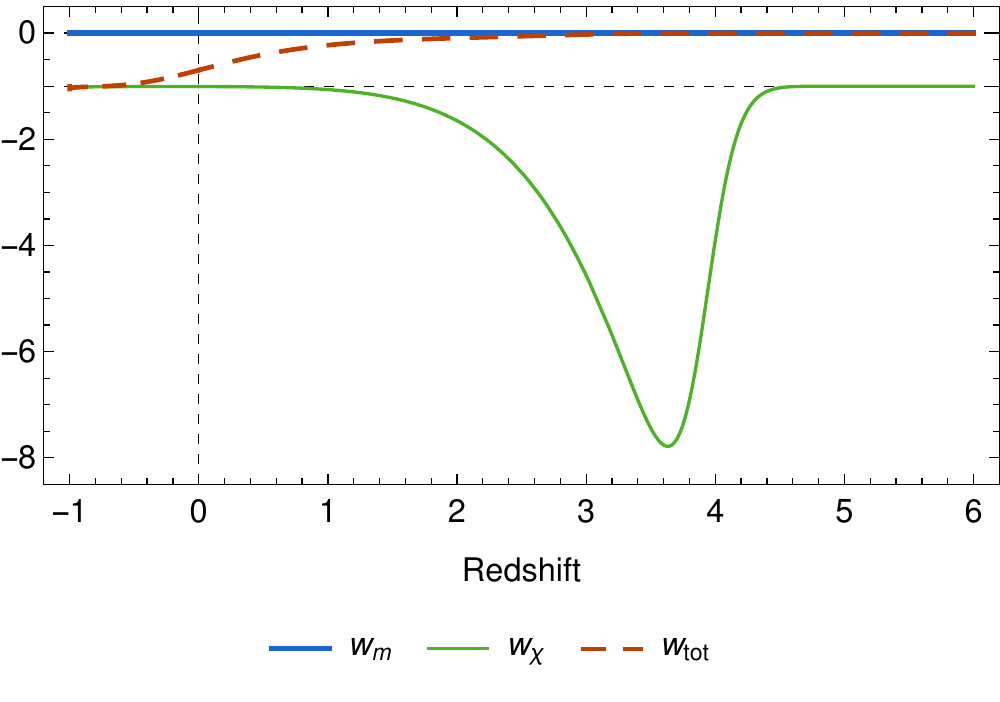}}

\caption{\label{EoS_NoInteraction}%
The evolution of the parameter of EoS of DM (blue thick line), DE (green thin line), and of the total fluid in terms of the redshift. The plot on the left represents the evolution of the system for the case $\mathcal{P}_{z}^{I}$, while the plot on the right represents the evolution for the case $\mathcal{P}_{z}^{II}$.
}

\end{figure}

%%%%%%%%%%%%%%%%%%%%%%%%%%%%

In this section, we analyse the statefinder parameters for the non-interacting
3-form DE model. We take into account the recent Planck 2015 results \cite{Ade:2015xua,Adam:2015rua} and use the values $\Omega_\mathrm{m}\simeq0.3065$ and $w_d\simeq-1.006$, obtained for a DE~model with constant parameter of EoS, $w$, as reference values for the present day values of the relative energy density of DM, $\Omega_{m,0}$, and of the parameter of EoS of the 3-form field, $w_{\chi,0}$. These can be related to the present day values of the dynamical variables of our model as
\begin{align}
	\label{initial condition values}
	\Omega_{m,0} = s_0^2 = 1- \left(y_0^2+z_0^2\right)
	\,,
	\qquad
	w_{\chi,0}=-1-\frac{2}{9}\xi\frac{z_0^{2}}{y_0^{2}+z_0^{2}}\tan^2\left(\frac{\pi}{2}u_0\right)
	\,.
\end{align}
These two expressions can be combined with Eq.~\eqref{q_of_uyz} to give the present day value of $A_2$:
\begin{align}
	A_{2,0} = - \frac{1}{2}(1+3w_{\chi,0}) + \frac{3}{2}\Omega_{m,0}w_{\chi,0}
	\simeq 0.5340
	\,.
\end{align}

For a fixed value of the potential parameter $\xi$, Eqs.~\eqref{initial condition values} provide only two conditions for the initial conditions of the trio of dynamical variables $(u,y,z)$.
{
In order to obtain a third condition, we define the parameter $\mathcal{P}_z\equiv z_0^2/(y_0^2+z_0^2)$, the fraction of the energy density of the 3-form at the present time that corresponds to potential energy, and $\mathrm{d}S_n^{(1)}=(\sum_{n=3}^5 (S_n^{(1)}-1)^2)^{1/2}$ as a measure of deviation from $\Lambda$CDM.
While maintaining the constant values $\Omega_{m,0}=0.3065$ and $w_{\chi,0}=-1.006$, we find two sets of initial conditions which minimise $\mathrm{d}S_n^{(1)}$. These correspond to the values $\mathcal{P}_z^{I}\simeq0.02051$ and $\mathcal{P}_z^{II}\simeq0.9721$, with $u_0$ and $y_0$ having the same sign%
 \footnote{We highlight this feature as for a given $(u_0,\,y_0,\,z_0)$ and $(-u_0,\,y_0,\,z_0)$ we get the same value of $\mathcal{P}_z$ but not the same value of $\mathrm{d}S_n$ (cf. the Appendix \ref{ statefinders_uyz} for the dependence of the statefinders in terms of the dynamical variables $(u,\,y,\,z)$).}%
, and verify $\mathrm{d}S_n^{(1)}(\mathcal{P}_z^{I})=0.02526$ and $\mathrm{d}S_n^{(1)}(\mathcal{P}_z^{I})=0.03455$. The present day values of the dynamical variables and respective statefinder parameters for these cases are displayed in Table~\ref{NoInteraction_2minima}.

In order to understand the sensitivity of the cosmological and statefinder parameters with respect to small perturbations of the initial conditions, we expand each parameter $f$ around the solution $(u_0,\, y_0,\,z_0)$ as
\begin{align}
	f(u_0+\delta u,\, y_0+\delta y,\,z_0+\delta_z)\approx &~
		f(u_0,\, y_0,\,z_0)
		+ \sum_{X=u,y,z}
		\left(\partial_X f\right)_{(u_0,\, y_0,\,z_0)}\delta X
		\nn\\
		= &~
		f(u_0,\, y_0,\,z_0)
		\left[
			1
			+\sum_{X=u,y,z}\left(\frac{\delta \log f}{\delta \log X}\right)_{(u_0,\, y_0,\,z_0)} \frac{\delta X}{X_0}
		\right]
		\,.
\end{align}
The higher the absolute value of $(\delta \log f/\log X)$ the more susceptible the parameter is relative to variations of the initial conditions.
As can be seen from Table~\ref{NoInteraction_2minima}, most cosmological parameters appear not to be very sensitive to changes of the initial conditions, indicating that nearby trajectories are also compatible with the current observations. In particular, by numerical investigation we found that as the value of $w_{\chi,0}$ is set closer to $-1$, the deviations in the values of the statefinders approach unity for a higher number of trajectories, as could be expected, and while all of them converge to the points $\pi_0^\pm$ in the past some approach the saddle point $p_0$ in an initial phase of evolution.
}%

For the two sets of initial conditions found previously we plot the trajectories in the 3-dimensional space $(u,\,y,\,z)$ on the left panel of Fig.~\ref{Trajectories}. The evolution of the relative energy densities $\Omega_{m}=s^2$, $\Omega_{\chi}=y^2+z^2$, $\Omega_\mathrm{Kin}=y^2$, and $\Omega_\textrm{Pot}=z^2$ are plotted in Fig.~\ref{Cosmology_NoInteraction} against the evolution of $\Omega_{m}$ and $\Omega_{\Lambda}$ of the $\Lambda$CDM model, while the evolution of the parameters of EoS of DM, DE and of the total fluid are plotted in Fig.~\ref{EoS_NoInteraction}. In the case of $\mathcal{P}_z^{I}$ the system starts from a matter era near the point $\pi_0^+$ and then evolves towards LSBR at $p_1^+$, while mimicking almost perfectly $\Lambda$CDM until the present time, as can be seen on the left panel of Fig.~\ref{Cosmology_NoInteraction}. In contrast, for the case of $\mathcal{P}_z^{II}$ the system also starts from a matter era near the point $\pi_0^+$ but initially evolves towards the fixed point $p_2$ before going to $p_1^+$ at late-time. Notice that in this case the deviations from $\Lambda$CDM become noticeable in the past evolution of $\Omega_m$, $\Omega_\chi$, and $w_\chi$, as can be seen on the right panel of Figs.~\ref{Cosmology_NoInteraction} and \ref{EoS_NoInteraction}. In both cases it appears that at late-time the trajectories approach LSBR fixed point by following the line $y^2+z^2=1$ and $u=(2/\pi)\arctan(y)$.

The preference for the kinetic dominated solution from the cosmological evolution is in concordance with the fact that a massless 3-form (no potential) behaves exactly like a cosmological constant \cite{Koivisto:2009fb}. However, once we go to higher order derivatives of the scale factor, the divergence between the 3-form DE model and a cosmological constant starts to become apparent. This can be seen in the evolution of the statefinder parameters $S_3^{(1)}$, $S_4^{(1)}$, and $S_5^{(1)}$ in Fig.~\ref{ statefinders_evolution} and in the statefinder diagnosis $\{S_3^{(1)},\,S_4^{(1)}\}$ and $\{S_3^{(1)},\,S_5^{(1)}\}$ in Fig~\ref{ statefinders_diagnosis}.

%%%%%%%%%%%%%%%%%%%%%%%%%%%

\begin{table}[t]
\centering
\begin{tabularx}{\textwidth}{ C{.6} C{1.4} C{1} C{1} C{1} C{1} C{1} C{1} C{1} C{1}}
\hline
	{\textbf{Inter.}}
	&
	{$\mathbf{ X}$}
	& 
	{$\mathbf{ S_3^{(1)} }$}
	& 
	{$\mathbf{ S_4^{(1)} }$}
	& 
	{$\mathbf{ S_5^{(1)} }$}
	& 
	{$\mathbf{\Omega_{m}}$}
	& 
	{$\mathbf{ w_{\chi} }$}
	& 
	{$\mathbf{ w_{\chi}^\mathrm{eff} }$}
	& 
	{$\mathbf{ w_{m}}$}
	& 
	{$\mathbf{ w_{m}^\mathrm{eff} }$}
\\ \hline
\footnotesize
%%%%%%%%%%%%% No adjust - IV
	\multirow{ 3}{\hsize}{\centering \textbf{IV}}
	&
	$u_0=0.5155$
	&
	\multirow{ 3}{\hsize}{\centering$0.9183$}
	&
	\multirow{ 3}{\hsize}{\centering$0.9752$}
	&
	\multirow{ 3}{\hsize}{\centering$0.6560$}
	&
	\multirow{ 3}{\hsize}{\centering$0.3065$}
	&
	\multirow{ 3}{\hsize}{\centering$-1.006$}
	&
	\multirow{ 3}{\hsize}{\centering$-0.9761$}
	&
	\multirow{ 3}{\hsize}{\centering$0$}
	&
	\multirow{ 3}{\hsize}{\centering$-0.06788$}
\\
%%%%%%%%%%%%%
	&
	$y_0=0.8223$
\\
%%%%%%%%%%%%%
	&
	$z_0=0.1319$
\\ \hline
%%%%%%%%%%%%% No adjust - V
	\multirow{ 3}{\hsize}{\centering \textbf{V}}
	&
	$u_0=0.5362$
	&
	\multirow{ 3}{\hsize}{\centering$0.9452$}
	&
	\multirow{ 3}{\hsize}{\centering$0.9297$}
	&
	\multirow{ 3}{\hsize}{\centering$0.8061$}
	&
	\multirow{ 3}{\hsize}{\centering$0.3065$}
	&
	\multirow{ 3}{\hsize}{\centering$-1.006$}
	&
	\multirow{ 3}{\hsize}{\centering$-0.9858$}
	&
	\multirow{ 3}{\hsize}{\centering$0$}
	&
	\multirow{ 3}{\hsize}{\centering$-0.04325$}
\\
%%%%%%%%%%%%%
	&
	$y_0=0.8238$
\\
%%%%%%%%%%%%%
	&
	$z_0=0.1218$
\\ \hline
\end{tabularx}
\caption{\label{Interactions_presentday_values}%
The values of the dynamical variables, the statefinder parameters, and the cosmological parameters $\Omega_m$, $w_\chi$, $w_\chi^\mathrm{eff}$, and $w_m^\mathrm{eff}$, obtained for interactions IV and V. We present these values both at $x=0$, the present time in the case of the no interaction, and at the moment when the energy density of DM reaches the value $0.3065$ \cite{Ade:2015xua,Adam:2015rua}. These results were obtained for $\alpha_\chi=\alpha_{\chi\chi}=-0.03$ and setting the initial conditions at redshift 6 using the values of the trajectory $\mathcal{P}_z^I$.}
\end{table}

%%%%%%%%%%%%%%%%%%%%%%%%%%%%%%

%%%%%%%%%%%%%%%%%%%%%%%%%%%%
\begin{figure}[t]
{\includegraphics[width=.48\hsize]{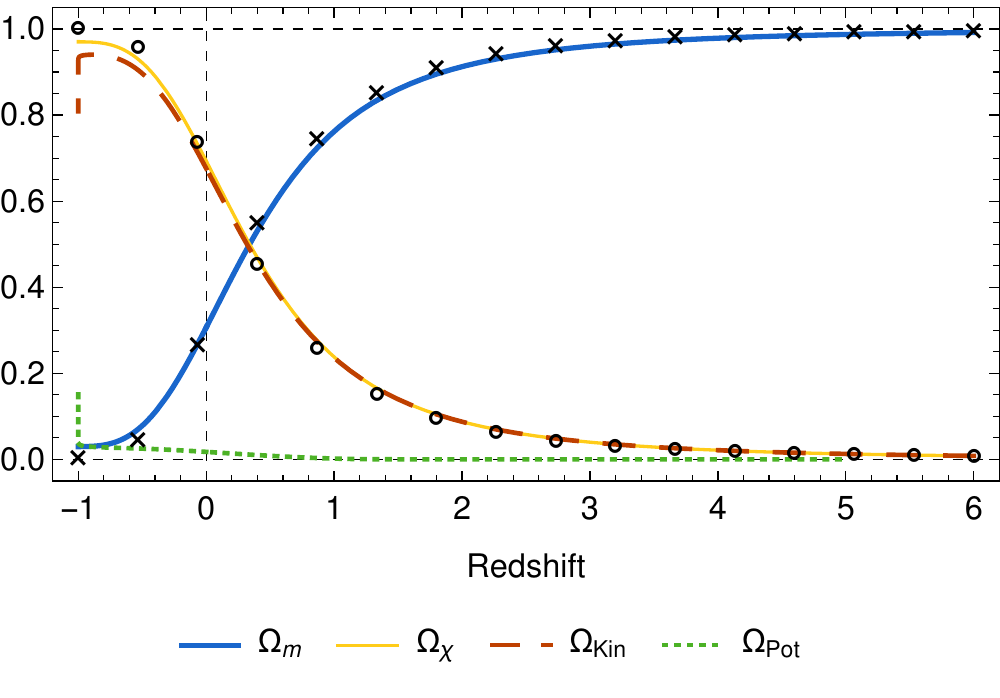}}
\hfill
{\includegraphics[width=.48\hsize]{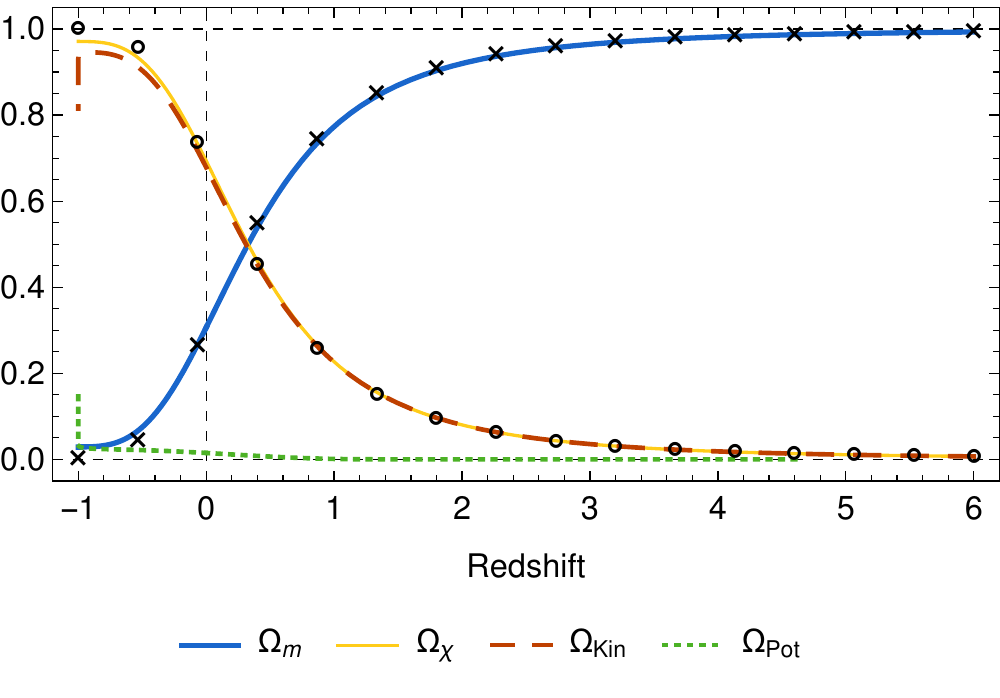}}

\caption{\label{Cosmology_interaction}%
The evolution of the relative energy densities of DM (blue line) and DE (pink line), and of the individual components that contribute to the energy density of DE: kinetic energy (red dashed line) and potential energy (green dotted line), in terms of the redshift. The plot on the left represents the evolution of the system in the case of interaction IV ($\alpha_\chi=-0.03$), while the plot on the right represents the evolution for the case of interaction V ($\alpha_{\chi\chi}=-0.03$). These results were obtained by setting the initial conditions at redshift 6 using the values of the trajectory $\mathcal{P}_z^I$.
For each case the moment of redshift 0 is defined as the moment when $\Omega_m=0.3065$ \cite{Ade:2015xua,Adam:2015rua}. The circles and crosses indicate, respectively, the values of $\Omega_\Lambda$ and $\Omega_m$ for the $\Lambda$CDM model.
}

\end{figure}

%%%%%%%%%%%%%%%%%%%%%%%%%%%%
\begin{figure}[ht]
{\includegraphics[width=.48\hsize]{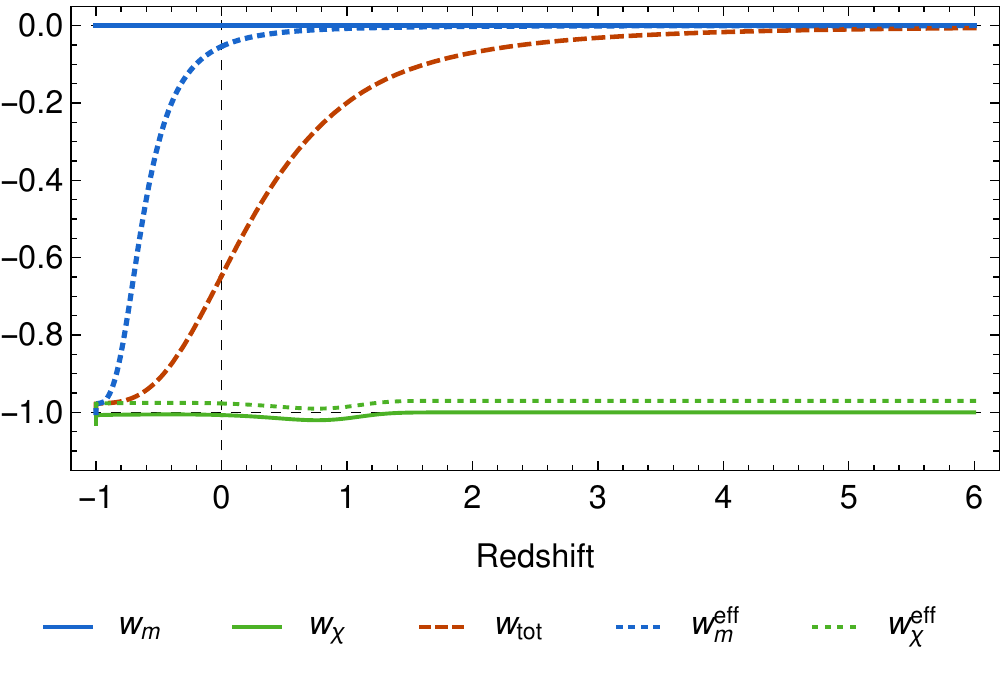}}
\hfill
{\includegraphics[width=.48\hsize]{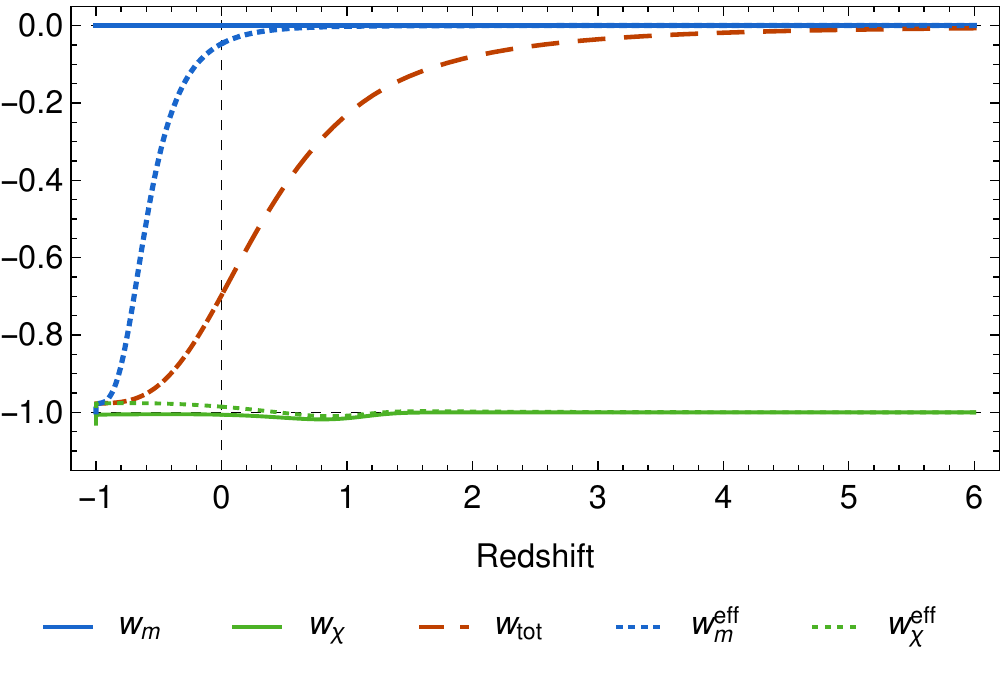}}

\caption{\label{EoS_Interaction}%
The evolution of the parameter of EoS of DM (blue thick line), DE (green thin line), and of the total fluid in terms of the redshift. The plot on the left represents the evolution of the system in the case of interaction IV ($\alpha_\chi=-0.03$), while the plot on the right represents the evolution for the case of interaction V ($\alpha_{\chi\chi}=-0.03$).
}

\end{figure}

%%%%%%%%%%%%%%%%%%%%%%%%%%%%

%%%%%%%%%%%%%%%%%%%%%%%%%%%%%%%%%%%%%%%%
%
%	 statefinder diagnosis for interacting 3-form DE model
%
%%%%%%%%%%%%%%%%%%%%%%%%%%%%%%%%%%%%%%%%

\subsection{Statefinder diagnosis for interacting 3-form DE model}

\label{statefinder_interactions}

%%%%%%%%%%%%%%%%%%%%%%%%%%%%%%
\begin{figure}[t]
{\includegraphics[width=.325\hsize]{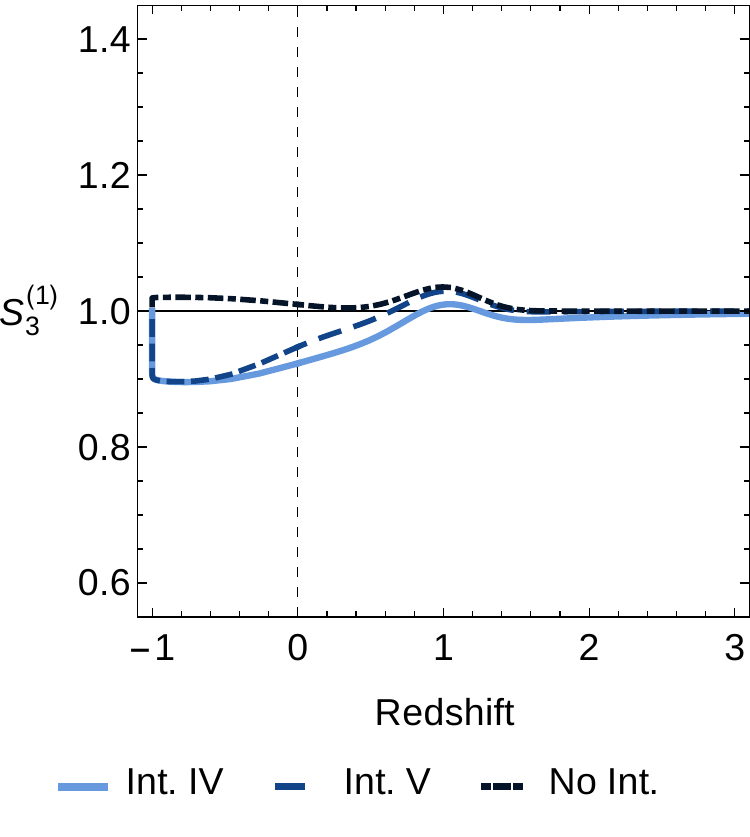}}
\hfill
{\includegraphics[width=.325\hsize]{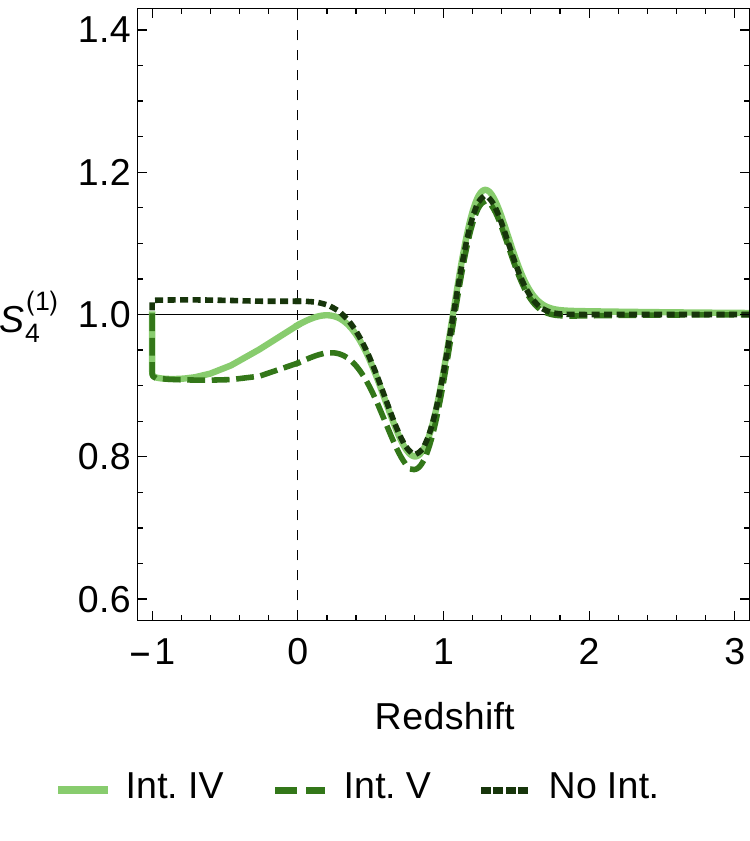}}
\hfill
{\includegraphics[width=.325\hsize]{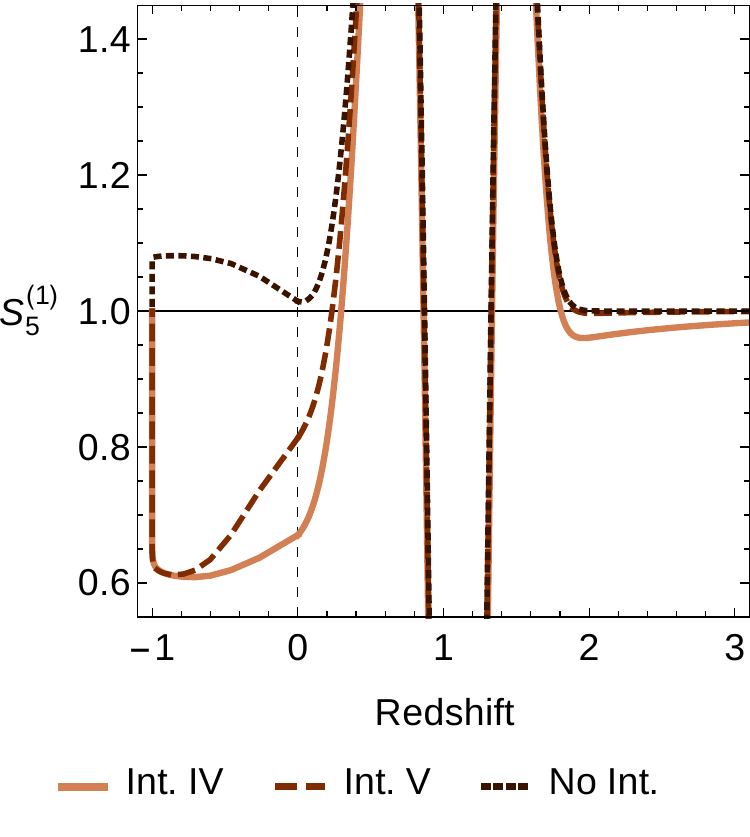}}

\caption{\label{ statefinders_evolution}%
The evolution of the statefinder parameters $S_3^{(1)}$, $S_4^{(1)}$, and $S_5^{(1)}$, in terms of the redshift for the cases of no interaction (dotted line), interaction IV (full line), and interaction V (dashed line). All three case have the same value of the dynamical variables at redshift 6.
}

\end{figure}

%%%%%%%%%%%%%%%%%%%%%%%%%%%%%

%%%%%%%%%%%%%%%%%%%%%%%%%%%%%

\begin{figure}[t]

\includegraphics[width=.48\hsize]{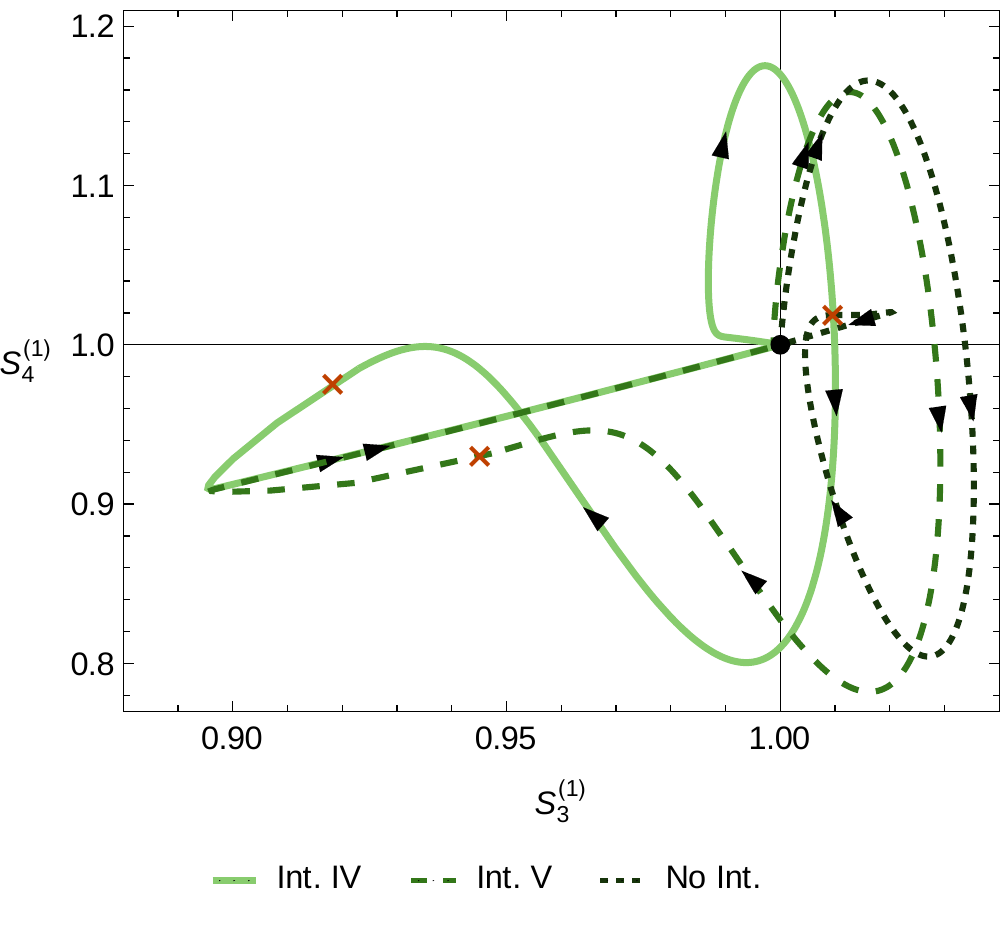}
\hfill
\includegraphics[width=.48\hsize]{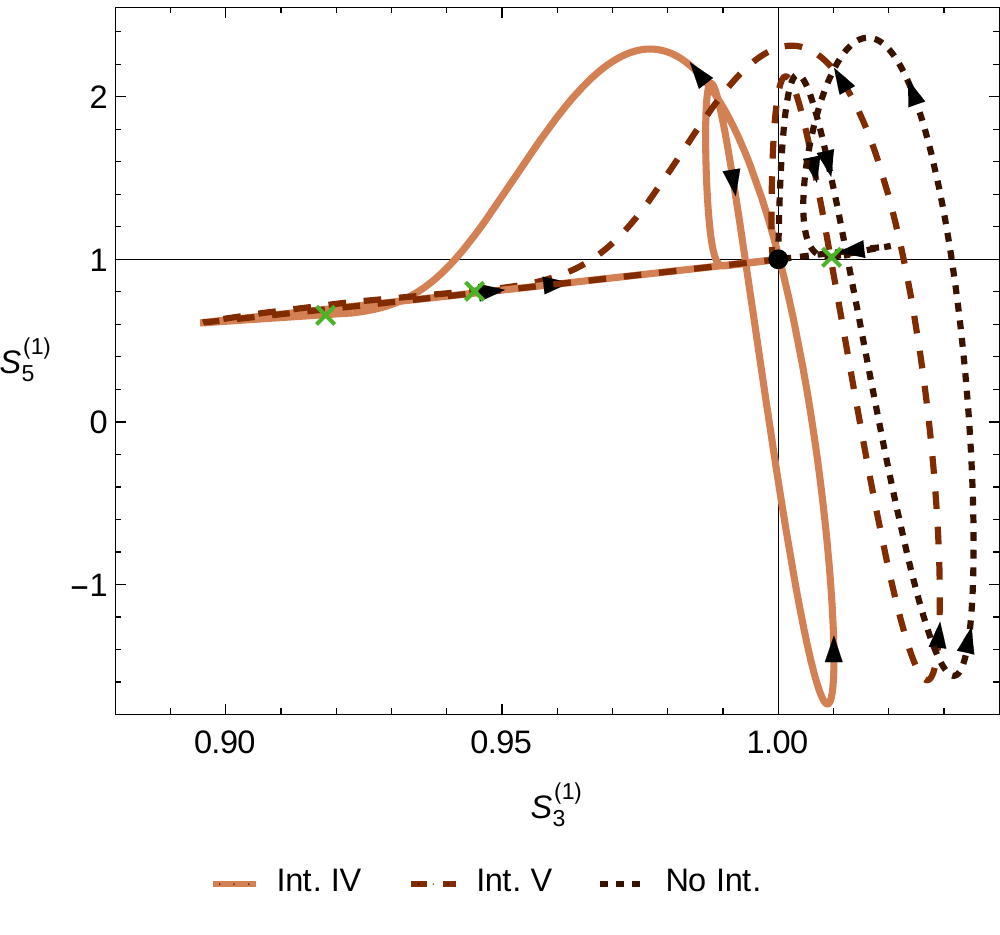}

\caption{\label{ statefinders_diagnosis}%
On this figure we present the statefinder diagnosis $\{S_3^{(1)},\,S_4^{(1)}\}$ (left panel) and $\{S_3^{(1)},\,S_5^{(1)}\}$ (fight panel) for the non-interacting 3-form DE model (dotted line), and when the interactions IV (full line) and V (dashed line) are turned on. The point $\{1,\,1\}$ indicates the $\Lambda$CDM model. The crosses indicate the values of the statefinder parameters at the present time.
}

\end{figure}

%%%%%%%%%%%%%%%%%%%%%%%%%%%%%

We now wish to employ the statefinder Hierarchy to distinguish the different interactions presented in Sect~\ref{Examples}. In particular we will focus our analysis in interactions IV and V as those are the ones that remove LSBR event in the future evolution of the Universe.
Following the results obtained in the previous section, cf. Table~\ref{Table_fixedpoints_II}, both interactions IV and V introduce two late-time attractors, $\tilde{p}_{1}^\pm$, into the system as long as the interaction parameters of each interaction satisfy the inequalities
\begin{align}
	\label{alpha_constraints}
	-\frac{\sqrt{1+2\xi/9}-1}{2} < \alpha_\chi <0
	\,,
	\qquad
	-\frac{\xi}{18}<\alpha_{\chi\chi}<0
	\,.
\end{align}
For the reference value $\xi=1$ that we have considered in the previous section, where we analysed the non-interacting model, these inequalities imply $-5.277\times10^{-2}<\alpha_\chi<0$ and $-5.556\times10^{-2}<\alpha_\chi<0$. As seen in Sect~\ref{Dynsys}, these points correspond to scaling solutions with DE dominance where the Universe enters a de Sitter epoch at late-time. The fraction of the total energy density that corresponds to DE, $\beta_-$, is given for each interaction by Eqs.~\eqref{Case_IV_beta_sol} and \eqref{Case_V_beta_sol}, respectively.

In order to have an imprint of the interaction on the cosmological evolution of the system, we use as initial conditions the values of $(u,\,y,\,z)$ for the trajectory $\mathcal{P}_z^{I}$ taken at redshift $6$ ($x_\mathrm{ini}=-\log7\simeq-1.946$). In the previous section we found that this trajectory is the one with smallest deviation from $\Lambda$CDM while verifying the observational values $\Omega_\mathrm{m}\simeq0.3065$ and $w_d\simeq-1.006$ \cite{Ade:2015xua,Adam:2015rua}. { In order} to compare the results coming from both interactions we choose the same value for both interaction coefficients: $\alpha_\chi=\alpha_{\chi\chi}=-0.03$. This value is sufficiently large, in modulus, for the interactions to have noticeable effects on the cosmological evolution while at the same time not saturating the constraints in \eqref{alpha_constraints}. Furthermore, this value is compatible with the results obtained in Refs.~\cite{Feng:2008fx,Costa:2016tpb}. For such value of the interaction coefficients the value of $\beta_-$ is $0.9709$ for the linear interaction and $0.9717$ for the quadratic one, meaning that in both cases the final state of the Universe will be almost completely dominated by DE, with less than $3\%$ of the total energy density corresponding to DM.

In Table~\ref{Interactions_presentday_values} we present the values of the dynamical system variables and several physical quantities evaluated both at present, $x=0$, which we define as the moment when, for each interaction, $\Omega_{m}$ matches the value $0.3065$ taken from the Planck 2015 data \cite{Ade:2015xua,Adam:2015rua}. Notice that due to the transfer of energy from DE to DM, the energy density of DM takes a longer time to decay than in the non-interacting case, in particular in the case of the linear interaction.
In addition, there is a substantial increase in the deviation to the $\Lambda$CDM model in comparison with the non-interaction case as $\mathrm{d}S_n^{(1)}(\textrm{Inter. IV})=0.3544$ and $\mathrm{d}S_n^{(1)}(\textrm{Inter. V})=0.2134$. Once more the deviation is bigger in the linear case.
This result is a consequence of the fact that, for equal values of $\alpha_\chi$ and $\alpha_{\chi\chi}$, the term $Q$ of the quadratic interaction is suppressed by a factor $\rho_\chi/\rho_\mathrm{tot}$ with respect to the linear case, which is particularly small during the matter dominated era.
For both interactions we plot in Fig.~\ref{Cosmology_interaction} the evolution of the relative energy densities $\Omega_{m}=s^2$, $\Omega_{\chi}=y^2+z^2$, $\Omega_\mathrm{Kin}=y^2$, and $\Omega_\textrm{Pot}=z^2$ against the $\Omega_{m}$ and $\Omega_{\Lambda}$ of $\Lambda$CDM model. In addition, we present in Fig.~\ref{EoS_Interaction} the evolution of the parameters of EoS of DM, DE and of the total fluid, as well as the effective parameters of EoS of DM and DE.

Despite the similarities with the non-interacting case in the cosmological evolution of the relative energy densities, cf. Figs~\ref{Cosmology_NoInteraction} and \ref{Cosmology_interaction}, they can be differentiated once the statefinder parameters are analysed in Figs~\ref{ statefinders_evolution} and \ref{ statefinders_diagnosis}. The differences in the three cases start to become noticeable in the recent past, i.e. for redshift smaller than 1. This coincides with the epoch when the interactions, whose strength depends on the relative energy density of DE, start to become important. One particular difference between the interacting and non-interacting models is that $S_3^{(1)}$ does not go to values smaller than unity in the non-interacting model, while for the two interactions considered it reaches values as low as $\sim0.9$. In the asymptotic de Sitter phase in the future, when $\Omega_\chi\approx1$, both interactions have approximately the same strength and the curves of the statefinders for the two cases become indistinguishable, cf. Figs~\ref{ statefinders_evolution} and \ref{ statefinders_diagnosis}. Nevertheless, at the present time the statefinder hierarchy serves as a good diagnosis to distinguish the different non-interacting and interacting models.

%%%%%%%%%%%%%%%%%%%%%%%%%%%
%
%		Growth rate of matter perturbations
%
%%%%%%%%%%%%%%%%%%%%%%%%%%%

\section{Growth rate of matter perturbations}

\label{perturbations}

\subsection{General framework}

In this section, we consider the theory of linear scalar perturbations around a FLRW background. A bar over a quantity indicates that we are referring to its background value, while perturbations are identified by a $\delta$ before the variable. We will work in the longitudinal (or Newtonian) gauge, where all non-diagonal terms of the metric vanish and the perturbed line element can be written as
\begin{align}
	\label{pert_ds2}
	ds^2 = -\left[1+2\Phi\left(t,\vec{x}\right)\right]dt^2 + a^2(t)\left[1-2\Psi\left(t,\vec{x}\right)\right]d\vec{x}^2
	\,.
\end{align}
Here, $\Phi$ and $\Psi$ are the gauge invariant Bardeen potentials \cite{Bardeen:1980kt}. Since neither DM nor the 3-form field introduces anisotropies at the linear level, the off-diagonal spatial components of Einstein equations imply the equality between the two potentials. In what follows, we will assume that equality.

The perturbation of the Einstein equations reads \cite{Malik:2004tf}
\begin{align}
	\label{pert_Gmunu}
	 \delta {G^\mu}_\nu = \kappa^2 \delta {T^\mu}_\nu
	 \,,
\end{align}
where $ \delta {G^\mu}_\nu$ is the perturbation of the Einstein tensor and $ \delta {T^\mu}_\nu$ is the perturbation of the energy-momentum tensor. Using Eq.~\eqref{pert_ds2} we can calculate the $(0-0)$, $(0-i)$, and $(i-i)$ components of $ \delta {G^\mu}_\nu$ and decompose Eq.\eqref{pert_Gmunu} as \cite{Malik:2004tf,Valiviita:2008iv}
\begin{align}
	\label{Pert_Einstein_00}
	3H\left(\dot\Phi + H\Phi\right) - \frac{\vec\nabla^2}{a^2}\Phi =&~ \frac{\kappa^2}{2} \delta {T^0}_0
	\,,
	\\
	\label{Pert_Einstein_0i}
	-\partial_i\left(\dot\Phi + H\Phi\right) =&~ \frac{\kappa^2}{2} \delta {T^0}_i
	\,,
	\\
	\label{Pert_Einstein_ii}
	\ddot\Phi 
	+4H\dot\Phi
	+\left(2\dot H + 3H^2\right)\Phi
	=&~
	\frac{\kappa^2}{6} \delta {T^i}_i
	\,,
\end{align}
where the Laplacian operator is defined as $\vec\nabla^2\equiv \delta^{ij}\partial_i\partial_j$. These equations define $\dot\Phi$ and $\Phi$ in terms of the perturbations of the energy-momentum tensor.

In the matter sector, at the perturbative level we assume that DE is smooth at all relevant scales, so that only perturbations of DM need to be considered. As a consequence, any energy transfer between DM and DE at the perturbative level is likewise disregarded. This approximation is validated by the fact that at early times the potential of the 3-form and the interactions (IV) and (V) that we are considering vanish, therefore the 3-form field behaves essentially as a cosmological constant. The perturbations on the right hand side of equations \eqref{Pert_Einstein_00}, \eqref{Pert_Einstein_0i}, and \eqref{Pert_Einstein_ii} can then be written as
\begin{align}
	\delta{T^0}_0^{}= -\bar\rho_m\delta_m
	\,,
	\qquad
	\delta{T^0}_i^{}=\bar\rho_m \partial_iv_m
	\,,
	\qquad
	\delta{T^i}_i^{}=0
	\,,
\end{align}
where $\delta_m$ is the fractional energy density perturbation and $v_m$ is the peculiar velocity potential \cite{Malik:2004tf,Valiviita:2008iv}. The evolution equations for $\delta_m$ and $v_m$ are obtained from the perturbed equations for the conservation of the energy-momentum tensor and read 
\begin{align}
	\label{deltam_evol}
	\dot\delta_m
	+ \frac{\vec\nabla^2}{a^2}v_m
	- 3\dot{\Phi} 
	=&~0
	\,,
	\\
	\label{Um_evol}
	\dot v_m
	+ \Phi 
	=&~ 0
	\,.
\end{align}

After a Fourier decomposition of the perturbation variables:
\begin{align}
	\delta_m\left(t,\vec x\right) =&~ \frac{1}{\left(2\pi\right)^{3/2}}\int \mathrm{d}^3\vec k \,\delta_m^{(k)}\left(t\right) e^{-i \vec k\cdot\vec x}
	\,,
	\\
	v_m\left(t,\vec x\right) =&~ \frac{1}{\left(2\pi\right)^{3/2}}\int \mathrm{d}^3\vec k \,v_m^{(k)}\left(t\right) e^{-i \vec k\cdot\vec x}
	\,,
	\\
	\Phi\left(t,\vec x\right) =&~ \frac{1}{\left(2\pi\right)^{3/2}}\int \mathrm{d}^3\vec k \,\Phi^{(k)}\left(t\right) e^{-i \vec k\cdot\vec x}
	\,,
\end{align}
we can combine Eqs.~\eqref{Pert_Einstein_00}, \eqref{Pert_Einstein_0i}, \eqref{Pert_Einstein_ii}, \eqref{deltam_evol}, and \eqref{Um_evol} to obtain a second order differential equation for $\delta_m^{(k)}$ \cite{delaCruzDombriz:2008cp}
\begin{align}
	\label{delta_m_total}
	\ddot{\delta}_m^{(k)} &~
	+ H\frac{4k^4 + 9\kappa^2a^2\rho_m \left[k^2 + a^2\left(H^2-\dot H\right)\right]}{2k^4 + 3\kappa^2a^2\rho_m\left(k^2 + 3a^2H^2\right)}\dot{\delta}_m ^{(k)}
	\nn\\&~
	-\kappa^2\rho_m\frac{2k^4 + 3k^2a^2\left(\kappa^2\rho_m + 2H^2 + 4\dot H\right) + 9\kappa^2a^4\rho_m\left(2\dot H + 3H^2\right) }{4k^4 + 6\kappa^2a^2\rho_m\left(k^2 + 3a^2h^2\right)}\delta_m^{(k)}=0
\end{align}

During the matter era, all relevant scales are much smaller than the Hubble horizon, i.e. $a^2H^2/k^2\ll1$. Within this approximation, in Eq.~\eqref{delta_m_total} we can disregard all but the leading terms in $k^2$, obtaining \cite{Linder:2003dr,Yin:2015pqa}
\begin{align}
	\label{matter_evolution_equation}
	\ddot\delta_m + 2H\dot\delta_m - \frac{\kappa^2}{2}\rho_m \delta_m =0
	\,.
\end{align}
Changing to $x$ as the time variable, and making use Eq.~\eqref{h1h}, we can re-write the previous equation in terms of the dynamical variables as
\begin{align}
	\label{final_evolution_equation}
	\delta_m'' + \frac{1}{2}\left[
		1 
		+3 \left(y^{2}+z^2\right) 
		+\frac{2}{3}\xi\, z^{2}\tan^2\left(\frac{\pi}{2}u\right)
	\right]\delta_m' - \frac{3}{2}\left[1-\left(y^2+z^2\right)\right] \delta_m = 0
	\,.
\end{align}

\subsection{The Growth Rate and the Composite Null Diagnosis}

%%%%%%%%%%%%%%%%%%%%%%%%%%%%
\begin{figure}[t]
\includegraphics[width=.48\hsize]{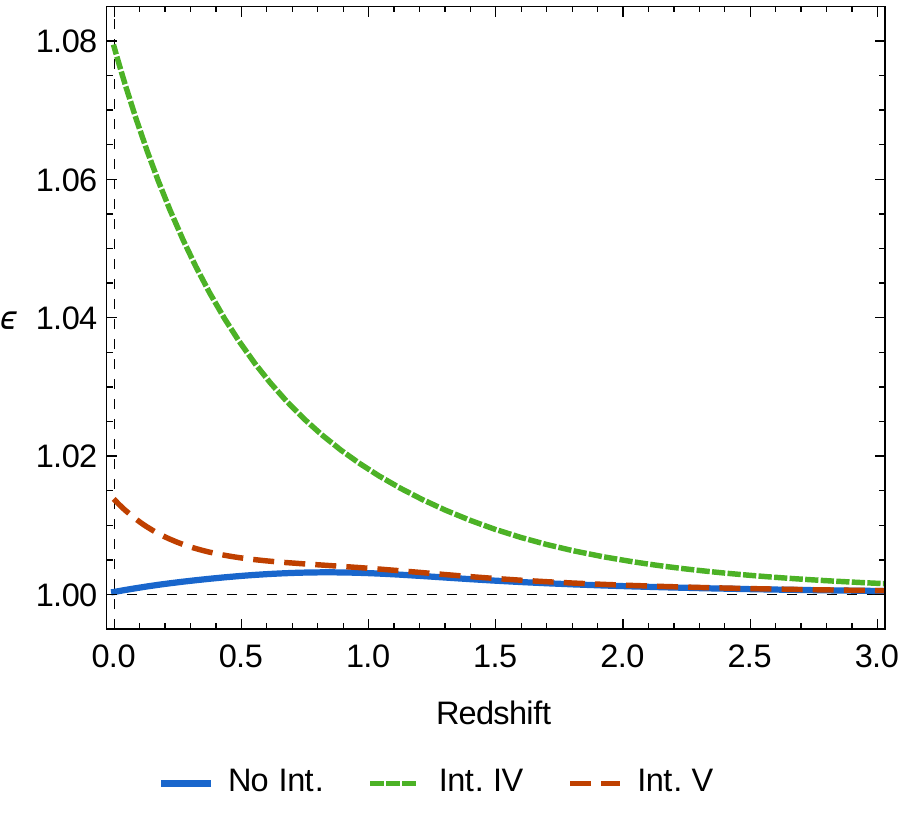}
\hfill
\includegraphics[width=.48\hsize]{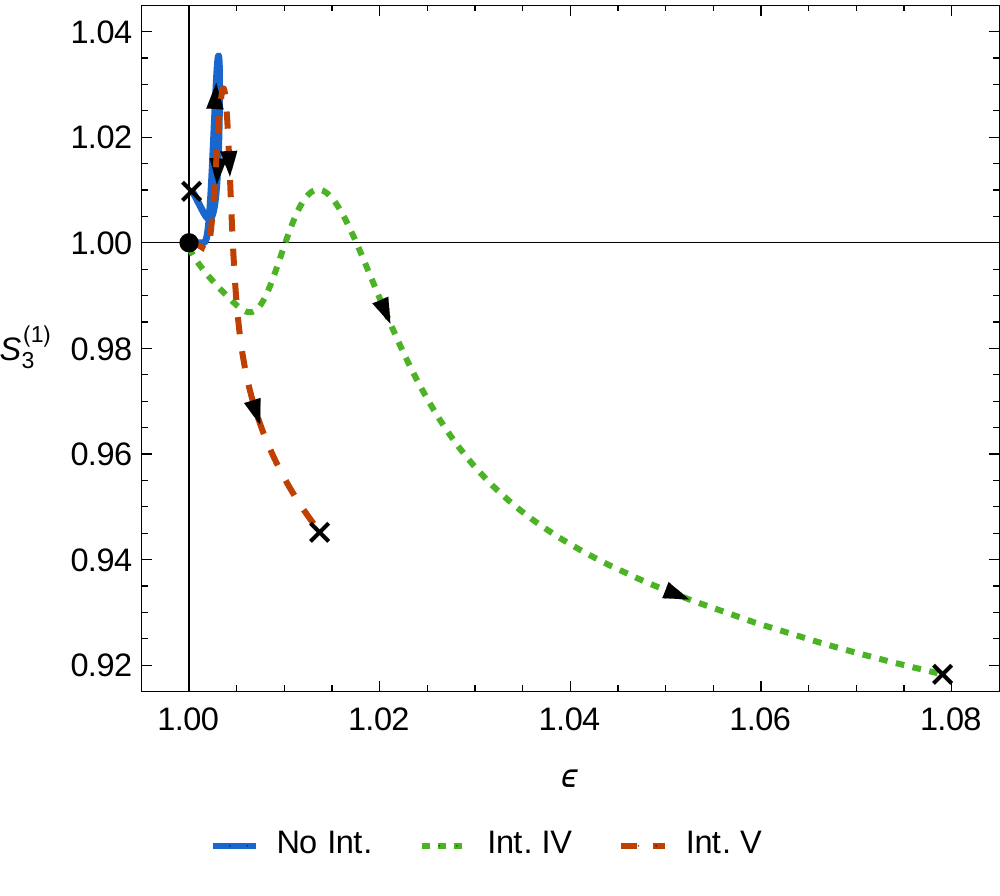}
\caption{\label{Perturbations}%
On the left panel we present the evolution of the fractional growth factor $\epsilon$ in terms of the redshift for the model with no interaction (solid blue line ) and for the models with the interactions IV (green dotted line) and V (red dashed line). On the right panel we present the CND $\{\epsilon,\,S_3^{(1)}\}$for the same models. The point $\{1,\,1\}$ indicates the $\Lambda$CDM model. The crosses indicate the values of the statefinder parameters at the present time.
}

\end{figure}

%%%%%%%%%%%%%%%%%%%%%%%%%%%

The evolution of $\delta\rho_m$ in any given DE model can be compared with the one in $\Lambda$CDM using the so called composite null diagnosis (CND) \cite{Yin:2015pqa}. This diagnosis maps the statefinders $S_n^{(1)}$ against the growth factor $\epsilon(x)$ 
\begin{align}
	\epsilon(x) \equiv \frac{f(x)}{f_{\Lambda \mathrm{CDM}}(x)}
	\,,
\end{align}
where the growth rate of structure $f(x)$ is defined as $f(x)\equiv (\partial \delta_m/\partial x)/\delta_m$.
For this section, we compute the matter perturbations for the interacting 3-form DE model and use
the CND $\{\epsilon,\,S_{3}^{(1)}\} $ \cite{Yin:2015pqa} to look for deviations of the model from $\Lambda$CDM. Notice that by construction, and similarly to what happens in the statefinder hierarchy, the $\Lambda$CDM model corresponds to the point $\{1,1\}$ in the CND mapping.

In order to compute the evolution of the growth factor for the models considered in the previous section, we integrate Eq.~\eqref{final_evolution_equation}, where the values of background quantities are given by the respective numerical solutions obtained in Sects.~\ref{statefinder_nointeraction} and \ref{statefinder_interactions}. The initial conditions are set at redshift 6, when the Universe is well inside the matter era and all relevant modes are inside the horizon. We assume that initially $\delta_m$ grows linearly with the scale factor.

The results obtained for the cosmological evolution of $\epsilon$ are presented on the left panel of Fig.~\eqref{Perturbations}, while the CND is presented on the right panel of the same figure. It can be seen that though the deviations from $\Lambda$CDM of the growth rate in the non-interacting case are small (less than 1\%), in the models with interactions IV and V the growth rate becomes increasingly high as the Universe evolves. This difference in the growth pattern near the present time is interpreted as a consequence of the increasing strength of the interactions at play. As the fraction of DE becomes higher, the interactions become noticeable and the behaviour of the background quantities in Eq.~\eqref{final_evolution_equation} in the interacting and non-interacting cases begins to differ. This leads to the different late-time evolution of the linear perturbations and to the distinct CND profiles observed on the two panels of Fig.~\eqref{Perturbations}. Consequently, we find that the CND can positively distinguish the three models from $\Lambda$CDM and in between themselves.
To conclude this section we mention the three points obtained from the SDSS III BOSS DR12 data \cite{Satpathy:2016tct}: $f(z_\textrm{eff}=0.38)=0.638\pm0.080$, $f(z_\textrm{eff}=0.51)=0.715\pm0.090$, and $f(z_\textrm{eff}=0.61)=0.753\pm0.088$. For all the values of redshift considered, the deviation of $\epsilon$ from unity is within the $5\%$ margin for all the three models, while the $1\sigma$ error for the measurements of $f$ is above the $10\%$ mark. This means that our results are compatible with the SDSS III results  \cite{Satpathy:2016tct} and corroborates the choice made in Sect.~\ref{statefinder_interactions} for the value of the interaction parameters $\alpha_\chi$ and $\alpha_{\chi\chi}$. A more thorough comparison of the theoretical predictions for the growth rate in this type of models will be presented elsewhere.

%%%%%%%%%%%%%%%%%%%%%%%%%%%%
\section{Conclusions}

\label{Conclusions}

 In this work we revisited the role of a 3-form field minimally coupled to gravity as a DE source to explain the current observed acceleration phase of the Universe expansion. 

We started by considering the simplest case, where the matter content of the model also includes DM in a non-interacting scenario with DE, and showed that the system could evolve towards a future abrupt event, namely the Little Sibling of the Big Rip (LSBR) \cite{Bouhmadi-Lopez:2014cca}. We selected a Gaussian self-interacting potential for the 3-form field, which embodies the correct behaviour in order to avoid ghost and Laplacian instabilities and proposed a general expression for the DM-DE interaction. The choice of this potential allowed us to obtain quantitative results without compromising the ability to draw more general considerations applicable to a broader class of suitable potentials. The precise form of this interaction finds part for its motivation in the generalisation of some similar cases studied in the past but also in the fact that it enables a simple mathematical translation of the problem in a dynamical system context. 
 
One of the main questions addressed in this work was which kind of interaction, extracted from the herein proposed general expression, cf. Eqs.~\eqref{Int} and \eqref{quadratic_int}, between DM and DE would avoid a future LSBR. The extensive dynamical analysis carried for the linear, quadratic and mixed DM-DE interactions, through the identification of fixed points and their corresponding stability, enabled us to conclude that only interactions that do not involve a DM dependence effectively avoid the evolution towards a LSBR, replacing it by a de Sitter inflationary era. In addition a class of strongly repulsive fixed points, corresponding to a past DM era that emerged in all the cases (with or without interaction), were identified in this work. This new class of fixed points inhabit regions of the dynamical phase space where the 3-form field is infinite and, despite the characterization given here, their special properties need a more detailed mathematical study to be carried in a future work \cite{ Bouhmadi:2016}.

Subsequently, we directed our attention on how to observationally distinguish the linear and quadratic DE interaction (the only cases were the LSBR is avoided). In order to tackle this problem we have applied the statefinder hierarchy and computed the growth rate of matter perturbations to distinguish between the aforementioned DE interactions, the non-interacting case and $\Lambda$CDM model. Particular attention was given to the statefinder diagnosis $\{S_3^{(1)},\,S_4^{(1)}\}$, $\{S_3^{(1)},\,S_5^{(1)}\}$ and to the composite null diagnostic CND $\{\epsilon,\,S_{3}^{(1)}\} $ as adequate tools that allowed to pinpoint relevant differences between the cases under scope. More importantly, we found sufficient evidences to discriminate, at present, between the linear and quadratic DE interaction. Moreover, the fact that the ongoing transfer of energy from DE to DM implies a longer time to decay for the energy density of DM, in the linear DE interaction case, explains the difference found. For all the models considered the results for the growth rate of the matter perturbations are within the observational constraints of the SDSS III data \cite{Satpathy:2016tct}.

Despite the strategy followed in this work, to focus our attention on the DM-DE interactions suitable to classically avoid the LSBR, the mixed DM-DE and exclusively DM dependent interactions are not simply dismissed as valid scenarios. The expected evolution towards a LSBR, in those cases, might imply the need to consider quantum corrections near the singular event \cite{Albarran:2015cda}. It would also be of interest to investigate the implications of these interactions between the 3-form DE and DM within a quantum field theoretical setting \cite{D'Amico:2016kqm,Marsh:2016ynw}. Nevertheless, this program, rising several new questions on its own, should be left for future appraisal.
 
%%%%%%%%%%%%%%%%%%%%%%%%%%%%

\section{Acknowledgements}

The Authors are grateful to Juan~M.~Aguirregabiria and C\'esar~Silva for enlightening discussions on dynamical system analysis.
The Authors also acknowledge Nelson~Nunes for helping us to understand several issues related to 3-forms.
The work of MBL is supported by the Portuguese Agency “Funda\c{c}\~ao para a Ci\^encia e Tecnologia” through an Investigador FCT Research contract, with reference IF/01442/2013/ CP1196/CT0001. She also wishes to acknowledge the partial support from the Basque government Grant No. IT592-13 (Spain) and FONDOS FEDER under grant FIS2014-57956-P (Spanish government).
JMorais is thankful to UPV/EHU for a PhD fellowship and UBI for hospitality during the completion of part of this work and acknowledges the support from the Basque government Grant No. IT592-13 (Spain) and FONDOS FEDER under grant FIS2014-57956-P (Spanish Government).
SK acknowledges for the support of grant SFRH/BD/51980/2012 from Portuguese ``Funda\c{c}\~ao para a Ci\^encia e Tecnologia''and is thankful to the hospitality of University of Basque country where part of this work was carried out.
YT wishes to acknowledge the financial support from INEF (Iran). He also thanks the Brazilian agencies CAPES and FAPES for partial financial support.
This research work is supported by the grant UID/MAT/00212/2013. The authors acknowledge the COST Action CA15117 (CANTATA).

%%%%%%%%%%%%%%%%%%%%%%%%%%%%%%%%%%%%%%%%%%%%%%%%%%%%
%
% 	Appendix
%
%%%%%%%%%%%%%%%%%%%%%%%%%%%%%%%%%%%%%%%%%%%%%%%%%%%%

\appendix

\section{ Statefinders as functions of $(u,\,y,\,z)$}
\label{ statefinders_uyz}

On this appendix we present the analytical expressions for the statefinder parameters in terms of the dynamical variables for the model presented in Sect~\ref{interacting 3form}.

\begin{align}
	\label{S3_of_uyz}
	S_3^{(1)} =&~ 1 + \xi z^2
	\left\{
		\tan\left(\frac{\pi}{2}u\right)y
		\left[
			2 
			- \frac{2\xi}{9}\tan^2\left(\frac{\pi}{2}u\right)
		\right]
		- \tan^2\left(\frac{\pi}{2}u\right)
		\left[
			1
			- \frac{2\xi}{9}\tan^2\left(\frac{\pi}{2}u\right)
		\right]
	\right\}
	\nn\\&~
	+\frac{9}{2}\left[
		\alpha_0
		+\alpha_1\left(y^2+z^2\right)
		+\alpha_2\left(y^2+z^2\right)^2 
	\right]
	\,.
\end{align}
\begin{align}
	\label{S4_of_uyz}
	S_4^{(1)} =&~ 1 
	+\frac{\xi z^2}{54}\bigg\{
		y^2\left[
			324
			-9\left(9+20\xi\right)\tan^2\left(\frac{\pi}{2}u\right)
			+2\xi\left(9+4\xi\right)\tan^4\left(\frac{\pi}{2}u\right)
		\right]
	\bigg.
	\nn\\&~~~~~~~~~~~
	\bigg.
		-4\tan\left(\frac{\pi}{2}u\right)y
		\left[
			135
			-87\xi \tan^2\left(\frac{\pi}{2}u\right)
			+4\xi^2\tan^4\left(\frac{\pi}{2}u\right)
		\right]
	\bigg.
	\nn\\&~~~~~~~~~~~
	\bigg.
		+\tan^2\left(\frac{\pi}{2}u\right)
		\left[
			351
			-186\xi \tan^2\left(\frac{\pi}{2}u\right)
			+8\xi^2\tan^4\left(\frac{\pi}{2}u\right)
			-\left(
				81-36\xi
				+4\xi^2\tan^2\left(\frac{\pi}{2}u\right)
				-4\xi^2\tan^4\left(\frac{\pi}{2}u\right)
			\right)z^2
		\right]
	\bigg\}
	\nn\\&~
	+\frac{1}{12y}
	\bigg\{
		\alpha_0\left[
			27y\left(-7+9y^2\right)
			+\left(
				2\xi\left[18-(9+\xi)\tan^2\left(\frac{\pi}{2}u\right)\right] \tan\left(\frac{\pi}{2}u\right)
				+27\left[9+2\xi\tan^2\left(\frac{\pi}{2}u\right)\right]y
			\right)z^2
		\right]
	\bigg.
	\nn\\&~~~~~~~~~~
	\bigg.
		+\alpha_1
		\Big[
			27y^3\left(-1+3y^2\right)
		\Big.
	\bigg.
	\nn\\&~~~~~~~~~~~~~~~~
	\bigg.
		\Big.
			+y
			\left\{
				9\left[
					-3
					+4\xi\tan^2\left(\frac{\pi}{2}u\right)
				\right]
				+2y\xi
				\left[
					18 
					- \left(9+\xi\right) \tan^2\left(\frac{\pi}{2}u\right)
				\right]\tan\left(\frac{\pi}{2}u\right)
				+18\left[
					9 
					+ \xi\tan^2\left(\frac{\pi}{2}u\right)
				\right]y^2
			\right\}z^2
		\Big.
	\bigg.
	\nn\\&~~~~~~~~~~~~~~~~
	\bigg.
		\Big.
			+\left\{
				2\xi\left[18 - \left(9+\xi\right)\tan^2\left(\frac{\pi}{2}u\right)\right]\tan\left(\frac{\pi}{2}u\right)
				+9\left[9+2\xi\tan^2\left(\frac{\pi}{2}u\right)\right]y
			\right\}z^4
		\Big]
	\bigg.
	\nn\\&~~~~~~~~~~
	\bigg.
		+\alpha_2
		\Big[
			27y^5\left(5-3y^2\right)
		\Big.
	\bigg.
	\nn\\&~~~~~~~~~~~~~~~~
	\bigg.
		\Big.
			+y^3
			\left\{
				270+72\xi\tan^2\left(\frac{\pi}{2}u\right)
				-4\xi\left[
					-9 + \xi\tan^2\left(\frac{\pi}{2}u\right)
				\right]\tan\left(\frac{\pi}{2}u\right)y
				-9\left[
					27 
					+ 2 \xi\tan^2\left(\frac{\pi}{2}u\right)
				\right]y^2
			\right\}z^2
		\Big.
	\bigg.
	\nn\\&~~~~~~~~~~~~~~~~
	\bigg.
		\Big.
			+\left\{
				9\left[
					15+8\xi\tan^2\left(\frac{\pi}{2}u\right)
				\right]
				+4\xi\left[
					18 - \left(9+\xi\right)\tan^2\left(\frac{\pi}{2}u\right)
				\right]\tan\left(\frac{\pi}{2}u\right)y
				-9\left[
					27
					+4\xi\tan^2\left(\frac{\pi}{2}u\right)
				\right]y^2
			\right\}z^4
		\Big.
	\bigg.
	\nn\\&~~~~~~~~~~~~~~~~
	\bigg.
		\Big.
			+\left\{
				2\xi\left[
					18 - \left(9+\xi\right)\tan^2\left(\frac{\pi}{2}u\right)
				\right]\tan\left(\frac{\pi}{2}u\right)
				-9\left[9+2\xi\tan^2\left(\frac{\pi}{2}u\right)\right]y
			\right\}z^6
		\Big]
	\bigg\}
	\nn\\&~
	+\frac{27}{2}
	\left[\alpha_1 + 2\alpha_2\left(y^2+z^2\right)\right]
	\left[\alpha_0 + \alpha_1\left(y^2+z^2\right) + \alpha_2\left(y^2+z^2\right)^2\right]
	\,.
\end{align}
For simplicity, we omit writing explicitly the expression for $S_5^{(1)}$ as it is even larger than the one found for $S_4^{(1)}$.

\section{Hurwitz criterion for cubic Polynomials}

\label{AppendixB}

Let $\mathcal{P}_3(z)$ be a polynomial of degree $3$ on $z$ with real coefficients $a_i$ ($i=0,1,\dots,3$) and $a_3\neq0$:
\begin{align}
	\mathcal{P}_3(z) = a_0 + a_1 z + a_2 z^2 + a_3 z^3
	\,.
\end{align}
 According to Hurwitz criterion \cite{Olver2010a}, all roots of $\mathcal{P}_3(z) $ have negative real parts if and only if for
	\begin{align}
		D_1 = a_1
		\,,
		\qquad
		D_2 = 
		\begin{vmatrix}
		a_1 & a_3\\
		a_0 & a_2
		\end{vmatrix}
		\,,
		\qquad
		D_3 = 
		\begin{vmatrix}
		a_1 & a_3 & 0\\
		a_0 & a_2 & 0\\
		0 & a_1 & a_3
		\end{vmatrix}
		\,,
	\end{align}
	we have $a_0\neq0$, $D_{2}>0$ and $\sign\, D_{1}=\sign\, D_{3}=\sign\, a_0$.

Let us now consider an autonomous dynamical system
\begin{align}
	\vec{\bf x}'
	=
	\vec{\bf f}\left(\vec{\bf z}\right)
	\,,
\end{align}
with at least a fixed point at $\vec{\bf x}=\vec{\bf x}_{fp}$. Let $J\equiv \nabla\cdot\vec{\bf f}$ be the Jacobian of the system and evaluated at the fixed point $\vec{\bf x}_{fp}$. The characteristic polynomial of $J$, $p_J(\gamma)$, is defined by
\begin{align}
	p_J(\gamma) = \det \left(J-\gamma \mathbb{I}_3\right) = a_0 + a_1 \gamma + a_2 \gamma^2 - \gamma^3
	\,,
\end{align}
where $ \mathbb{I}_3$ is the $3\times3$ identity matrix.
Following Hurwitz's criterion, we find that $p_J$ is stable if
\begin{align}
	a_0\neq0
	\,,
	\qquad
	\sign\,a_1=\sign\,a_0
	\,,
	\qquad
	a_1a_2+a_0 >0
	\,.
\end{align}

\bibliography{References}

\bibliographystyle{utphys}

\end{document}